\documentclass[11pt]{article}
\usepackage[british,UKenglish,USenglish,english,american]{babel}
\usepackage[utf8]{inputenc}
\usepackage{latexsym}
\usepackage{graphicx}
\usepackage{amsfonts}
\usepackage{amssymb}
\usepackage{amsmath}
\usepackage{mathrsfs}
\usepackage{dsfont}
\usepackage{amsthm}
\usepackage{mdframed}
\usepackage{lipsum}
\usepackage{bbm}
\usepackage{color}
\usepackage{capt-of}
\usepackage[format=hang,font=small,labelfont=bf]{caption}
\usepackage{booktabs}
\usepackage{xcolor}
\usepackage{hyperref}

\usepackage{subcaption}
\newtheorem{Prop}{Proposition}
\newtheorem{Lem}{Lemma}
\numberwithin{equation}{section}

\makeatletter
\newcommand*{\rom}[1]{\expandafter\@slowromancap\romannumeral #1@}
\makeatother
\title{Valuing Tradeability in Exponential Lévy Models}
\author{\sc \Large Ludovic Mathys\footnote{Email: ludovic.mathys@bf.uzh.ch} \vspace{0.5em} \\
       {\it Department of Banking and Finance, University of Zurich, Switzerland.}\vspace{0.5em}
       }
\date{}
\providecommand{\keywords}[1]{\textbf{Keywords:} #1}
\providecommand{\mscclass}[2]{\textbf{MSC (2010) Classification:} #1}
\providecommand{\jelclass}[3]{\textbf{JEL Classification:} #1}
\providecommand{\acknow}[4]{\textbf{Acknowledgements:} #1}

\begin{document}

\maketitle

\thispagestyle{empty}
% The abstract

\begin{abstract}
\noindent The present article provides a novel theoretical way to evaluate tradeability in markets of ordinary exponential Lévy type. We consider non-tradeability as a particular type of market illiquidity and investigate its impact on the price of the assets. Starting from an adaption of the continuous-time optional asset replacement problem initiated by McDonald and Siegel (cf.~\cite{ms}), we derive tradeability premiums and subsequently characterize them in terms of free-boundary problems. This provides a simple way to compute non-tradeability values, e.g.~by means of standard numerical techniques, and, in particular, to express the price of a non-tradeable asset as a percentage of the price of a tradeable equivalent. Our approach is illustrated via numerical examples where we discuss various properties of the tradeability premiums.
\end{abstract}
$\;$ \vspace{2.3em} \\
\noindent \keywords{Tradeability, Liquidity, Exponential Lévy Processes, Real Options, Maturity-Randomization, Optimal Stopping, Free-Boundary Problems.} \vspace{0.5em} \\ % Write down at least 3 Keywords
\noindent \mscclass{91B70, 91G20, 91G80.}{} \vspace{0.5em}\\
\noindent \jelclass{C32, G12, G13.} \vspace{-0.5em} \\
\section{Introduction}
Market liquidity and related risks have played an important role since the emergence of financial markets and their relevance for various types of financial activities has been noticed by academics since many years. Recently, the financial crisis has made it again clear how valuable and important market liquidity is. While trading costs rose for many assets dramatically, other assets could not be even traded for several months. Under such circumstances, liquidating an open position either became prohibitively expensive or was just impossible so that many investors were forced to sit on their positions and accumulated losses. In view of these incidents, it is not surprising that investors apprehend liquidity-related issues and usually demand a price discount when purchasing illiquid assets. This behavior is well documented by a vast body of empirical literature that started with the seminal articles of Amihud and Mendelson (cf.~\cite{am86} and \cite{am89}). \vspace{1em} \\
\noindent In the literature, market liquidity usually either refers to the possibility to sell and buy -- thus just to trade -- financial assets on their respective markets or to the ability to trade them without initiating significant changes on the market. Although these two concepts are quite close to each other, there is an essential difference between them. While the first view merely understands liquidity in the sense of absolute tradeability, the second approach includes the effects that trading may trigger on the markets. For this reason, considering liquidity in the sense of the second approach generally offers more modeling flexibility than focussing on the first view. This could possibly explain why only little theoretical work analyzes the impact of non-tradeability on asset prices.\footnote{cf.~\cite{ls95}, \cite{lo17} and \cite{ck12} for examples of articles tackling these issues and additional explanations on the challenges encountered when modeling non-tradeability.}~Indeed, despite the importance of non-tradeability issues, most theoretical models focus on the second view and capture (il-)liquidity by modeling the costs associated with trading the assets. Examples include the financial economics models\footnote{cf.~\cite{am05} for a survey of this literature.} of Amihud and Mendelson (cf.~\cite{am86}) and Acharya and Pedersen (cf.~\cite{ap05}) as well as many articles in the mathematical literature on liquidity, such as \cite{ja94}, \cite{cj04} and \cite{cr07} just to name a few.\footnote{cf.~\cite{so11} for a survey of the mathematical literature on liquidity.}~In addition to the scarcity of the literature on tradeability, theoretical articles dealing with non-tradeability issues mostly derive premiums based on optimal selling strategies and could, therefore, only offer limited explanations for the existence and, in particular, the size of tradeability premiums. This includes the works of Longstaff (cf.~\cite{ls95} and \cite{lo17}), as well as the articles of Koziol and Sauerbier (cf.~\cite{ks07}) and of Chesney and Kempf (cf.~\cite{ck12}). For these reasons, there is a clear need for alternative models that complement this literature and its current approaches. Such an alternative is proposed in the present article.\vspace{1em} \\
\noindent We propose a novel theoretical way to analyze the impact of non-tradeability on the price of assets in exponential Lévy markets. As we shall see, our framework starts from an adaption of the continuous-time optional asset replacement problem initiated in the seminal paper of McDonald and Siegel (cf.~\cite{ms}). Considering an investor that holds an asset of (ordinary) exponential Lévy type and that faces the decision to replace it with an alternative investment project allows us to analyze two different tradeability scenarios for the asset: A fully liquid and a fully illiquid scenario. By assuming that the investor acts optimally in any of these scenarios, we derive absolute tradeability premiums as differences between the value of the replacement option in the respective scenarios and subsequently provide a free-boundary characterization of the latter premiums. This finally gives us a way to compute non-tradeability values, e.g.~by means of standard numerical techniques, and, in particular, to express the price of an illiquid asset as a percentage of the price of a tradeable equivalent. \vspace{1em} \\
\noindent Our method has some similarities with the approaches taken in \cite{ls95} and \cite{lo17}, \cite{ks07} and \cite{ck12}. As in these articles, valuing tradeability is linked to the opportunity costs of holding the asset. However, there are essential differences in the way the worthiness of tradeability is triggered. For instance, while the value of tradeability arises in \cite{ck12} from the ability of traders to exploit temporary pricing inefficiencies in the market, tradeability enables one, in our model, to take advantage of the continuous possibility to invest in an alternative project. Therefore, instead of valuing tradeability merely out of optimal selling strategies, our approach considers reinvestment opportunities. In this sense, our model has a higher degree of completeness and provides more realistic bounds for the (individual) valuation of tradeability. \vspace{1em} \\
\noindent The remaining of this paper is structured as follows: In Section \ref{SEC1}, we establish the general framework in which we model tradeability. This section essentially focuses on a proper introduction of the broad model as well as of the notation used in the rest of the paper. For this reason, the discussion therein does not include any tradeability aspects and the latter are only introduced in Sections \ref{SECC} and \ref{SEC4new}. Sections \ref{SECC} and \ref{SEC4new} are both divided into two parts. While the first part introduces our tradeability modeling approaches, the second part deals with partial integro-differential equations (PIDEs) and ordinary integro-differential equations (OIDEs) for tradeability valuation. Here, our main results are Proposition~\ref{prop3} and Proposition~\ref{prop6} where free-boundary characterizations of the (absolute) tradeability premiums are provided. The importance of these propositions is illustrated in Section \ref{numres} where the respective free-boundary problems are solved for a particular model and numerical results are discussed. The paper concludes with Section \ref{SEC6}. All proofs and complementary results are presented in the Appendices (Appendix A, B, C and D).
\section{General Framework and Notation}
\label{SEC1}
\noindent We start with a setting similar to that of the investment problem introduced in the seminal paper of McDonald and Siegel (cf.~\cite{ms}): We consider the investment decision of an investor that holds an asset $(S_t)_{t \geq 0}$ and that has the option to replace it with an investment alternative. At any time $t\geq 0$, the investor can pay $S_{t}$ to enter (or acquire a corresponding share of) an investment project that generates positive, net instantaneous cash-flow per unit of investment $(C_{u})_{u \geq t} $ and has to make the decision to either continue holding the asset or to switch to the alternative project. As in \cite{ms}, this asset replacement is understood as a continuous-time and irreversible decision to be taken. Whether or not the investment project is fully owned by the investor will not play any role in our analysis.\footnote{We assume that the project's remuneration is proportional to the investment and, in particular, that the cash-flow generated out of the project does not depend on the type of ownership.}
\subsection{Dynamics of the Initial Asset}
\noindent We denote by $r$ the risk-free interest rate, fix with $ ( \Omega, \mathcal{F}, \mathbf{F}, \mathbb{Q})$ a filtered probability space -- a chosen risk-neutral probability space\footnote{It is well-known that exponential Lévy markets are incomplete as defined by Harrison and Pliska (cf.~\cite{hp81}). Specifying or discussing a particular choice of risk-neutral measure is not the sake of this article. Instead, we assume that a pricing measure under which our model has the required dynamics was previously fixed.} -- and assume that the filtration $ \mathbf{F} = ( \mathcal{F}_{t})_{t \geq 0}$ satisfies the usual conditions. Determining the properties of the asset replacement involves a complete description of its components, the initial asset and the alternative project. We start by characterizing the investor's initial investment: We assume that the investor's initial asset $(S_{t})_{t \geq 0}$ trades on a usual market that is described, under the risk neutral measure $ \mathbb{Q}$, by an (ordinary) exponential Lévy model, i.e.~we assume that the price dynamics of the asset are given by
\begin{equation}
\label{eq1}
S_{t} = S_{0} e^{ X_{t}}, \hspace{1.5em} S_{0} > 0, \, t \geq 0.
\end{equation}
\noindent Here, the process $(X_{t})_{t \geq 0}$ is an $\mathbf{F}$-Lévy process associated with a triplet $(b_{X}, \sigma_{X}^{2}, \Pi_{X})$, i.e.~a càdlàg (right-continuous with left limits) process having independent and stationary increments and Lévy-exponent $\Psi_{X}(\cdot)$ defined, for $\theta \in \mathbb{R}$, by
\begin{equation}
\Psi_{X}( \theta ) := - \log \left( \mathbb{E}^{\mathbb{Q}} \left[ e^{i \theta X_{1}} \right] \right)  =  -ib_{X} \theta + \frac{1}{2} \sigma_{X}^{2}\theta^{2} + \int \limits_{ \mathbb{R}} (1 - e^{i \theta y} + i \theta y \mathds{1}_{\{ | y | \leq 1\}}) \Pi_{X}( dy),
\end{equation}
\noindent where $\mathbb{E}^{\mathbb{Q}}[ \cdot ]$ refers to expectation with respect to the measure~$\mathbb{Q}$. Applying the well-known Lévy-Itô decomposition theorem (cf.~\cite{sa}, \cite{Ap}) allows one to separate $(X_{t})_{t \geq 0}$ into its diffusion and jump parts: Indeed, there exists an $\mathbf{F}$-Brownian motion $(W_{t}^{X})_{t \geq 0}$ and an independent Poisson random measure $N_{X}$ on $[0,\infty) \times \mathbb{R} \setminus \{ 0\}$ having intensity measure $\Pi_{X}$, such that
\begin{equation}
X_{t} = b_{X}t + \sigma_{X} W_{t}^{X} + \int \limits_{\mathbb{R}} y \; \bar{N}_{X}(t,dy), \hspace{1.5em}  t \geq 0,
\label{bzbzeq}
\end{equation}
\noindent where we use for $t \geq 0$ and any Borel set $A \in \mathcal{B}(\mathbb{R}\setminus\{0 \})$ the notation
\begin{align*}
N_{X}(t,A) & := N_{X}((0,t]\times A), \\
\tilde{N}_{X}(dt,dy) & :=  N_{X}(dt,dy) - \Pi_{X}(dy)dt ,\\
\bar{N}_{X}(dt,dy)&:= \left \{ \begin{array}{cc}
\tilde{N}_{X}(dt,dy), & \mbox{if} \; |y|\leq 1, \\
N_{X}(dt,dy), & \mbox{if} \; |y| > 1. \end{array}
\right. 
\end{align*}
\noindent This directly gives a corresponding factorization of the price dynamics $(S_t)_{t \geq 0}$ into exponentials of the diffusion and jump parts of $(X_{t})_{t \geq 0}$. \vspace{1em} \\
\noindent Additionally, the Laplace exponent of the Lévy process $(X_t)_{t \geq 0}$ is defined, for any $ \theta \in \mathbb{R}$ satisfying the condition $ \mathbb{E}^{\mathbb{Q}} \left[ e^{\theta X_{1}} \right] < \infty$, by the following identity:
\begin{equation}
\Phi_{X}( \theta )  :=  \log \left( \mathbb{E}^{\mathbb{Q}} \left[ e^{ \theta X_{1}} \right] \right)  =  b_{X} \theta + \frac{1}{2} \sigma_{X}^{2} \theta^{2} - \int \limits_{ \mathbb{R}} (1 - e^{ \theta y} +  \theta y \mathds{1}_{\{ | y | \leq 1\}}) \Pi_{X}( dy).
\end{equation}
\noindent In the sequel, we will assume that this quantity is at least for $\theta =1$ well defined, i.e.~that $\mathbb{E}^{\mathbb{Q}} \left[ e^{X_{1}} \right] < \infty$, and finally require that $\Phi_{X}(1) \leq r$. The latter condition has an important feature: It is well-known that discounted, exponential Lévy models of the form of (\ref{eq1}) have the martingale property if and only if the usual integrability condition\footnote{$ \mathbb{E}^{\mathbb{Q}} \left[ e^{X_{t}} \right] < \infty$ or, equivalently, $\int \limits_{\{|y|> 1\}} e^{y} \; \Pi_{X}(dy) < \infty$ (cf.~\cite{sa}, Theorem 25.3). This is clearly satisfied by our assumptions.} and additionally $\Phi_{X}(1) = r$ are satisfied (cf.~\cite{jy06}, \cite{Ap}). Hence, requiring $\Phi_{X}(1) \leq r$ to hold under the measure $\mathbb{Q}$ allows the asset to pay a (continuous) dividend and the discounted asset dynamics have the martingale structure only under a lower, adjusted discount factor $r-\tilde{r}$. Such dynamics are typically found in foreign exchange markets, where $\tilde{r}$ represents the foreign risk-free interest rate (cf.~\cite{gk83}, \cite{jc02}).
\subsection{Dynamics of the Investment Alternative}
\noindent We next turn to the investor's investment alternative. As we shall see in a moment, characterizing the investor's investment project reduces to specifying the dynamics of the process $(C_{t})_{t \geq 0}$, the net instantaneous cash-flow generated out of a one-unit investment in the project. Indeed, once this process is specified the project's value can be easily recovered by computing the expected net present value of the project's future cash-flows. Therefore, we start by determining the dynamics of the cash-flow process and assume that $(C_{t})_{t \geq 0}$ follows under $\mathbb{Q}$ another exponential Lévy model of the form
%% \noindent The investor has now the opportunity to switch and invest into an alternative investment project whose value process\footnote{Depending on the nature of the project, the investment value may be understood in different ways: While, for instance, the value of a market-traded asset is just given by its market value, assigning a value to any OTC-investment is rather done using the concept of net worth.} per unit of investment, $(V_{t})_{t \in [0,\infty)}$, follows under $\mathbb{Q}$ another exponential Lévy process of the form
\begin{equation}
\label{eq2}
{C}_{t} =  {C}_{0} e^{Y_{t}}, \hspace{1.5em} C_{0} > 0, \, t \geq 0,
\end{equation}
\noindent where $(Y_{t})_{t \geq 0}$ denotes an $\mathbf{F}$-Lévy process with Lévy triplet $(b_{Y}, \sigma_{Y}^{2}, \Pi_{Y})$. As for $(X_t)_{t \geq 0}$, one obtains (by means of the Lévy-Itô decomposition theorem) a separation of $(Y_{t})_{t \geq 0}$ into its diffusion and jump parts of the form
\begin{equation}
Y_{t} = b_{Y}t + \sigma_{Y} W_{t}^{Y} + \int \limits_{\mathbb{R}} y \; \bar{N}_{Y}(t,dy), \hspace{1.5em}  t \geq 0,
\end{equation}
\noindent where $(W_{t}^{Y})_{t \geq 0}$ denotes an $\mathbf{F}$-Brownian motion and $N_{Y}$ a corresponding Poisson random measure on $[0,\infty) \times \mathbb{R} \setminus \{ 0\}$ that is independent of $(W_{t}^{Y})_{t \geq 0}$. The dependence structure between the two processes $(X_{t})_{t \geq 0}$ and $(Y_{t})_{t \geq 0}$ (and so between both exponential Lévy models $(S_{t})_{t \geq 0}$ and $(C_{t})_{t \geq 0}$) is additionally fixed by assuming that the Poisson random measures $N_{X}$ and $N_{Y}$ are independent and that the Brownian parts $(W_{t}^{X})_{t \geq 0}$ and $(W_{t}^{Y})_{t \geq 0}$ have correlation coefficient $|\rho| \leq 1$, i.e.~that $\left[ W^{X},W^{Y} \right]_{t} = \rho t$. As earlier, we require the existence of the Laplace exponent $\Phi_{Y}(1)$ and demand that $\Phi_{Y}(1)<r$.
\subsection{Asset Replacement Dynamics}
\noindent To finally derive the time-$t$ value of the asset replacement, we first compute for any $t \geq 0$ the expected net present value of the future cash-flow generated out of a one-unit investment in the project, $E_{t}$: Using Fubini's theorem for conditional expectation and the dynamics (\ref{eq2}), one obtains that
\begin{align}
E_{t} & = \mathbb{E}^{\mathbb{Q}} \Bigg[ \int \limits_{t}^{\infty} e^{-r(u-t)} {C}_{u} \, du  \bigg | \mathcal{F}_{t} \Bigg]  =  \int \limits_{t}^{\infty} e^{-r(u-t)} \, \mathbb{E}^{\mathbb{Q}} \left[ {C}_{u}   | \mathcal{F}_{t} \right] du \nonumber \\
& =  \int \limits_{t}^{\infty} e^{-r(u-t)}  {C}_{t}  e^{(u-t) \Phi_{Y}(1)} \, du  =  {C}_{t} \, \int \limits_{t}^{\infty} e^{-(r-\Phi_{Y}(1))(u-t)} \, du = \frac{{C}_{t}}{r- \Phi_{Y}(1)}. \label{eq222}
\end{align}
\noindent Hence, the dynamics of $(E_{t})_{t \geq 0}$ are proportional to those of $({C}_{t})_{t \geq 0}$ and $E_{t}$ equals, at any $t \geq 0$,
$$ E_{t} = E_{0} e^{Y_{t}}, \; \; \; \; \; \; \; E_{0} = \frac{{C}_{0}}{r- \Phi_{Y}(1)} .$$
\noindent The time-$t$ value of a one-unit investment in the project, $V_{t}$, is now easily deduced. Clearly, this value corresponds to the difference of the expected net present value of the future cash-flows generated out of a one-unit investment in the project, $E_{t}$, and $1$, the costs of such an investment. As a consequence, we obtain by (\ref{eq222}) that
\begin{equation}
\label{eq32}
V_{t} =  E_{t} - 1 =  \frac{{C}_{t}}{r- \Phi_{Y}(1)}-1.
\end{equation}
\noindent At any possible switching date $t \geq 0$, the investor holds the option to sell his asset and to reinvest its full proceeds in the alternative project. Hence, the investor's possible time-$t$ level of investment corresponds to the value $S_{t}$ of the asset currently held. This finally gives that the time-$t$ value of the asset replacement, $V_{t}^{S}$, equals
\begin{equation}
\label{eq4}
V_{t}^{S}  =  S_{t} \cdot V_{t} =  S_{t} \left( E_{t} -1 \right).
\end{equation}
\noindent \underline{\bf Remark 1.} \noindent
\begin{itemize} \setlength \itemsep{-0.1em}
\item[i)] Equation (\ref{eq4}) describes a version of the asset replacement that is scaled to one unit of the initial asset. However, looking at more general holdings does not change the replacement problem significantly and any such problem can be easily reduced to the one-unit situation.
\item[ii)] Notice that we did not make any assumption on the exclusiveness of the investment project: The project may represent an investment opportunity that is linked to the investor -- if one thinks of the investor as a company, this could represent for instance a company's internal project -- and so that is unique and not necessarily available (at least not in the exact same conditions) to any other competitor. But also more standard and open investment alternatives could be considered. In this context, any evaluation of the investment alternative under the risk-neutral measure $\mathbb{Q}$ does not correspond to a real pricing attempt but merely serves as an assessment of the project from the point of view of a  ``typical investor'' within the market.
\end{itemize}
\hspace{45em} \scalebox{0.75}{$\blacklozenge$}
\newpage
\section{Valuing Tradeability: Deterministic Illiquidity Horizon}
\label{SECC}
\subsection{Generalities}
\label{secgenal}
\noindent Up to this point, our general framework did not include any element that aimed to model differences in tradeability. This should be addressed next. To this end, we fix a (deterministic) time horizon $T_{D}>0$ and consider variants of the optional asset replacement problem introduced in Section \ref{SEC1} on the time interval $[0,T_{D}]$.\footnote{Although $T_{D} = \infty$ could also be considered, it is not very meaningful. Therefore, we implicitly understand $T_{D}$ to be finite and consider finite analogues of the optional asset replacement problem introduced in Section \ref{SEC1}.}~We assume that the investment project is available at any date $t \in [0,T_{D}]$ and derive tradeability premiums by varying the marketability of the initial asset $(S_t)_{t \geq 0}$ on $[0,T_{D}]$ and analyzing the behavior of an investor that acts optimally in the resulting asset replacement problem. Hereby we compare two scenarios:
\begin{itemize} \setlength \itemsep{-0.1em}
\item[1.] An illiquid scenario, where any attempt to sell the asset $(S_t)_{t \geq 0}$ at time $t \in [0,T_{D})$ fails and the investor has to make a new decision at $T_{D}$. Hence, $\mathcal{T}:=T_{D}-t$ is interpreted as illiquidity horizon.
\item[2.] A liquid scenario, where the tradeability of the investor's asset is guaranteed at any date $t \in [0,T_{D}]$. \vspace{-0.2em} \\
\end{itemize}
\noindent \underline{\bf Remark 2.} \vspace{0.2em} \\
\noindent It is important to note that the present tradeability valuation approach is in line with \cite{ls95}, \cite{lo17}, and \cite{ck12}, and therefore understands tradeability to only occur at very few points in time. Under this assumption, restricting the analysis to the first illiquidity interval $[0,T_{D}]$ already provides sensible results while keeping a certain degree of tractability. Nevertheless, we emphasize that other approaches could be considered. As an example, analyzing a situation where non-tradeability is a temporary state beyond which the asset remains fully tradeable could be addressed as part of future research. \\
\mbox{ }\hspace{44.6em} \scalebox{0.75}{$\blacklozenge$} \\
\subsubsection{Illiquid Scenario}
\noindent We start by analyzing the investor's trading behavior in the illiquid scenario. Being modeled by ordinary exponential Lévy models, the processes $(S_{t})_{t \geq 0}$ and $(E_{t})_{t \geq 0}$ are assumed to be efficient. Hence, the investor cannot anticipate future fluctuations and base his decision at any time $t \in [0,T_{D}]$ on his current information $ \mathcal{F}_{t}$. At any time $t \in [0,T_{D}]$ at which $V_{t}^{S} > 0$, the investment project is more valuable than the asset and switching from $S_{t}$ to $S_{t} E_{t}$, i.e.~investing $S_{t}$ in the project, provides an immediate increase in wealth in the amount of $V_{t}^{S} >0$. Since the investor can only switch, in the illiquid scenario, at $t=T_{D}$, he will do so if and only if $V_{T_{D}}^{S} > 0 $. As a consequence, the time-$t$ value of this switching option $\mathfrak{C}_{{\bf E}}(\cdot)$ is obtained as
\begin{equation}
\mathfrak{C}_{{\bf E}}(\mathcal{T},S_{t},E_{t}) := \mathbb{E}^{\mathbb{Q}}_{S_{t},E_{t}} \big[ e^{-r\mathcal{T}} \left( V_{\mathcal{T}}^{S} \vee 0 \right) \big] =\mathbb{E}^{\mathbb{Q}}_{S_{t},E_{t}} \big[ e^{-r\mathcal{T}} S_{\mathcal{T}} \left( E_{\mathcal{T}} - 1 \right)^{+} \big],
\label{eqeurop}
\end{equation}
\noindent where we denote by $\mathbb{E}_{s_{0},e_{0}}^{\mathbb{Q}}[\cdot]$ the expectation under $\mathbb{Q}_{s_{0},e_{0}}$, the probability measure under which $(S_{t})_{t \geq 0}$ and $(E_{t})_{t \geq 0}$ start at $S_{0} = s_{0}$ and $E_{0}=e_{0}$, respectively. This corresponds to the time-$t$ value of a European exchange option.
\subsubsection{Liquid Scenario}
\noindent Deriving the investor's trading behavior in the liquid scenario can be done by the very same arguments. However, since the initial asset is now perfectly tradeable there are no restrictions on the investor's switching possibilities. Hence, the investor will choose a switching rule that maximizes his immediate increase in wealth in expectation. As a consequence, evaluating the switching option in the liquid scenario reduces to valuing an American exchange option $\mathfrak{C}_{{\bf A}}(\cdot)$ of the form
\begin{equation}
\mathfrak{C}_{{\bf A}}(\mathcal{T},S_{t}, E_{t}) := \sup \limits_{ \tau \in \mathfrak{T}_{[0,\mathcal{T}]} } \mathbb{E}^{\mathbb{Q}}_{S_{t},E_{t}} \big[ e^{-r \tau} \left( V_{\tau}^{S} \vee 0\right) \big] = \sup \limits_{ \tau \in \mathfrak{T}_{[0,\mathcal{T}]} } \mathbb{E}^{\mathbb{Q}}_{S_{t},E_{t}} \big[ e^{-r \tau} S_{\tau} \left( E_{\tau}-1 \right)^{+} \big],
\label{eqamer}
\end{equation} 
\noindent where $\mathfrak{T}_{[0,\mathcal{T}]}$ denotes the set of stopping times that take values in the time interval $[0,\mathcal{T}]$.
\subsubsection{Tradeability Premium and Transformation}
\noindent The above optimal trading strategies can now be used to value tradeability: Both options $\mathfrak{C}_{{\bf E}}(\cdot)$ and $\mathfrak{C}_{{\bf A}}(\cdot)$ yield a monetization of the benefits that can be generated out of the exchange opportunity within the respective tradeability scenarios. Since the asset's tradeability is the only changing parameter, any inequality in these benefits must be a consequence of its variation. Therefore, we identify the (absolute) time-$t$ tradeability/liquidity\footnote{As emphasized in the introduction, we understand liquidity in the sense of absolute tradeability and will use, from now on, both terms interchangeably.}~premium $ \mathfrak{L}(\cdot)$ with the difference of $\mathfrak{C}_{{\bf A}}(\cdot)$ and $\mathfrak{C}_{{\bf E}}(\cdot)$, i.e. we set
\begin{equation}
\mathfrak{L}(\mathcal{T},S_{t},E_{t}) := \mathfrak{C}_{{\bf A}}(\mathcal{T},S_{t},E_{t}) - \mathfrak{C}_{{\bf E}}(\mathcal{T},S_{t},E_{t}).
\label{liqprem}
\end{equation}
\noindent At this point, we already notice a few properties of the tradeability premium (\ref{liqprem}). First, it is clear that our tradeability premium substantially depends on the dynamics of the alternative project. Since the dynamics and characteristics of available projects depend themselves on the investor's relations, resources, etc., our tradeability premium results in an individual value.\footnote{Remember that we evaluate the investment alternative under the risk-neutral measure. Therefore, the resulting tradeability premium provides an individual, though market-weighted value.}~In addition, this value provides a theoretical lower bound for the (individual) valuation of tradeability. Indeed, our setting examines investment alternatives that are irreversible, at least during the time horizon $[0,T_{D}]$ considered. However, reversible investment possibilities clearly exist in practice. Therefore, extending the analysis to investment projects that can be themselves exchanged against others would provide more accuracy in our valuation approach. This extension is left out and could be part of future research.  \vspace{1em} \\
\noindent \underline{\bf Remark 3.} \vspace{0.2em} \\
\noindent Instead of considering absolute values, it is often more informative to look at relative quantities. For this reason, our numerical results in Section \ref{numres} will focus on figures related to the relative time-$t$ tradeability premium, defined as
$$ \mathfrak{L}_{Rel.}(\mathcal{T},S_{t},E_{t}) := \frac{\mathfrak{L}(\mathcal{T},S_{t},E_{t})}{\mathfrak{C}_{{\bf E}}(\mathcal{T},S_{t},E_{t})}= \frac{\mathfrak{C}_{{\bf A}}(\mathcal{T},S_{t},E_{t})}{\mathfrak{C}_{{\bf E}}(\mathcal{T},S_{t},E_{t})}-1 .$$
\hspace{45em} \scalebox{0.75}{$\blacklozenge$} \vspace{1em} \\
 \noindent Valuing both switching options $\mathfrak{C}_{{\bf E}}(\cdot)$ and $\mathfrak{C}_{{\bf A}}(\cdot)$ and so the tradeability premium $\mathfrak{L}(\cdot)$ under $\mathbb{Q}$, i.e.~from the point of view of a ``typical investor'' in the market, reduces to the usual pricing procedure: First, we introduce, for any stopping time $\tau \in \mathfrak{T}_{[0,\infty)} \cup \{\infty \}$,\footnote{At $t = \infty$ we set $S_{t}:=E_{t}:=0$. This is just for the sake of accuracy as it will not play a real role in this article.}~the following notations
\begin{equation}
\mathfrak{C}(\tau,S_{t},E_{t}) := \mathbb{E}^{\mathbb{Q}}_{S_{t},E_{t}} \big[ e^{-r \tau} \left(V_{\tau}^{S} \vee 0 \right) \big] ,
\label{TOOL1}
\end{equation}
\begin{equation}
\mathfrak{C}^{\star}(\tau,E_{t}) := \mathfrak{C}(\tau,1,E_{t}) ,
\label{TOOL2}
\end{equation}
and show in Appendix A that, under the new measure $\mathbb{Q}^{(1)}$ defined by the ($1$-)Esscher transform\footnote{\noindent The Esscher transform was first introduced 1932 by Esscher and later established in the theory of option pricing by Gerber and Shiu (cf.~\cite{gs94}). An economic interpretation of this pricing technique in the continuous-time framework can be found in \cite{gs94}.}
\begin{align}
\left. \frac{d \mathbb{Q}^{(1)}}{d \mathbb{Q}} \right|_{\mathcal{F}_{t}} & :=  \frac{e^{1 \cdot X_{t}}}{\mathbb{E}^{\mathbb{Q}}\left[ e^{1 \cdot X_{t}}\right]} =  e^{X_{t} - t \Phi_{X}(1)},
\label{bubup}
\end{align}
\noindent the process $(Y_{t})_{t \in [0,T]}$ is, for any finite time horizon $T>0$, again a Lévy process with Lévy-exponent $\Psi_{Y}^{(1)}(\cdot)$ having the form
\begin{align}
\Psi_{Y}^{(1)}(\theta)= -i(b_{Y}+\rho \sigma_{X} \sigma_{Y}) \theta + \frac{1}{2} \sigma_{Y}^{2} \theta^{2} + \int \limits_{ \mathbb{R}} (1 - e^{i \theta y} + i \theta y \mathds{1}_{\{ | y | \leq 1\}}) \Pi_{Y}( dy) .
\label{dynQ}
\end{align}
\noindent Then, rewriting $\mathfrak{C}(\cdot)$ under the change of measure (\ref{bubup}) while bearing in mind the dynamics (\ref{eq1}), (\ref{eq4}) and (\ref{bubup}) readily provides, for any $T>0$, the expression
\begin{equation}
\mathfrak{C}(T \wedge \tau,S_{t},E_{t})  =  S_{t} \cdot \mathfrak{C}^{\star}(T \wedge \tau,E_{t}) = S_{t} \cdot \mathbb{E}^{\mathbb{Q}^{(1)}}_{E_{t}} \big[ e^{-(r-\Phi_{X}(1)) (T \wedge \tau)} \left(E_{T \wedge \tau}-1\right)^{+}\big], 
\label{bibip}
\end{equation}
\noindent where $\mathbb{E}_{e_{0}}^{\mathbb{Q}^{(1)}}[\cdot]$ denotes expectation under $\mathbb{Q}_{e_{0}}^{(1)}$, the probability measure (associated to $\mathbb{Q}^{(1)}$ and) under which $(E_{t})_{t \geq 0}$ starts at $E_{0}=e_{0}$. This latter equation substantially simplifies the valuation problem for both $\mathfrak{C}_{{\bf E}}(\cdot)$ and $\mathfrak{C}_{{\bf A}}(\cdot)$. Indeed, combining the relations
\begin{equation}
\mathfrak{C}_{{\bf E}}(\mathcal{T},S_{t},E_{t})  =  \mathfrak{C}(\mathcal{T},S_{t},E_{t}) \hspace{1.5em} \mbox{and} \hspace{1.7em} \mathfrak{C}_{{\bf A}}(\mathcal{T},S_{t},E_{t})  = \sup \limits_{\tau \in \mathfrak{T}_{[0,\mathcal{T}]}} \mathfrak{C}(\tau,S_{t},E_{t})
\end{equation}
\noindent with (\ref{bibip}) while introducing the notations $\mathfrak{C}_{{\bf E}}^{\star}(\mathcal{T}, E_{t}) := \mathfrak{C}_{{\bf E}}(\mathcal{T},1, E_{t})$ and $\mathfrak{C}_{{\bf A}}^{\star}(\mathcal{T},E_{t}) := \mathfrak{C}_{{\bf A}}(\mathcal{T},1,E_{t})$ allows us to rewrite
\begin{align}
\mathfrak{C}_{{\bf E}}(\mathcal{T},S_{t}, E_{t}) & =  S_{t} \cdot \mathfrak{C}_{{\bf E}}^{\star}(\mathcal{T}, E_{t})  =  S_{t} \cdot \mathbb{E}^{\mathbb{Q}^{(1)}}_{E_{t}} \big[ e^{-(r-\Phi_{X}(1)) \mathcal{T}} \left(E_{\mathcal{T}}-1\right)^{+}\big] , \\
\mathfrak{C}_{{\bf A}}(\mathcal{T},S_{t}, E_{t}) & =  S_{t} \cdot \mathfrak{C}_{{\bf A}}^{\star}(\mathcal{T}, E_{t})  =  S_{t} \cdot \sup \limits_{\tau \in \mathfrak{T}_{[0,\mathcal{T}]}} \mathbb{E}^{\mathbb{Q}^{(1)}}_{E_{t}} \big[ e^{-(r-\Phi_{X}(1)) \tau} \left(E_{\tau}-1\right)^{+}\big]. 
\end{align}
\noindent Hence, both switching options are linear in the value $S_{t}$ of the asset initially held. Furthermore, the scaled versions $\mathfrak{C}_{{\bf E}}^{\star}(\cdot)$ and $\mathfrak{C}_{{\bf A}}^{\star}(\cdot)$ correspond to simple European and American-type options written on the exponential Lévy process $(E_{t})_{t \in [0,T_{D}]}$. Consequently, valuing -- under $\mathbb{Q}$ -- any of the switching options $\mathfrak{C}_{{\bf E}}(\cdot)$ and $\mathfrak{C}_{{\bf A}}(\cdot)$ reduces -- under $\mathbb{Q}^{(1)}$ -- to the consideration of corresponding valuation problems for simple options on the exponential Lévy model
$$ E_{t} = E_{0} e^{Y_{t}}, \hspace{1.5em} E_{0}>0, \, t \in [0,T_{D}],$$
\noindent with Lévy exponent $\Psi_{Y}^{(1)}(\cdot)$ defined as in (\ref{dynQ}), risk-free interest rate $\tilde{r} := r - \Phi_{X}(1)$, and strike price $K:=1$.
\subsection{PIDEs for Tradeability Valuation}
\label{sec22}
\noindent Our next goal consists in deriving partial integro-differential equations that can be used to value tradeability. As argued in the previous section, we focus from now on on the respective valuation problems for $\mathfrak{C}_{{\bf E}}^{\star}(\cdot)$ and $\mathfrak{C}_{{\bf A}}^{\star}(\cdot)$. We then define
\begin{equation}
\mathfrak{L}^{\star}(\mathcal{T},E_{t}) := \mathfrak{L}(\mathcal{T},1,E_{t}) = \mathfrak{C}_{{\bf A}}^{\star} (\mathcal{T},E_{t}) - \mathfrak{C}_{{\bf E}}^{\star}(\mathcal{T},E_{t})
\end{equation}
\noindent and recover $\mathfrak{L}(\cdot)$ from its scaled version $\mathfrak{L}^{\star}(\cdot)$ by means of the obvious relation
\begin{equation}
\mathfrak{L}(\mathcal{T},S_{t}, E_{t}) = S_{t} \cdot \mathfrak{L}^{\star}(\mathcal{T},E_{t}) .
\end{equation}
\noindent \underline{\bf Remark 4.} \vspace{0.2em} \\
\noindent Note that we can also express the relative time-$t$ tradeability premium, using the above notation, as
$$ \mathfrak{L}_{Rel.}(\mathcal{T},S_{t},E_{t}) = \mathfrak{L}_{Rel.}(\mathcal{T},1,E_{t}) = \frac{\mathfrak{L}^{\star}(\mathcal{T},E_{t})}{\mathfrak{C}_{{\bf E}}^{\star}(\mathcal{T},E_{t})}= \frac{\mathfrak{C}_{{\bf A}}^{\star} (\mathcal{T},E_{t})}{\mathfrak{C}_{{\bf E}}^{\star}(\mathcal{T},E_{t})}-1 .$$
\hspace{45em} \scalebox{0.75}{$\blacklozenge$} \vspace{1em} \\
\noindent In what follows, we will always assume that the second moment of the ($\mathbb{Q}^{(1)}$-)Lévy model $(E_{t})_{t \in [0,T_{D}]}$ exists, or equivalently (cf.~\cite{sa}, Theorem 25.3) that
\begin{equation}
\int \limits_{\{|y|> 1\}} e^{2y} \; \Pi_{Y}(dy) < \infty ,
\label{condidi}
\end{equation}
\noindent and note that this is a weak assumption that could be even relaxed (cf.~\cite{cv05}). We start by determining the dynamics of the process $(E_{t})_{t \in [0,T_{D}]}$ under the measure $\mathbb{Q}^{(1)}$. This is done using Itô's Lemma and readily gives that
\begin{equation}
dE_{t} =  E_{t-} \bigg( \Phi_{Y}^{(1)}(1) dt + \sigma_{Y} d\tilde{W}_{t}^{Y} + \int \limits_{ \mathbb{R}} (e^{y} -1) \tilde{N}_{Y}(dt,dy) \bigg),
\label{didy}
\end{equation}
\noindent where $(\tilde{W}_{t}^{Y})_{t \in [0,T_{D}]}$ denotes a $\mathbb{Q}^{(1)}$-Brownian motion (cf.~Appendix A) and $\Phi_{Y}^{(1)}(\cdot)$ refers to the Laplace-exponent of $(Y_{t})_{t \in [0,T_{D}]}$ under $\mathbb{Q}^{(1)}$.\footnote{Note that the existence of $\Phi_{Y}^{(1)}(1)$ directly follows from our initial assumptions, since the measure change defined by (\ref{bubup}) does not alter the jump component of $(Y_t)_{t \geq 0}$ and we initially assumed that $\int \limits_{\{|y|> 1\}} e^{y} \; \Pi_{Y}(dy) < \infty$.} Therefore, whenever well-defined, its infinitesimal generator is a partial integro-differential operator obtained, for $V: [0,T_{D}] \times \mathbb{R} \rightarrow \mathbb{R}$, by
\begin{align}
\mathcal{A}_{E} V(\mathcal{T},x) & :=  \lim \limits_{t \downarrow 0} \; \frac{\mathbb{E}^{\mathbb{Q}^{(1)}}_{x} \big[ V(\mathcal{T},E_{t})\big] - V(\mathcal{T},x)}{t} \nonumber \\
& \hspace{0.25em} =  \frac{1}{2} \sigma^{2}_{Y} x^{2} \partial_{x}^{2} V(\mathcal{T},x)  +  \Phi_{Y}^{(1)}(1) x \partial_{x} V(\mathcal{T},x) \nonumber \\
&  \hspace{5em} +  \int \limits_{\mathbb{R}} \big[ V(\mathcal{T},xe^{y}) - V(\mathcal{T},x) - x(e^{y}-1)\partial_{x} V(\mathcal{T},x) \big] \Pi_{Y}(dy). 
\end{align}
\subsubsection{PIDE \rom{1}: Illiquid Scenario}
\noindent We first deal with the illiquid scenario and rewrite, for $ (\mathcal{T} , x) \in [0,T_{D}] \times [0,\infty)$, the European-type switching option in the form
\begin{equation}
\mathfrak{C}_{{\bf E}}^{\star}(\mathcal{T}, x) = \mathbb{E}^{\mathbb{Q}^{(1)}}_{x} \big[ \left(\bar{E}_{\mathcal{T}}-1\right)^{+}\big],
\end{equation}
\noindent where $(\bar{E}_{t})_{t \in [0,T_{D}]}$ refers to the (strong) Markov process\footnote{It is well-known (cf.~\cite{pe06}) that the process $(\bar{E}_{t})_{t \in [0,T_{D}]}$ defined this way preserves the (strong) Markov property of the underlying process $(E_{t})_{t \in [0,T_{D}]}$.} obtained by ``killing'' the sample path of $(E_{t})_{t \in [0,T_{D}]}$ at the proportional rate $\tilde{r} := r - \Phi_{X}(1)$. The process' transition probabilities are then given by
\begin{equation}
\mathbb{Q}_{x}^{(1)} \left( \bar{E}_{t} \in A \right) = \mathbb{E}_{x}^{\mathbb{Q}^{(1)}} \left[ e^{- \tilde{r} t} \, \mathds{1}_{A}(E_{t}) \right]
\label{TPKilling_Bup}
\end{equation}
\noindent and we identify its cemetery state, without loss of generality, with $\partial \equiv 0$. Therefore, for any initial value $z= (\mathbf{t},x) \in [0,T_{D}] \times [0,\infty) $, the process $(Z_{t})_{t \in [0,\mathbf{t}]}$ defined via $Z_{t}:= (\mathbf{t}-t, \bar{E}_{t})$, $\bar{E}_{0}=x$, is a strong Markov process with state domain given by $\mathcal{D}_{\mathbf{t}}:= [0,\mathbf{t}] \times [0,\infty)$. Additionally, $\mathfrak{C}_{{\bf E}}^{\star}(\cdot)$ can be re-expressed as
\begin{equation}
\mathfrak{C}_{{\bf E}}^{\star}(\mathcal{T}, x) = V_{E}\big((\mathcal{T},x)\big),
\label{IMeq1_Bup}
\end{equation}
\noindent where the value function $V_{E}(\cdot)$ has the following representation under the measure $\mathbb{Q}_{z}^{(1),Z}$ having initial distribution $Z_{0} = z$:
\begin{align}
V_{E}(z) : = \mathbb{E}^{\mathbb{Q}^{(1),Z}}_{z} \big[ G(Z_{\tau_{\mathcal{S}}}) \big], \hspace{1.5em}  G(z)  := (x -1)^{+},
\end{align}
\noindent and $\tau_{\mathcal{S}} := \inf \{ t \geq 0: Z_{t} \in \mathcal{S} \}$, $\mathcal{S} := \big(\{0 \} \times [0,\infty) \big) \cup \big( [0,\mathbf{t}] \times \{0 \} \big)$, is a stopping time that satisfies $\tau_{\mathcal{S}} \leq \mathbf{t}$, under $\mathbb{Q}_{z}^{(1),Z}$ with $z = (\mathbf{t},x)$. Furthermore, the stopping region $\mathcal{S}$ is for any $\mathbf{t} \in [0,T_{D}]$ a closed set in $\mathcal{D}_{\mathbf{t}}$. Therefore, standard arguments based on the strong Markov property of $(Z_{t})_{t \in [0,\mathbf{t}]}$ (cf.~\cite{pe06}) imply that $V_{E}(\cdot)$ satisfies the following problem
\begin{align}
\mathcal{A}_{Z} V_{E}(z) & = 0, \hspace{2em} \mbox{on} \,\,  \mathcal{D}_{T_{D}} \setminus \mathcal{S}, \\
V_{E}(z) & = G(z), \hspace{1.5em} \mbox{on} \,\,  \mathcal{S},
\end{align}
\noindent where $\mathcal{A}_{Z}$ denotes the infinitesimal generator of the process $(Z_{t})_{t \in [0,\mathbf{t}]}$. To complete the proof, we note that (for any suitable function $V: \mathcal{D}_{\mathbf{t}} \rightarrow \mathbb{R}$) the infinitesimal generator $\mathcal{A}_{Z}$ can be re-expressed as
\begin{align}
\mathcal{A}_{Z} V\big((\mathbf{t},x) \big) & = -\partial_{\mathbf{t}} V\big((\mathbf{t},x) \big) + \mathcal{A}_{\bar{E}} V\big((\mathbf{t},x) \big) \nonumber \\
& = -\partial_{\mathbf{t}} V\big((\mathbf{t},x) \big) + \mathcal{A}_{E} V\big((\mathbf{t},x) \big) - \tilde{r} V\big((\mathbf{t},x) \big).
\label{IgEnE_Bup}
\end{align}
\noindent Consequently, recovering $\mathfrak{C}_{{\bf E}}^{\star}(\cdot)$ via (\ref{IMeq1_Bup}) finally gives the following PIDE:
\begin{align}
- \partial_{\mathcal{T}} \mathfrak{C}_{{\bf E}}^{\star}(\mathcal{T},x) + \mathcal{A}_{E} \mathfrak{C}_{{\bf E}}^{\star}(\mathcal{T},x) -\tilde{r} \mathfrak{C}_{{\bf E}}^{\star}(\mathcal{T},x) & =  0, \; \; \; \; \; \; \; \; \; \; \; \; \;\; \; \; \; \; \; \; \; \;  \; \mbox{on} \; (0,T_{D}] \times (0,\infty), \label{PIDE1}\\
\mathfrak{C}_{{\bf E}}^{\star}(0,x) & =  (x-1 )^{+}, \hspace{4.5em}   x \in [0,\infty). \label{PIDE2}
\end{align}
\noindent Under few additional assumptions,\footnote{Numerous Lévy models considered in the financial literature as well as the model considered in Section \ref{numres} satisfy these assumptions.}~smoothness of the European-type switching option can be additionally shown. This is the content of the following proposition.
\begin{Prop}
\label{Prop1}
Assume that 
\begin{equation}
\mbox{either} \; \; \; \; \; \; \; \sigma_{Y} \neq 0 \; \;\; \; \; \; \;  \mbox{or} \; \; \; \; \; \;\; \; \exists \alpha \in (0,2): \; \;\; \;  \liminf \limits_{ \epsilon \downarrow 0} \frac{1}{\epsilon^{2-\alpha}} \int \limits_{-\epsilon}^{\epsilon} |y|^{2} \; \Pi_{Y}(dy) > 0.
\label{bediwi}
\end{equation}
\noindent Then, the value of the European-type switching option under deterministic illiquidity horizon, $\mathfrak{C}_{{\bf E}}^{\star}(\cdot)$, is continuous on $[0,T_{D}] \times [0,\infty)$, $C^{1,2}$ on $(0,T_{D}) \times (0, \infty)$ and solves the partial integro-differential equation
\begin{equation}
- \partial_{\mathcal{T}} \mathfrak{C}_{{\bf E}}^{\star}(\mathcal{T},x) + \mathcal{A}_{E} \mathfrak{C}_{{\bf E}}^{\star}(\mathcal{T},x) -\tilde{r} \mathfrak{C}_{{\bf E}}^{\star}(\mathcal{T},x)  =   0 \label{PIDE1+}
\end{equation}
\noindent on $(0,T_{D}] \times (0,\infty)$ with initial condition
\begin{equation}
\mathfrak{C}_{{\bf E}}^{\star}(0,x)  =  (x-1 )^{+}, \hspace{1.5em}  x \in [0,\infty). \label{PIDE2+}
\end{equation}
\end{Prop}
\noindent The proof of Proposition \ref{Prop1} is similar to that of Proposition 2 in \cite{cv05}. In this article, the authors work with exponential Lévy processes that have the martingale property. However, since $(E_{t})_{t \in [0,T_{D}]}$ does not necessarily satisfy this property, we provide in Appendix B an adaption of their proof that works in our more general context. Parts of the proof that do not involve any martingale argument will be directly referred to~\cite{cv05}.
\subsubsection{PIDE \rom{2}: Liquid Scenario}
\noindent We next turn to the liquid scenario. As in the illiquid scenario, we derive a characterization of the American-type switching option $\mathfrak{C}_{{\bf A}}^{\star}(\cdot)$ by adapting well-established results for standard American options on exponential Lévy models. This leads to the next proposition.
\begin{Prop}
\label{propp2}
The value of the American-type switching option under deterministic illiquidity horizon, $\mathfrak{C}_{{\bf A}}^{\star}(\cdot)$, is continuous on $[0,T_{D}] \times [0,\infty)$ and solves the non-linear Hamilton-Jacobi-Bellman (HJB) equation
\begin{equation}
\max \bigg \{ - \partial_{\mathcal{T}} \mathfrak{C}_{{\bf A}}^{\star}(\mathcal{T},x) + \mathcal{A}_{E} \mathfrak{C}_{{\bf A}}^{\star}(\mathcal{T},x) -\tilde{r} \mathfrak{C}_{{\bf A}}^{\star}(\mathcal{T},x), \; (x-1)^{+} - \mathfrak{C}_{{\bf A}}^{\star}(\mathcal{T},x) \bigg\}  = 0,
\end{equation}
\noindent on $(0,T_{D}] \times [0,\infty)$ with initial condition
\begin{equation}
\mathfrak{C}_{{\bf A}}^{\star}(0,x)  =  (x-1)^{+} ,\hspace{1.5em} x \in [0,\infty).
\end{equation}
\end{Prop}
\noindent Proposition \ref{propp2} is due to Pham (cf.~\cite{ph97}, \cite{ph98}), who proved it in greater generality. The proof can be found in his seminal article \cite{ph98}. Alternatively, we note that Proposition \ref{propp2} could be derived via similar techniques as the ones used in the proof of the upcoming Proposition \ref{Prop5} (cf.~Appendix C).
\subsubsection{PIDE \rom{3}: Free-Boundary Characterization}
\label{BectionIII}
\noindent Motivated by the theory of early exercise premiums in classical American option settings, we finally aim to derive a free-boundary characterization of the (absolute) tradeability premium $\mathfrak{L}^{\star} (\cdot)$. We start by collecting in Lemma \ref{lemma1} a few useful properties of $\mathfrak{C}_{\bf A}^{\star}(\cdot) $ that essentially follow from the (strong) Markov property of Lévy processes. A proof of this result is provided in Appendix B. 
\begin{Lem}
\label{lemma1}
The American-type switching option $\mathfrak{C}_{\bf A}^{\star}(\cdot)$ satisfies the following properties:
\begin{itemize} \setlength \itemsep{-0.1em}
\item[$a)$] For every $\mathcal{T} \in [0,T_{D}]$, the function $x \mapsto \mathfrak{C}_{\bf A}^{\star}(\mathcal{T},x)$ is non-decreasing and convex on $[0,\infty)$.
\item[$b)$] For every $x \in [0,\infty)$, the function $\mathcal{T} \mapsto \mathfrak{C}_{\bf A}^{\star}(\mathcal{T},x)$ is non-decreasing on $[0,T_{D}]$.
\item[$c)$] For every $\mathcal{T} \in [0,T_{D}]$, we have that
$$| \mathfrak{C}_{{\bf A}}^{\star}(\mathcal{T},x) - \mathfrak{C}_{{\bf A}}^{\star}(\mathcal{T},y) |  \leq C | x -y |, \; \; \; \; \; \; \forall x,y \in  [0,\infty),$$
\noindent with $C=1$ whenever $\tilde{r} \geq \Phi_{Y}^{(1)}(1)$. 
\end{itemize}
\end{Lem}
\noindent As in the classical theory of American options, we next decompose the domain $(0,T_{D}] \times [0,\infty)$ into two regions, the holding region $\mathfrak{D}_{h}$ and the switching region $\mathfrak{D}_{s}$. First, combining the results in Lemma \ref{lemma1} with Proposition \ref{propp2} ensures that by defining
\begin{align}
\mathfrak{D}_{h} & :=  \Big \{ (\mathcal{T},x) \in (0,T_{D}] \times [0,\infty): \, \mathfrak{C}_{{\bf A}}^{\star}(\mathcal{T},x) > (x-1)^{+}\Big \}, \\
\mathfrak{D}_{s} & :=  \Big \{ (\mathcal{T},x) \in (0,T_{D}] \times [0,\infty): \,  \mathfrak{C}_{{\bf A}}^{\star}(\mathcal{T},x) = (x-1)^{+}\Big \},
\end{align}
\noindent we obtain $\mathfrak{D}_{h} \dot{\cup}\hspace{0.15em}\mathfrak{D}_{s}=(0,T_{D}] \times [0, \infty)$. At this point, one should note that nothing has been said about these sets. In fact, while it is easily seen that $\mathfrak{D}_{h}$ is non-empty, $\mathfrak{D}_{s} = \emptyset$ could still hold. Looking at Lemma~\ref{lemma1}.$c)$ already suggests that this may depend on the sign of $\tilde{r} - \Phi_{Y}^{(1)}(1)$. Indeed, for $ \tilde{r} \leq  \Phi_{Y}^{(1)}(1)$ we obtain that $\mathfrak{D}_{s} = \emptyset$ and the American-type switching option $\mathfrak{C}_{\bf A}^{\star}(\cdot)$ reduces to its European counterpart $\mathfrak{C}_{\bf E}^{\star}(\cdot)$. This follows since, under $\tilde{r}  \leq \Phi_{Y}^{(1)}(1)$, the process $ (e^{-\tilde{r}t} E_{t})_{t \in [0,T_{D}]} $ is a ($\mathbb{Q}^{(1)}$-)submartingale.\footnote{For $ \tilde{r} =  \Phi_{Y}^{(1)}(1)$ it is actually a martingale. However, the submartingale property is for our purpose sufficient (cf.~\cite{jy06}).}~Hence, for $\tilde{r}  \leq \Phi_{Y}^{(1)}(1)$ the tradeability premium is zero and we focus in the following on the case where $\tilde{r}  > \Phi_{Y}^{(1)}(1)$. Here, we show that for any $\mathcal{T} \in (0,T_{D}]$ there exists a switching boundary $\mathfrak{b}_{s}(\mathcal{T})$ above which switching to the alternative project is optimal and that it is defined by $\mathfrak{b}_{s}(\mathcal{T}) := \inf \mathfrak{D}_{s,\mathcal{T}}$, where
$$ \mathfrak{D}_{s,\mathcal{T}} := \Big \{ x \in [0,\infty): \, \mathfrak{C}_{\bf A}^{\star}(\mathcal{T},x) = (x-1)^{+} \Big \} .$$
\noindent To prove the existence of such boundary, we start by proving that, for any $\mathcal{T} \in (0,T_{D}]$, the set $\mathfrak{D}_{s,\mathcal{T}}$ is non-empty. This is done via similar techniques to the ones used in \cite{fk18} and \cite{dk17}. First, we compute, for $x \in (0,\infty)$ and $f(x):=(x-1)^{+}$, the instantaneous benefit of waiting to switch, $H(x):= (\mathcal{A}_{E}f -\tilde{r}f)(x)$, and obtain that
\begin{equation}
H(x) = \bigg( \Phi_{Y}^{(1)}(1) x - \tilde{r}(x-1) -x  \int \limits_{\mathbb{R}} (e^{y} -1) \Pi_{Y}(dy) \bigg) \mathds{1}_{\{x \geq 1 \}} + \int \limits_{\mathbb{R}} \big( f(xe^{y}) - f(x) \big) \Pi_{Y}(dy).
\end{equation}
Then, using Peskir's generalized change-of-variable formula (cf.~\cite{Pe07}), we obtain that, for any stopping-time $\tau \in \mathfrak{T}_{[0,\mathcal{T}]}$ and $x_{0} \in (0,\infty)$,
\begin{equation}            
\mathbb{E}_{x_{0}}^{\mathbb{Q}^{(1)}} \left[ e^{-\tilde{r} \tau} \big(E_{\tau} - 1)^{+} \right] = (x_{0}-1 )^{+} \; + \; \mathbb{E}_{x_{0}}^{\mathbb{Q}^{(1)}} \left[ \int \limits_{0}^{\tau} e^{-\tilde{r} s} H(E_{s}) ds \right] + \frac{1}{2} \; \mathbb{E}^{\mathbb{Q}^{(1)}}_{x_{0}} \left[ \int \limits_{0}^{\tau} e^{-\tilde{r}s} \mathds{1}_{\{E_{s-}=1, \; E_{s}=1 \}} \;d\ell_{s}^{1}(E)\right].  \label{IMPEQ}
\end{equation}
\noindent Here, $\left( \ell_{t}^{1}(E)\right)_{t \in [0,\mathcal{T}]}$ is the local time of $(E_{t})_{t \in [0,\mathcal{T}]}$ at the level $1$ which is defined, for $t \in [0,\mathcal{T}]$, by means of the equation
\begin{equation}
\big|E_{t}-1\big| = \big|E_{0}-1\big| + \int \limits_{0}^{t} \mbox{sgn}(E_{s-}-1) dE_{s}  + \ell_{t}^{1}(E) + \sum \limits_{0<s\leq t} \left( \big|E_{s}-1\big|-\big|E_{s-}-1\big|-\mbox{sgn}(E_{s-}-1 ) \Delta E_{s}\right),
\end{equation}
\noindent where $\mbox{sgn}(0):=0$, and $d \ell_{s}^{1}(E)$ refers to integration with respect to the continuous increasing function $s \mapsto \ell_{s}^{1}(E)$. We claim that Equation (\ref{IMPEQ}) already gives that $\mathfrak{D}_{s,\mathcal{T}} \neq \emptyset$. Indeed, one first obtains that the local time term goes to zero as $x_{0}$ becomes large. At the same time, as $x \uparrow \infty$ we have that $H(x) \downarrow - \infty$. This can be seen by combining the condition $\tilde{r} > \Phi_{Y}^{(1)}(1)$ with the fact that, for any $x> e^{1}$,
\begin{equation}
\bigg| \int \limits_{\mathbb{R}} f(xe^{y}) -f(x) -x( e^{y}-1) \Pi_{Y}(dy) \bigg| \; \leq \;  \Pi_{Y}\big(\{ | y | > 1\} \big) < \infty
\end{equation}
holds, since for such $x$ we have that 
$$\int \limits_{\{ | y | \leq  1\}} \big(f(xe^{y}) -f(x) -x( e^{y}-1) \big) \Pi_{Y}(dy) \; = \; 0 $$
\noindent and, for $x \in (0,\infty)$, the function $x \mapsto |f(xe^{y})-f(x)-x(e^{y}-1) |$ is bounded by $1$ (in general, by the strike~$K$), uniformly in $y$. Due to the lack of time to compensate for the very negative $H(\cdot)$, it is therefore optimal to stop for large $x_{0}$ at once. Consequently, $(x_{0}-1)^{+} = \mathfrak{C}_{\bf A}^{\star}(\mathcal{T},x_{0})$ must be true for some $x_{0} \in (0,\infty)$. This gives that $\mathfrak{D}_{s,\mathcal{T}} \neq \emptyset$. To see that, for any $\mathcal{T} \in (0,T_{D}]$, $\mathfrak{b}_{s}(\mathcal{T}) := \inf \mathfrak{D}_{s,\mathcal{T}}$ gives a boundary with the required properties, we use Lemma \ref{lemma1}. Indeed, combining Properties $a)$ and $c)$ of Lemma \ref{lemma1} we obtain that whenever $(x-1)^{+}= \mathfrak{C}_{\bf A}^{\star}(\mathcal{T},x)$ is satisfied for some $x \in [0,\infty)$, we must also have for $y \geq x$ that $(y-1)^{+}= \mathfrak{C}_{\bf A}^{\star}(\mathcal{T},y)$. This implies that, for any $\mathcal{T} \in (0,T_{D}]$, $\mathfrak{D}_{s,\mathcal{T}}$ is an up-connected set and that it can be written as $ \mathfrak{D}_{s,\mathcal{T}} = [\inf \mathfrak{D}_{s,\mathcal{T}}, \infty)$, which gives the required properties. \vspace{1em} \\
\noindent The previous discussion provides an alternative expression for the holding and switching regions, as 
\begin{align}
\mathfrak{D}_{h} & = \big \{(\mathcal{T},x) \in (0,T_{D}] \times [0,\infty): \; x < \mathfrak{b}_{s}(\mathcal{T}) \big \}, \\
\mathfrak{D}_{s} & = \big \{(\mathcal{T},x) \in (0,T_{D}] \times [0,\infty): \; x \geq  \mathfrak{b}_{s}(\mathcal{T}) \big \} .
\end{align}
\noindent Together with an appropriate smooth-fit property (cf.~Appendix B), these results finally lead to the following free-boundary characterization of the (absolute) tradeability premium $\mathfrak{L}^{\star}(\cdot) $.
\begin{Prop}
\label{prop3}
Assume that $\sigma_{Y} \neq 0$. Then, we have the following properties:
\begin{itemize} \setlength \itemsep{-0.5em}
\item[1.] If $\tilde{r} \leq \Phi_{Y}^{(1)}(1)$, the (absolute) tradeability premium $\mathfrak{L}^{\star}(\cdot)$ satisfies
$$ \mathfrak{L}^{\star}(\mathcal{T},x) = 0, \hspace{2em} \forall (\mathcal{T},x) \in [0,T_{D}] \times [0,\infty) .$$
\item[2.] If $\tilde{r} > \Phi_{Y}^{(1)}(1)$, the pair $\big(\mathfrak{L}^{\star}(\cdot),\mathfrak{b}_{s}(\cdot)\big)$ solves the following free-boundary problem:
\begin{equation}
-\partial_{\mathcal{T}} \mathfrak{L}^{\star}(\mathcal{T},x) + \mathcal{A}_{E} \mathfrak{L}^{\star} (\mathcal{T},x) - \tilde{r} \mathfrak{L}^{\star}(\mathcal{T},x)  =  0 , \; \; \; \; x \in (0,\mathfrak{b}_{s}(\mathcal{T})), \; \mathcal{T} \in (0,T_{D}],
\end{equation}
\noindent subject to the boundary conditions
\begin{align}
\mathfrak{L}^{\star}(\mathcal{T}, \mathfrak{b}_{s}(\mathcal{T})) & =  \mathfrak{b}_{s}(\mathcal{T}) - 1 - \mathfrak{C}_{{\bf E}}^{\star}(\mathcal{T}, \mathfrak{b}_{s}(\mathcal{T})), \; \; \; \; \mathcal{T} \in (0,T_{D}], \\
\partial_{x} \mathfrak{L}^{\star}( \mathcal{T}, \mathfrak{b}_{s}(\mathcal{T})) & =  1 - \partial_{x} \mathfrak{C}_{{\bf E}}^{\star} (\mathcal{T}, \mathfrak{b}_{s}(\mathcal{T})), \; \; \; \; \mathcal{T} \in (0,T_{D}],  \label{equazzzion} \\
\mathfrak{L}^{\star}(\mathcal{T},0) & = 0, \; \; \; \; \mathcal{T} \in (0,T_{D}],
\end{align}
\noindent and initial condition
\begin{equation}
\mathfrak{L}^{\star}(0, x)  = 0, \; \; \; \; x \in (0,\mathfrak{b}_{s}(\mathcal{T})).
\end{equation}
\end{itemize} 
\end{Prop}
\noindent \underline{\bf Remark 5.}
\begin{itemize} \setlength \itemsep{-0.1em}
\item[i)] Proposition \ref{prop3} is of great practical importance. Although we did not obtain an analytical expression for the (absolute) tradeability premium $\mathfrak{L}^{\star}(\cdot)$, there exist several well-established numerical methods that deal with free-boundary problems in the form of Proposition \ref{prop3}. Using such methods, our tradeability valuation problem can be solved for any model that satisfy our (very few) assumptions.
\item[ii)] We have just seen that the tradeability premium reduces to zero whenever $\tilde{r}\leq \Phi_{Y}^{(1)}(1)$, or equivalently whenever $-\big(\Phi_{Y}(1)-r\big) \leq \Phi_{X}(1) +\rho \sigma_{X} \sigma_{Y}$. From a financial perspective this condition is very intuitive. Indeed, in view of Equation (\ref{didy}) (and of its derivation), one first obtains that the Laplace exponents $\Phi_{X}(1)$ and $\Phi_{Y}(1)$ describe the growth rate of the corresponding processes $(S_{t})_{t \geq 0}$ and $(C_{t})_{t \geq 0}$, i.e.~of the initial asset and of the net instantaneous cash-flow generated out of a one-unit investment in the project, respectively. With this understanding, the above condition has the following meaning: It demands that the growth rate of asset $(S_{t})_{t \geq 0}$ adjusted for covariance effects across the dynamics of the asset and of the alternative investment exceeds the negative growth, i.e.~the loss in terms of the discounted cash-flow level, incurred while waiting to switch to the alternative project.
\end{itemize}
\hspace{45em} \scalebox{0.75}{$\blacklozenge$}
\section{Valuing Tradeability: Stochastic Illiquidity Horizon}
\label{SEC4new}
\subsection{Generalities}
\label{SecR11}
\noindent In Section \ref{SECC}, we provided a characterization of tradeability premiums when the illiquidity horizon is fully known in advance. Although this characterization already allows for an efficient evaluation of tradeability, starting from a deterministic illiquidity horizon is clearly not a realistic assumption: In practice, agents do not usually know the exact duration of a non-tradeability period and fixing ahead a deterministic illiquidity horizon $T_{D}$ may seem very simplistic. For this reason, we next extend the previous analysis to the case of a stochastic illiquidity horizon $T_{R} > 0$. We assume that $T_{R}$ is exponentially distributed with rate $\vartheta > 0$ and derive tradeability premiums by analyzing randomized versions of the original scenarios:
\begin{itemize} \setlength \itemsep{-0.1em}
\item[1.] A randomized illiquid scenario, where any attempt to sell the asset $(S_{t})_{t \geq 0}$ at time $t \in [0,T_{R})$ fails and the investor has to make a new decision at $T_{R}$.
\item[2.] A randomized liquid scenario, where the tradeability of the investor's asset is guaranteed at any date $t \in [0,T_{R}]$.
\end{itemize}
\noindent Ideally, we would like to allow for any possible dependency between $T_{R}$ and the processes $(S_{t})_{t \geq 0}$ and $(E_{t})_{t \geq 0}$ characterizing the asset replacement. However, dealing with a general stochastic illiquidity horizon can quickly become cumbersome. For this reason, we assume in the sequel that $T_{R}$ is independent of $(V^{S}_{t})_{t \geq 0}$. Extending our model to allow for a more general dependency structure between $T_{R}$ and $(V^{S}_{t})_{t \geq 0}$ could be part of future research.
\subsubsection{Tradeability Premium: Definition}
\noindent Analyzing both the (randomized) illiquid and liquid scenario can be done via similar arguments to the ones used in their deterministic version and leads to comparable switching options. However, due to the memoryless property of the exponential distribution, the passage of time has no effect on either of the resulting switching options. Consequently, the time-$t$ value of these options is not time-dependent anymore and this leads to the following time-$t$ representations of the investor's trading behavior under stochastic illiquidity horizon:
\begin{align}
\mathfrak{C}^{R}_{{\bf E}}(S_{t}, E_{t}) := \mathbb{E}^{\mathbb{Q}}_{S_{t},E_{t}} \big[ e^{-r T_{R}} \left( V_{T_{R}}^{S} \vee 0\right) \big] =  \mathbb{E}^{\mathbb{Q}}_{S_{t},E_{t}} \big[ e^{-r T_{R}} S_{T_{R}} \left( E_{T_{R}}-1 \right)^{+} \big],
\end{align}
\begin{align}
\mathfrak{C}^{R}_{{\bf A}}(S_{t}, E_{t}) := \sup \limits_{ \tau \in \mathfrak{T}_{[0,\infty)} } \mathbb{E}^{\mathbb{Q}}_{S_{t},E_{t}} \big[ e^{-r (T_{R} \wedge \tau )} \left( V_{T_{R} \wedge \tau}^{S} \vee 0\right) \big] = \sup \limits_{ \tau \in \mathfrak{T}_{[0,\infty)} } \mathbb{E}^{\mathbb{Q}}_{S_{t},E_{t}} \big[ e^{-r ( T_{R} \wedge \tau)} S_{ T_{R} \wedge \tau} \left( E_{ T_{R} \wedge \tau}-1 \right)^{+} \big].
\end{align}
\noindent We therefore identify the (absolute) time-$t$ tradeability premium under stochastic illiquidity horizon $\mathfrak{L}^{R}(\cdot)$ by means of the relation
\begin{equation}
\mathfrak{L}^{R}(S_{t},E_{t}) := \mathfrak{C}_{{\bf A}}^{R}(S_{t},E_{t}) - \mathfrak{C}_{{\bf E}}^{R}(S_{t},E_{t}),
\label{LiVaEq}
\end{equation}
\noindent and finally note that its relative counterpart is defined accordingly, as 
\begin{equation}
\mathfrak{L}^{R}_{Rel.}(S_{t},E_{t}) := \frac{\mathfrak{L}^{R}(S_{t},E_{t})}{\mathfrak{C}_{{\bf E}}^{R}(S_{t},E_{t})} = \frac{\mathfrak{C}_{{\bf A}}^{R}(S_{t},E_{t})}{\mathfrak{C}_{{\bf E}}^{R}(S_{t},E_{t})} - 1 . \nonumber
\end{equation}

\subsubsection{Tradeability Premium: Transformation}
\noindent Following the steps taken in the deterministic version of the problem, we next transform the tradeability valuation equation (\ref{LiVaEq}) into a more tractable expression. First, we introduce, for any $\tau \in \mathfrak{T}_{[0,\infty)} \cup \{ \infty \}$, the following notation
\begin{equation}
\mathfrak{C}^{R}( \tau, S_{t}, E_{t}) := \mathbb{E}^{\mathbb{Q}}_{S_{t},E_{t}} \big[ e^{-r (T_{R} \wedge \tau)} \left( V_{T_{R} \wedge \tau}^{S} \vee 0\right) \big],
\end{equation}
\begin{equation}
\mathfrak{C}^{R,\star}( \tau, E_{t}) : = \mathfrak{C}^{R}( \tau, 1 , E_{t}),
\end{equation}
\noindent and note that both $\mathfrak{C}^{R}_{\bf E}(\cdot)$ and $ \mathfrak{C}_{\bf A}^{R}(\cdot)$ can be expressed in terms of $\mathfrak{C}^{R}(\cdot)$ as
\begin{equation}
\mathfrak{C}_{\bf E}^{R}(S_{t},E_{t}) = \mathfrak{C}^{R}(\infty, S_{t},E_{t}) \hspace{1.5em} \mbox{and} \hspace{1.7em} \mathfrak{C}_{{\bf A}}^{R}(S_{t},E_{t})  = \sup \limits_{\tau \in \mathfrak{T}_{[0,\infty)}} \mathfrak{C}^{R}(\tau,S_{t},E_{t}). 
\label{GOadRel}
\end{equation}
\noindent Then, conditioning on the random time $T_{R}$, allows us to write
\begin{equation}
\mathfrak{C}^{R}(\tau,S_{t},E_{t}) = S_{t} \cdot \mathfrak{C}^{R,\star}(\tau,E_{t}) = S_{t} \cdot \int \limits_{0}^{\infty} \vartheta e^{-\vartheta t_{R}} \, \mathfrak{C}^{\star}(t_{R} \wedge \tau, E_{t}) \, dt_{R},
\end{equation}
\noindent which implies via Relation (\ref{GOadRel}) and with $\mathfrak{C}_{\bf E}^{R,\star}(E_{t}) := \mathfrak{C}_{\bf E}^{R}(1,E_{t})$ and $\mathfrak{C}_{\bf A}^{R,\star}(E_{t}) := \mathfrak{C}_{\bf A}^{R}(1,E_{t})$ that
\begin{align}
\mathfrak{C}_{{\bf E}}^{R}(S_{t}, E_{t})  =  S_{t} \cdot \mathfrak{C}_{{\bf E}}^{R,\star}& (E_{t}) = S_{t} \cdot  \int \limits_{0}^{\infty} \vartheta e^{-\vartheta t_{R}} \,\mathfrak{C}^{\star}_{\bf E}(t_{R}, E_{t}) \, dt_{R}, \label{morca1} \\
\mathfrak{C}_{{\bf A}}^{R}(S_{t}, E_{t})  =  S_{t} \cdot \mathfrak{C}_{{\bf A}}^{R,\star}(E_{t}) & = S_{t} \cdot \sup \limits_{\tau \in \mathfrak{T}_{[0,\infty)}}  \int \limits_{0}^{\infty} \vartheta e^{-\vartheta t_{R}} \, \mathfrak{C}^{\star}(t_{R} \wedge \tau, E_{t}) \, dt_{R}. \label{morca2}
\end{align}
\noindent Therefore, we focus in the sequel on the valuation of $\mathfrak{C}_{\bf E}^{R,\star}(\cdot)$ and $\mathfrak{C}_{\bf A}^{R,\star}(\cdot)$ and solve these valuation problems by relying on results for their deterministic versions $\mathfrak{C}_{\bf E}^{\star}(\cdot)$ and $\mathfrak{C}_{\bf A}^{\star}(\cdot)$. \vspace{1em} \\
\noindent At this point, we should notice that, in general, the switching options $\mathfrak{C}_{\bf E}^{R,\star}(\cdot)$ and $\mathfrak{C}_{\bf A}^{R,\star}(\cdot)$ may have infinite value for certain parameter choices. To avoid this to happen, we assume in the sequel that the following condition is satisfied
\begin{equation}
\vartheta + \tilde{r} - \Phi_{Y}^{(1)}(1) > 0.
\label{CONDinf}
\end{equation}
\noindent That this condition indeed rules out infinite values for $\mathfrak{C}_{\bf E}^{R,\star}(\cdot)$ and $\mathfrak{C}_{\bf A}^{R,\star}(\cdot)$ can be seen by combining Representation (\ref{morca2}) with Theorem 1 in \cite{mo02}, the submartingale property of the process $(E_{t})_{t \geq 0}$ under $\tilde{r} \leq \Phi_{Y}^{(1)}(1)$, and the well-known representation
\begin{equation}
\mathfrak{C}_{\bf E}^{\star}(t_{R},x) = x e^{-(\tilde{r}-\Phi_{Y}^{(1)}(1)) t_{R}} \; \tilde{\mathbb{Q}}^{(1)}\big(E_{t_{R}} \geq 1 \big) - e^{-\tilde{r} t_{R}} \; \mathbb{Q}^{(1)}\big(E_{t_{R}} \geq 1 \big) ,
\label{Zutzut}
\end{equation}
\noindent where
$$ \left. \frac{d \tilde{\mathbb{Q}}^{(1)}}{d \mathbb{Q}^{(1)}} \right|_{\mathcal{F}_{t}} :=  e^{Y_{t} - t \Phi_{Y}^{(1)}(1)}.$$
\subsection{OIDE for Tradeability Valuation}
\noindent We now turn to the derivation of ordinary integro-differential equations (OIDEs) that can be used for tradeability valuation. As earlier, we assume Condition (\ref{condidi}) and focus on the corresponding valuation problems for $\mathfrak{C}_{\bf A}^{R,\star}(\cdot)$ and $\mathfrak{C}_{\bf E}^{R,\star}(\cdot)$.
We therefore set
\begin{equation}
\mathfrak{L}^{R,\star} (E_{t}) :=  \mathfrak{L}^{R}(1,E_{t}) = \mathfrak{C}_{\bf A}^{R,\star}(E_{t}) - \mathfrak{C}_{\bf E}^{R,\star}(E_{t})
\end{equation}
\noindent and note, as in Remark 4, that
$$ \mathfrak{L}_{Rel.}^{R}(S_{t},E_{t}) = \mathfrak{L}_{Rel.}^{R}(1,E_{t}) = \frac{\mathfrak{L}^{R,\star}(E_{t})}{\mathfrak{C}_{\bf E}^{R,\star}(E_{t})} = \frac{\mathfrak{C}_{\bf A}^{R,\star}(E_{t})}{\mathfrak{C}_{\bf E}^{R,\star}(E_{t})} -1 .$$
\subsubsection{OIDE \rom{1}: Illiquid Scenario}
\noindent To tackle the illiquid scenario we use Representation (\ref{morca1}) and relevant results for the deterministic valuation problem. Indeed, combining few integrability results with Proposition \ref{Prop1} and (strong) Markovian arguments leads to the next proposition. A proof is provided in Appendix C.
\begin{Prop}
\label{Prop4}
Assume that Conditions (\ref{CONDinf}) and (\ref{bediwi}) hold. Then, the value of the European-type switching option under stochastic illiquidity horizon, $\mathfrak{C}_{{\bf E}}^{R,\star}(\cdot)$, is continuous on $[0,\infty)$, $C^{1}$ on $(0, \infty)$ and solves the ordinary integro-differential equation
\begin{equation}
\vartheta \left( \left( x -1 \right)^{+} - \mathfrak{C}_{\bf E}^{R,\star}(x) \right) +  \mathcal{A}_{E} \mathfrak{C}_{\bf E}^{R,\star}(x) - \tilde{r}  \mathfrak{C}_{\bf E}^{R,\star}(x)  =   0 \label{OIDE1}
\end{equation}
\noindent on $(0,\infty)$ with initial condition
\begin{equation}
\mathfrak{C}_{{\bf E}}^{R,\star}(0)  =  0 . \label{OIDE2}
\end{equation}
\end{Prop}
\subsubsection{OIDE \rom{2}: Liquid Scenario}
\noindent To deal with the liquid scenario, we start by collecting few properties of the American-type switching option $\mathfrak{C}_{\bf A}^{R,\star}(\cdot)$ in Lemma \ref{lemma2}. These results are analogues of the properties presented in Lemma \ref{lemma1} and their proof does not substantially differ from the proof provided, for $\mathfrak{C}_{\bf A}^{\star}(\cdot)$, in Appendix B. Therefore, we only state the results here. \vspace{1em} \\
\begin{Lem}
\label{lemma2}
The following properties hold:
\begin{itemize} \setlength \itemsep{-0.1em}
\item[$a)$] The American-type switching option $x \mapsto \mathfrak{C}_{\bf A}^{R,\star}(x)$ is non-decreasing and convex on $[0,\infty)$.
\item[$b)$] For $\tilde{r} \geq \Phi_{Y}^{(1)}(1)$ we have that
$$| \mathfrak{C}_{{\bf A}}^{R,\star}(x) - \mathfrak{C}_{{\bf A}}^{R,\star}(y) |  \leq | x -y |, \; \; \; \; \; \; \forall x,y \in  [0,\infty).$$
\end{itemize}
\end{Lem}
\noindent Combining Lemma \ref{lemma2} with well-known results for perpetual American options under exponential Lévy models (cf.~\cite{mo02}) allows us to derive the next proposition, which is the analogue of Proposition \ref{propp2} under stochastic illiquidity horizon. This result extends the findings obtained in \cite{ca98} in the classical Black \& Scholes model. A proof is provided in Appendix C.
\begin{Prop}
\label{Prop5}
Assume that Condition (\ref{CONDinf}) holds. Then, the value of the American-type switching option under stochastic illiquidity horizon, $\mathfrak{C}_{{\bf A}}^{R,\star}(\cdot)$, is continuous on $[0,\infty)$ and satisfies the following problem
\begin{align}
 \vartheta \left( \left( x -1 \right)^{+} - \mathfrak{C}_{\bf A}^{R,\star}(x) \right) +  \mathcal{A}_{E} \mathfrak{C}_{\bf A}^{R,\star}(x) - \tilde{r}  \mathfrak{C}_{\bf A}^{R,\star}(x) & =  0, \hspace{4.5em} x \in (0,\mathfrak{b}_{s}^{R}), \hspace{3.5em}\\
\mathfrak{C}_{\bf A}^{R,\star}(x) & = (x-1)^{+}, \hspace{1.5em} x \in [\mathfrak{b}_{s}^{R}, \infty) ,
\end{align}
\noindent with (unknown) boundary $\mathfrak{b}_{s}^{R} > 0$ and initial condition
\begin{equation}
\mathfrak{C}_{{\bf A}}^{R,\star}(0)   =  0 .
\label{OIDEinI}
\end{equation}
\end{Prop}

\subsubsection{OIDE \rom{3}: Free-Boundary Characterization}
\noindent Deriving a free-boundary characterization for the (absolute) tradeability premium under stochastic illiquidity horizon can now be done by relying on the previous results and proofs. First, the proof of Proposition \ref{Prop5} reveals that, for $\tilde{r} \leq \Phi_{Y}^{(1)}(1) $, the American-type switching option $\mathfrak{C}_{\bf A}^{R,\star}(\cdot)$ reduces to its European counterpart $\mathfrak{C}_{\bf E}^{R,\star}(\cdot)$. Secondly, combining the latter proof with Lemma \ref{lemma2} allows us to derive a representation of the holding and switching regions as
\begin{align}
\mathfrak{D}_{h} & := \big \{ x \in [0,\infty) : \,  \mathfrak{C}_{\bf A}^{R,\star}(x) > \left(x-1 \right)^{+} \big \} = \big[0,\mathfrak{b}_{s}^{R} \big), \\
\mathfrak{D}_{s} & := \big \{ x \in [0,\infty) : \, \mathfrak{C}_{\bf A}^{R,\star}(x) = \left(x-1 \right)^{+} \big \} = \big[\mathfrak{b}_{s}^{R},\infty \big).
\end{align}
\noindent Since the smooth-fit property can be obtained using the same methods as in the deterministic version of the problem (cf.~Appendix B), these results finally lead to the following free-boundary characterization of the (absolute) tradeability premium $\mathfrak{L}^{R,\star}(\cdot) $.
\begin{Prop}
\label{prop6}
Assume that $\sigma_{Y} \neq 0 $ holds. Then, we have the following properties:
\begin{itemize} \setlength \itemsep{-0.5em}
\item[1.] If $\tilde{r} \leq \Phi_{Y}^{(1)}(1)$, the (absolute) tradeability premium $\mathfrak{L}^{R,\star}(\cdot)$ satisfies
$$ \mathfrak{L}^{R,\star}(x) = 0, \hspace{2em} \forall x \in [0,\infty) .$$
\item[2.] If $\tilde{r} > \Phi_{Y}^{(1)}(1)$, the pair $\big(\mathfrak{L}^{R,\star}(\cdot),\mathfrak{b}_{s}^{R}\big)$ solves the following free-boundary problem:
\begin{equation}
 \mathcal{A}_{E} \mathfrak{L}^{R,\star} (x) - \big( \tilde{r} + \vartheta \big) \mathfrak{L}^{R,\star}(x)  =  0 , \; \; \; \; x \in (0,\mathfrak{b}_{s}^{R}), 
\end{equation}
\noindent subject to the boundary conditions
\begin{align}
\mathfrak{L}^{R,\star}(\mathfrak{b}_{s}^{R}) & = \mathfrak{b}_{s}^{R} - 1 - \mathfrak{C}_{{\bf E}}^{R,\star}(\mathfrak{b}_{s}^{R}), \\
\partial_{x} \mathfrak{L}^{R,\star}(\mathfrak{b}_{s}^R) & = 1 - \partial_{x} \mathfrak{C}_{{\bf E}}^{R,\star} (\mathfrak{b}_{s}^R),  \label{eqqquazzzion} \\
\mathfrak{L}^{R,\star}(0) & =  0, 
\end{align}
\end{itemize}
\end{Prop}
\noindent \underline{\bf Remark 6.}
\begin{itemize} \setlength \itemsep{-0.1em}
\item[i)] Proposition \ref{prop6} is the analogue of Proposition \ref{prop3} under stochastic illiquidity horizon. As such, it allows for an easy derivation of tradeability values via the application of well-established numerical methods and is therefore of great practical importance.
\item[ii)] Recall the financial interpretation of Condition $\tilde{r} \leq \Phi_{Y}^{(1)}(1)$ from Remark 5.ii).
\end{itemize}
\hspace{45em} \scalebox{0.75}{$\blacklozenge$}
\section{Numerical Results}
\label{numres}
\noindent To illustrate our approach, we finally derive tradeability premiums by combining the results from Section~\ref{SECC} and Section~\ref{SEC4new} with the algorithm developed in \cite{Ma18} and Appendix D.
\subsection{Model Consideration and Illiquidity Factor}
\noindent We consider the general asset dynamics defined by (\ref{eq1})-(\ref{bzbzeq}), i.e.~we assume that the initial asset dynamics $(S_{t})_{t\geq 0}$ are described (under $\mathbb{Q}$) by
\begin{equation}
\label{salope}
dS_{t} =  S_{t-} \bigg( \Phi_{X}(1) dt + \sigma_{X} dW_{t}^{X} + \int \limits_{ \mathbb{R}} (e^{y} -1) \tilde{N}_{X}(dt,dy) \bigg),
\end{equation}
\noindent and let the cash-flow process $(C_{t})_{t \geq 0}$ evolve (under $\mathbb{Q}$) according to (\ref{eq2}) with $(Y_{t})_{t \geq 0}$ specified by
\begin{equation}
Y_{t} := \big( b-\lambda(e^{\varphi}-1) - \frac{1}{2} \sigma^2\big) t + \sigma W_{t}^{Y} + \varphi N_{t}, \hspace{1.5em} t \geq 0.
\label{Dzzyn1}
\end{equation}
\noindent As in Section \ref{SEC1}, $(W_{t}^{X})_{t\geq 0}$ and $(W_{t}^{Y})_{t \geq 0}$ are correlated Brownian motions with correlation $\rho \in [-1,1]$ and $(N_{t})_{t \geq 0}$ denotes a Poisson process with deterministic intensity $\lambda > 0$ and that is independent of the Poisson random measure $N_{X}$. We emphasize that (m)any more advanced models could be considered for the dynamics of the cash-flow process $(C_{t})_{t \geq 0}$. In particular, the algorithm used in the computation of the liquidity premiums under deterministic illiquidity horizon could be analogously applied under Merton's model as well as under any hyper-exponential jump-diffusion model (cf.~\cite{Ma18}, \cite{ck11}, \cite{cs14}). Neveretheless, we stick for simplicity of the exposition with Dynamics (\ref{Dzzyn1}). We will determine (the range of) the relevant parameters in a moment. For now, we just note that $\Phi_{Y}(1)=b$. \vspace{1em} \\
\noindent Instead of considering absolute tradeability premiums, we next rely on relative quantities. Additionally, we slightly change our approach: While the relative tradeability premiums $\mathfrak{L}_{Rel.}(\cdot)$ and $\mathfrak{L}_{Rel.}^{R}(\cdot)$ provide a simple way to evaluate a tradeable asset based on the value of an illiquid equivalent,\footnote{The (time-$t$) value of a tradeable asset under deterministic and stochastic illiquidity horizon can be readily obtained by multiplying the value of its illiquid equivalent with the factor $\big(1+\mathfrak{L}_{Rel.}(\mathcal{T},1,E_{0})\big)$ and $\big(1+\mathfrak{L}_{Rel.}^{R}(1,E_{0})\big)$, respectively.}~one is more often interested in the reverse, i.e.~in evaluating an illiquid asset given the value of a tradeable equivalent. This motivates the consideration of corresponding time-$t$ illiquidity factors $\mathfrak{I}_{Rel.}(\cdot)$ and $\mathfrak{I}_{Rel.}^{R}(\cdot)$, defined via
\begin{align}
\mathfrak{I}_{Rel.}(\mathcal{T},S_{t},E_{t})  := \mathfrak{I}_{Rel.}(\mathcal{T}&,1,E_{t})  := \big(1+\mathfrak{L}_{Rel.}(\mathcal{T},1,E_{t}) \big)^{-1}, \label{relidi1}\\
\mathfrak{I}_{Rel.}^{R}(S_{t},E_{t}) :=  \mathfrak{I}_{Rel.}^{R}(&1,E_{t}) := \big(1+\mathfrak{L}_{Rel.}^{R}(1,E_{t}) \big)^{-1}.
\label{relidi}
\end{align}
\noindent Our numerical results will focus on these quantities, i.e.~we will always express the value of an illiquid asset as percentage of the value of a liquid equivalent. However, as should be clear from (\ref{relidi1}) and (\ref{relidi}), relative tradeability premiums and illiquidity factors are dual objects. We will therefore always compute illiquidity factors by means of Relations (\ref{relidi1}), (\ref{relidi}) and of the tradeability valuation approach discussed in the previous sections.
\subsection{Parameter Specification}
\noindent We next specify the parameters in our model: First, we note from the discussion in Section~\ref{SECC} and Section~\ref{SEC4new} that dynamics (\ref{salope}) only influences the relative tradeability premium via the value of its parameters $\Phi_{X}(1)$, $\sigma_{X}$ and $\rho$. Therefore, (time-$t$) illiquidity factors can be computed (by means of relative tradeability premiums), once the following parameters are specified: $\mathcal{T}$, $\vartheta$, $r$, $\Phi_{X}(1)$, $\sigma_{X}$, $\rho$, $b$, $\sigma$, $\varphi$, $\lambda$, $C_{0}$. \vspace{1em} \\
\noindent We determine these parameters by adjusting the parameter choice in \cite{ms} to current (US-)market data. For instance, all our numerical experiments assume a risk-free rate of $2.25\%$, which corresponds to a rough average of the US treasury yields with maturity $\mathcal{T} \in \{0.5, 1, 2, 5\}$ as of the end of March 2018.\footnote{The following values were extracted from Bloomberg, as of Friday 30 March 2018: 6-month US treasury yield, $1.91\%$; 1-year US treasury yield, $2.08\%$; 2-year US treasury yield, $2.27\%$; 5-year treasury yield, $2.56\%$.}~Since our numerical experiments consider the following illiquidity horizons and rates of arrival 
$$ \mathcal{T} \in \{0.5,1.5,2.5,5\} \hspace{2em} \mbox{and} \hspace{2em} \vartheta \in \Big\{ \frac{1}{\mathcal{T}}: \; \mathcal{T} \in \{0.5,1.5,2.5,5\} \Big\} ,$$
\noindent this risk-free rate seems to be a sensible choice. In analogy to the typical choices made in the option pricing literature, we take the volatility of the initial asset to be either $\sigma_{X}=20\%$ or $\sigma_{X}=40\%$. Additionally, we set $\Phi_{X}(1)=0.005$ and allow this way for a dividend rate of $\delta:= r- \Phi_{X}(1) = 1.75\%$. For the project's cash-flow dynamics, we take three different jump parameters (no jump, negative jump of $15\%$ and negative jump of $30\%$) and assume that jumps occur on average every $2$ years ($\lambda=0.5$). The volatility of the project is specified by $\sigma=20\%$. This parameter was already used in \cite{ms} where it represents the average standard deviation for unlevered equity in the US. The authors obtained it based on the average standard deviation of stocks on the New York Stock Exchange while assuming a debt to value ratio of $1/3$ (cf.~\cite{ms}). For the correlation, we take three generic correlation coefficients ($\rho=0.5$, $\rho=0$ and $\rho=-0.5$) that were similarly used in \cite{ms}. Finally, instead of specifying $C_{0}$, we express the results in terms of $E_{0}$, the expected net present value of the future cash-flow generated out of a one-unit investment in the project, and take $E_{0} \in \{0.9,1.0,1.1,1.2\}$. This is only done for the sake of simpler presentation and does not constitute a restriction. Indeed, $C_{0}$ can be easily recovered, for each set of parameter, out of $E_{0}$ via the relation $E_{0} = \frac{{C}_{0}}{r- b}$ (cf.~Section \ref{SEC1} with $\Phi_{Y}(1)=b$).
\subsection{Numerical Results: Deterministic Illiquidity Horizon}
\label{NUDE}
\noindent We first consider the illiquidity factor under deterministic illiquidity horizon, $\mathfrak{I}_{Rel.}(\cdot)$, and derive numerical results by combining Proposition \ref{prop3} with the algorithm developed in \cite{Ma18}. The results are displayed for $b=0.00$ in Table~\ref{table 8} and for $b=-0.04$ in Table~\ref{table 9}. \vspace{1em} \\
\noindent As seen from Table~\ref{table 8} and Table~\ref{table 9}, the (relative) tradeability premium substantially depends on the parameter choices and can become very large. Additionally, several properties of the (relative) tradeability premium can be extracted from these tables: As expected, one first sees that the discount for illiquidity, and hence the (relative) tradeability premium, increases with increasing illiquidity horizon $ \mathcal{T}$. Moreover, increasing the initial value of the alternative project $E_{0}$ (or, equivalently, the initial cash-flow level $C_{0}$), increases the discount for illiquidity. Secondly, we notice that diminishing the growth rate of the cash-flow process (i.e. diminishing $b$) seems to have a positive impact on the value of tradeability. This is intuitively clear, since reducing the growth rate of the cash-flow process induces a reduction of the project's expected value as time increases. When holding an illiquid asset the investor is forced to keep its position until tradeability (i.e.~time $T_{D}$) and its final exchange decision will have, in expectation, less value than before. \newpage
\begin{center}
\captionof{table}{Theoretical illiquidity factor, $\mathfrak{I}_{Rel.}(\mathcal{T},1,E_{0})$, for $b=0.00$, $\sigma=0.2$, $\lambda=0.5$ and $\Phi_{X}(1)=0.005$.}
\label{table 8}
\scalebox{0.75}{
\begin{tabular}{lrcccccccccccc}  
\toprule
\multicolumn{14}{c}{\it Illiquidity Factor $\mathfrak{I}_{Rel.}(\mathcal{T},1,E_{0})$} \\
\cmidrule{1-14}
  &         &      & \multicolumn{3}{c}{\it No Jump}   &  & \multicolumn{3}{c}{\it Jumps: $\varphi=\log(0.85)$} &  & \multicolumn{3}{c}{\it Jumps: $\varphi=\log(0.7)$}\\
\cmidrule{4-6} \cmidrule{8-10} \cmidrule{12-14}
 \multicolumn{2}{c}{\it Parameters}   &  \multicolumn{12}{c}{\it Correlation $\rho$}  \\
\cmidrule(r){1-2} \cmidrule{4-14} 
  &    $E_{0}$     & &    $\rho=0.5$    &    $\rho=0$  &   $\rho=-0.5$  & &  $\rho=0.5$   & $\rho=0$  & $\rho=-0.5$  & &   $\rho=0.5$      &   $\rho=0$ &  $\rho=-0.5$  \\
\midrule
(1.1.) & $0.9$ & & $1.000$ & $0.998$ & $0.986$ & & $1.000 $ & $0.999$ & $0.993$ & & $1.000$ & $0.999$ & $0.999$ \\
$r=2.25\%$ & $1.0$ & & $1.000$ & $0.998$ & $0.982$ & & $1.000$ & $0.999$ & $0.990$ & & $1.000$ & $0.999$ & $0.999$ \\
$\sigma_{X}=20\%$ & $1.1$ & & $1.000$ & $0.997$ & $0.975$ & & $1.000$ & $0.998$ & $0.985$ & & $1.000$ & $0.999$ & $0.997$ \\
$\mathcal{T}=0.5$ & $1.2$ & & $1.000$ & $0.996$ & $0.965$ & &  $1.000$ & $0.998$ & $0.977$ & & $1.000$ & $0.999$ & $0.995$ \\
\\
(1.2.) & $0.9$ & & $1.000$ & $0.994$ & $0.966$ & & $1.000$ & $0.996$ & $0.975$ & & $1.000$ & $0.998$ & $0.991$ \\
$r=2.25\%$ & $1.0$ & & $1.000$ & $0.994$ & $0.958$ & & $1.000$ & $0.995$ & $0.968$ & & $1.000$ & $0.998$ & $0.987$ \\
$\sigma_{X}=20\%$ & $1.1$ & & $1.000$ & $0.993$ & $0.947$ & & $1.000$ & $0.994$ & $0.960$ & & $1.000$ & $0.997$ & $0.983$ \\
$\mathcal{T}=1.5$ & $1.2$ & & $1.000$ & $0.991$ & $0.936$ & &  $1.000$ & $0.993$ & $0.951$ & & $1.000$ & $0.997$ & $0.978$ \\
\\
(1.3.) & $0.9$ & & $1.000$ & $0.990$ & $0.946$ & & $1.000$ & $0.992$ & $0.957$ & & $1.000$ & $0.995$ & $0.979$ \\
$r=2.25\%$ & $1.0$ & & $1.000$ & $0.989$ & $0.935$ & & $1.000$ & $0.991$ & $0.948$ & & $1.000$ & $0.994$ & $0.973$ \\
$\sigma_{X}=20\%$ & $1.1$ & & $1.000$ & $0.987$ & $0.922$ & & $1.000$ & $0.989$ & $0.938$ & & $1.000$ & $0.994$ & $0.966$ \\
$\mathcal{T}=2.5$ & $1.2$ & & $1.000$ & $0.985$ & $0.909$ & &  $1.000$ & $0.987$ & $0.927$ & & $1.000$ & $0.992$ & $0.959$ \\
\\
(1.4.) & $0.9$ & & $1.000$ & $0.979$ & $0.897$ & & $1.000$ & $0.981$ & $0.912$ & & $1.000$ & $0.986$ & $0.942$ \\
$r=2.25\%$ & $1.0$ & & $1.000$ & $0.976$ & $0.880$ & & $1.000$ & $0.978$ & $0.898$ & & $1.000$ & $0.984$ & $0.932$ \\
$\sigma_{X}=20\%$ & $1.1$ & & $1.000$ & $0.973$ & $0.863$ & & $1.000$ & $0.976$ & $0.884$ & & $1.000$ & $0.982$ & $0.922$ \\
$\mathcal{T}=5$ & $1.2$ & & $1.000$ & $0.969$ & $0.846$ & &  $1.000$ & $0.972$ & $0.869$ & & $1.000$ & $0.980$ & $0.913$ \\
\midrule
\midrule
(2.1.) & $0.9$ & & $1.000$ & $0.998$ & $0.971$ & & $1.000$ & $0.999$ & $0.984$ & & $1.000$ & $0.999$ & $0.999$ \\
$r=2.25\%$ & $1.0$ & & $1.000$ & $0.998$ & $0.960$ & & $1.000$ & $0.999$ & $0.975$ & & $1.000$ & $0.999$ & $0.999$ \\
$\sigma_{X}=40\%$ & $1.1$ & & $1.000$ & $0.997$ & $0.943$ & & $1.000$ & $0.998$ & $0.962$ & & $1.000$ & $0.999$ & $0.993$ \\
$\mathcal{T}=0.5$ & $1.2$ & & $1.000$ & $0.996$ & $0.921$ & &  $1.000$ & $0.998$ & $0.945$ & & $1.000$ & $0.999$ & $0.985$ \\
\\
(2.2.) & $0.9$ & & $1.000$ & $0.994$ & $0.927$ & & $1.000$ & $0.996$ & $0.944$ & & $1.000$ & $0.998$ & $0.979$ \\
$r=2.25\%$ & $1.0$ & & $1.000$ & $0.994$ & $0.906$ & & $1.000$ & $0.995$ & $0.927$ & & $1.000$ & $0.998$ & $0.968$ \\
$\sigma_{X}=40\%$ & $1.1$ & & $1.000$ & $0.993$ & $0.883$ & & $1.000$ & $0.994$ & $0.909$ & & $1.000$ & $0.997$ & $0.957$ \\
$\mathcal{T}=1.5$ & $1.2$ & & $1.000$ & $0.991$ & $0.857$ & &  $1.000$ & $0.993$ & $0.888$ & & $1.000$ & $0.997$ & $0.945$ \\
\\
(2.3.) & $0.9$ & & $1.000$ & $0.990$ & $0.883$ & & $1.000$ & $0.992$ & $0.906$ & & $1.000$ & $0.995$ & $0.951$ \\
$r=2.25\%$ & $1.0$ & & $1.000$ & $0.989$ & $0.857$ & & $1.000$ & $0.991$ & $0.884$ & & $1.000$ & $0.994$ & $0.937$ \\
$\sigma_{X}=40\%$ & $1.1$ & & $1.000$ & $0.987$ & $0.830$ & & $1.000$ & $0.989$ & $0.862$ & & $1.000$ & $0.993$ & $0.922$ \\
$\mathcal{T}=2.5$ & $1.2$ & & $1.000$ & $0.985$ & $0.802$ & &  $1.000$ & $0.987$ & $0.839$ & & $1.000$ & $0.992$ & $0.907$ \\
\\
(2.4.) & $0.9$ & & $1.000$ & $0.979$ & $0.779$ & & $1.000$ & $0.981$ & $0.811$ & & $1.000$ & $0.986$ & $0.874$ \\
$r=2.25\%$ & $1.0$ & & $1.000$ & $0.976$ & $0.747$ & & $1.000$ & $0.978$ & $0.783$ & & $1.000$ & $0.984$ & $0.854$ \\
$\sigma_{X}=40\%$ & $1.1$ & & $1.000$ & $0.973$ & $0.715$ & & $1.000$ & $0.976$ & $0.756$ & & $1.000$ & $0.982$ & $0.835$ \\
$\mathcal{T}=5$ & $1.2$ & & $1.000$ & $0.969$ & $0.683$ & &  $1.000$ & $0.972$ & $0.730$ & & $1.000$ & $0.980$ & $0.816$ \\
\bottomrule
\end{tabular}
}
\end{center}
$\mbox{}$ \vspace{0.7em} \\
Next, looking at the illiquidity factor when varying both the correlation coefficient $\rho$ and the asset's volatility $\sigma_{X}$ leads to other interesting properties.\footnote{These properties can be also formally derived by combining the representation of $\mathfrak{L}_{Rel.}(\cdot)$ with Relations (\ref{dynQ}) and (\ref{didy}).}~First, we note that any increase in correlation leads to a decrease in the discount for illiquidity. However, an increase in the asset's volatility can have various effects on the value of tradeability. Indeed, while an increase in the asset's volatility has no impact on the illiquidity factor, and so on the discount for illiquidity, when the initial asset and the alternative project are uncorrelated ($\rho=0$), a non-zero correlation can lead to either an increase or a decrease in the discount for illiquidity. In fact, the effect mainly depends on the sign of the correlation coefficient $\rho$. While an increase in the asset's volatility leads, for~$\rho>0$, to a reduction in the value of tradeability (higher illiquidity factor), the same increase will lead to a higher tradeability premium (lower illiquidity factor), if the correlation coefficient is negative, i.e.~if~$\rho<0$.
\begin{center}
\captionof{table}{Theoretical illiquidity factor, $\mathfrak{I}_{Rel.}(\mathcal{T},1,E_{0})$, for $b=-0.04$, $\sigma=0.2$, $\lambda=0.5$ and $\Phi_{X}(1)=0.005$.}
\label{table 9}
\scalebox{0.75}{
\begin{tabular}{lrcccccccccccc}  
\toprule
\multicolumn{14}{c}{\it Illiquidity Factor $\mathfrak{I}_{Rel.}(\mathcal{T},1,E_{0})$} \\
\cmidrule{1-14}
  &         &      & \multicolumn{3}{c}{\it No Jump}   &  & \multicolumn{3}{c}{\it Jumps: $\varphi=\log(0.85)$} &  & \multicolumn{3}{c}{\it Jumps: $\varphi=\log(0.7)$}\\
\cmidrule{4-6} \cmidrule{8-10} \cmidrule{12-14}
 \multicolumn{2}{c}{\it Parameters}   &  \multicolumn{12}{c}{\it Correlation $\rho$}  \\
\cmidrule(r){1-2} \cmidrule{4-14} 
  &    $E_{0}$     & &    $\rho=0.5$    &    $\rho=0$  &   $\rho=-0.5$  & &  $\rho=0.5$   & $\rho=0$  & $\rho=-0.5$  & &   $\rho=0.5$      &   $\rho=0$ &  $\rho=-0.5$  \\
\midrule
(1.1.) & $0.9$ & & $0.986$ & $0.971$ & $0.954$ & & $0.993$ & $0.984$ & $0.973$ & & $1.000$ & $1.000$ & $1.000$ \\
$r=2.25\%$ & $1.0$ & & $0.982$ & $0.960$ & $0.932$ & & $0.990$ & $0.975$ & $0.955$ & & $0.999$ & $0.999$ & $0.996$ \\
$\sigma_{X}=20\%$ & $1.1$ & & $0.975$ & $0.943$ & $0.903$ & & $0.985$ & $0.962$ & $0.932$ & & $0.997$ & $0.993$ & $0.985$ \\
$\mathcal{T}=0.5$ & $1.2$ & & $0.965$ & $0.921$ & $0.870$ & &  $0.977$ & $0.945$ & $0.902$ & & $0.995$ & $0.985$ & $0.969$ \\
\\
(1.2.) & $0.9$ & & $0.966$ & $0.927$ & $0.878$ & & $0.975$ & $0.944$ & $0.905$ & & $0.991$ & $0.979$ & $0.961$ \\
$r=2.25\%$ & $1.0$ & & $0.958$ & $0.906$ & $0.843$ & & $0.968$ & $0.928$ & $0.876$ & & $0.987$ & $0.968$ & $0.941$ \\
$\sigma_{X}=20\%$ & $1.1$ & & $0.947$ & $0.883$ & $0.805$ & & $0.960$ & $0.909$ & $0.845$ & & $0.983$ & $0.957$ & $0.921$ \\
$\mathcal{T}=1.5$ & $1.2$ & & $0.936$ & $0.857$ & $0.763$ & &  $0.951$ & $0.888$ & $0.811$ & & $0.978$ & $0.945$ & $0.900$ \\
\\
(1.3.) & $0.9$ & & $0.946$ & $0.883$ & $0.805$ & & $0.957$ & $0.906$ & $0.840$ & & $0.979$ & $0.951$ & $0.912$ \\
$r=2.25\%$ & $1.0$ & & $0.935$ & $0.857$ & $0.764$ & & $0.948$ & $0.884$ & $0.806$ & & $0.973$ & $0.937$ & $0.888$ \\
$\sigma_{X}=20\%$ & $1.1$ & & $0.922$ & $0.830$ & $0.723$ & & $0.938$ & $0.862$ & $0.771$ & & $0.966$ & $0.922$ & $0.863$ \\
$\mathcal{T}=2.5$ & $1.2$ & & $0.909$ & $0.802$ & $0.680$ & &  $0.927$ & $0.839$ & $0.735$ & & $0.959$ & $0.907$ & $0.839$ \\
\\
(1.4.) & $0.9$ & & $0.897$ & $0.779$ & $0.642$ & & $0.912$ & $0.811$ & $0.691$ & & $0.942$ & $0.874$ & $0.788$ \\
$r=2.25\%$ & $1.0$ & & $0.880$ & $0.747$ & $0.599$ & & $0.898$ & $0.783$ & $0.652$ & & $0.932$ & $0.854$ & $0.758$ \\
$\sigma_{X}=20\%$ & $1.1$ & & $0.863$ & $0.715$ & $0.556$ & & $0.884$ & $0.756$ & $0.615$ & & $0.922$ & $0.835$ & $0.730$ \\
$\mathcal{T}=5$ & $1.2$ & & $0.846$ & $0.683$ & $0.515$ & &  $0.869$ & $0.730$ & $0.579$ & & $0.913$ & $0.816$ & $0.703$ \\
\midrule
\midrule
(2.1.) & $0.9$ & & $0.998$ & $0.971$ & $0.935$ & & $0.999$ & $0.984$ & $0.958$ & & $1.000$ & $1.000$ & $1.000$ \\
$r=2.25\%$ & $1.0$ & & $0.998$ & $0.960$ & $0.901$ & & $0.999$ & $0.975$ & $0.930$ & & $0.999$ & $0.999$ & $0.989$ \\
$\sigma_{X}=40\%$ & $1.1$ & & $0.997$ & $0.943$ & $0.858$ & & $0.999$ & $0.962$ & $0.896$ & & $0.999$ & $0.993$ & $0.970$ \\
$\mathcal{T}=0.5$ & $1.2$ & & $0.996$ & $0.921$ & $0.820$ & &  $0.998$ & $0.945$ & $0.854$ & & $0.999$ & $0.985$ & $0.944$ \\
\\
(2.2.) & $0.9$ & & $0.994$ & $0.927$ & $0.820$ & & $0.996$ & $0.944$ & $0.857$ & & $0.998$ & $0.979$ & $0.935$ \\
$r=2.25\%$ & $1.0$ & & $0.994$ & $0.906$ & $0.771$ & & $0.995$ & $0.928$ & $0.816$ & & $0.998$ & $0.968$ & $0.906$ \\
$\sigma_{X}=40\%$ & $1.1$ & & $0.993$ & $0.883$ & $0.718$ & & $0.994$ & $0.909$ & $0.772$ & & $0.997$ & $0.957$ & $0.876$ \\
$\mathcal{T}=1.5$ & $1.2$ & & $0.991$ & $0.857$ & $0.662$ & &  $0.993$ & $0.888$ & $0.725$ & & $0.996$ & $0.945$ & $0.844$ \\
\\
(2.3.) & $0.9$ & & $0.990$ & $0.883$ & $0.715$ & & $0.992$ & $0.906$ & $0.763$ & & $0.995$ & $0.951$ & $0.862$ \\
$r=2.25\%$ & $1.0$ & & $0.989$ & $0.857$ & $0.663$ & & $0.991$ & $0.884$ & $0.717$ & & $0.994$ & $0.937$ & $0.828$ \\
$\sigma_{X}=40\%$ & $1.1$ & & $0.987$ & $0.830$ & $0.609$ & & $0.989$ & $0.862$ & $0.671$ & & $0.993$ & $0.922$ & $0.793$ \\
$\mathcal{T}=2.5$ & $1.2$ & & $0.985$ & $0.802$ & $0.554$ & &  $0.987$ & $0.839$ & $0.624$ & & $0.992$ & $0.907$ & $0.759$ \\
\\
(2.4.) & $0.9$ & & $0.979$ & $0.779$ & $0.505$ & & $0.981$ & $0.811$ & $0.565$ & & $0.986$ & $0.874$ & $0.691$ \\
$r=2.25\%$ & $1.0$ & & $0.976$ & $0.747$ & $0.456$ & & $0.978$ & $0.783$ & $0.520$ & & $0.984$ & $0.854$ & $0.653$ \\
$\sigma_{X}=40\%$ & $1.1$ & & $0.973$ & $0.715$ & $0.408$ & & $0.976$ & $0.756$ & $0.477$ & & $0.982$ & $0.835$ & $0.618$ \\
$\mathcal{T}=5$ & $1.2$ & & $0.969$ & $0.683$ & $0.364$ & &  $0.972$ & $0.730$ & $0.436$ & & $0.980$ & $0.816$ & $0.584$ \\
\bottomrule
\end{tabular}
}
\end{center}
$\mbox{}$ \vspace{0.7em} \\
\noindent Finally, we look at the effect of negative jumps on the size of the tradeability premium. Here, we note that the discount for illiquidity seems to decrease with increasing jump size. Indeed, although negative jumps lead to an abrupt devaluation of the project, they also have a positive effect on the risk-adjusted drift in Dynamics (\ref{Dzzyn1}). While these effects neutralize each other in expectation for a fixed time $T_{D}$, jumps may substantially affect the value of earlier investments and therefore lead to a decrease in the value of~tradeability.

\subsection{Numerical Results: Stochastic Illiquidity Horizon}
\label{NUAp}
\noindent We next consider the illiquidity factor under stochastic illiquidity horizon, $\mathfrak{I}_{Rel.}^{R}(\cdot)$. As shown in Appendix~D, the tradeability premium $\mathfrak{L}^{R,\star}(\cdot)$ is now available in semi-closed form. Using these results as well as Relation~(\ref{relidi}), we derive corresponding illiquidity factors for $b=0.00$ and $b=-0.04$. The results are summarized in Table~\ref{table 10} and Table~\ref{table 11}, respectively.

\begin{center}
\captionof{table}{Theoretical illiquidity factor, $\mathfrak{I}_{Rel.}^{R}(1,E_{0})$, for $b=0.00$, $\sigma=0.2$, $\lambda=0.5$ and $\Phi_{X}(1)=0.005$.}
\label{table 10}
\scalebox{0.75}{
\begin{tabular}{lrcccccccccccc}  
\toprule
\multicolumn{14}{c}{\it Illiquidity Factor $\mathfrak{I}_{Rel.}^{R}(1,E_{0})$} \\
\cmidrule{1-14}
  &         &      & \multicolumn{3}{c}{\it No Jump}   &  & \multicolumn{3}{c}{\it Jumps: $\varphi=\log(0.85)$} &  & \multicolumn{3}{c}{\it Jumps: $\varphi=\log(0.7)$}\\
\cmidrule{4-6} \cmidrule{8-10} \cmidrule{12-14}
 \multicolumn{2}{c}{\it Parameters}   &  \multicolumn{12}{c}{\it Correlation $\rho$}  \\
\cmidrule(r){1-2} \cmidrule{4-14} 
  &    $E_{0}$     & &    $\rho=0.5$    &    $\rho=0$  &   $\rho=-0.5$  & &  $\rho=0.5$   & $\rho=0$  & $\rho=-0.5$  & &   $\rho=0.5$      &   $\rho=0$ &  $\rho=-0.5$  \\
\midrule
(1.1.) & $0.9$ & & $1.000$ & $0.999$ & $0.986$ & & $1.000$ & $0.977$ & $0.967$ & & $1.000$ & $0.937$ & $0.928$ \\
$r=2.25\%$ & $1.0$ & & $1.000$ & $0.999$ & $0.986$ & & $1.000$ & $0.977$ & $0.967$ & & $1.000$ & $0.937$ & $0.928$ \\
$\sigma_{X}=20\%$ & $1.1$ & & $1.000$ & $0.999$ & $0.984$ & & $1.000$ & $0.993$ & $0.983$ & & $1.000$ & $0.978$ & $0.972$ \\
$\mathcal{T}=0.5$ & $1.2$ & & $1.000$ & $0.998$ & $0.977$ & &  $1.000$ & $0.997$ & $0.984$ & & $1.000$ & $0.996$ & $0.991$ \\
\\
(1.2.) & $0.9$ & & $1.000$ & $0.995$ & $0.957$ & & $1.000$ & $0.987$ & $0.957$ & & $1.000$ & $0.974$ & $0.954$ \\
$r=2.25\%$ & $1.0$ & & $1.000$ & $0.995$ & $0.957$ & & $1.000$ & $0.987$ & $0.957$ & & $1.000$ & $0.974$ & $0.954$ \\
$\sigma_{X}=20\%$ & $1.1$ & & $1.000$ & $0.994$ & $0.955$ & & $1.000$ & $0.991$ & $0.960$ & & $1.000$ & $0.986$ & $0.967$ \\
$\mathcal{T}=1.5$ & $1.2$ & & $1.000$ & $0.993$ & $0.949$ & &  $1.000$ & $0.993$ & $0.958$ & & $1.000$ & $0.993$ & $0.974$ \\
\\
(1.3.) & $0.9$ & & $1.000$ & $0.989$ & $0.928$ & & $1.000$ & $0.984$ & $0.934$ & & $1.000$ & $0.978$ & $0.944$ \\
$r=2.25\%$ & $1.0$ & & $1.000$ & $0.989$ & $0.928$ & & $1.000$ & $0.984$ & $0.934$ & & $1.000$ & $0.978$ & $0.944$ \\
$\sigma_{X}=20\%$ & $1.1$ & & $1.000$ & $0.988$ & $0.926$ & & $1.000$ & $0.987$ & $0.935$ & & $1.000$ & $0.984$ & $0.951$ \\
$\mathcal{T}=2.5$ & $1.2$ & & $1.000$ & $0.987$ & $0.921$ & &  $1.000$ & $0.987$ & $0.933$ & & $1.000$ & $0.988$ & $0.954$ \\
\\
(1.4.) & $0.9$ & & $1.000$ & $0.970$ & $0.861$ & & $1.000$ & $0.968$ & $0.874$ & & $1.000$ & $0.968$ & $0.898$ \\
$r=2.25\%$ & $1.0$ & & $1.000$ & $0.970$ & $0.861$ & & $1.000$ & $0.968$ & $0.874$ & & $1.000$ & $0.968$ & $0.898$ \\
$\sigma_{X}=20\%$ & $1.1$ & & $1.000$ & $0.969$ & $0.860$ & & $1.000$ & $0.969$ & $0.874$ & & $1.000$ & $0.970$ & $0.901$ \\
$\mathcal{T}=5$ & $1.2$ & & $1.000$ & $0.969$ & $0.854$ & &  $1.000$ & $0.969$ & $0.871$ & & $1.000$ & $0.972$ & $0.901$ \\
\midrule
\midrule
(2.1.) & $0.9$ & & $1.000$ & $0.999$ & $0.963$ & & $1.000$ & $0.977$ & $0.949$ & & $1.000$ & $0.937$ & $0.915$ \\
$r=2.25\%$ & $1.0$ & & $1.000$ & $0.999$ & $0.963$ & & $1.000$ & $0.977$ & $0.949$ & & $1.000$ & $0.937$ & $0.915$ \\
$\sigma_{X}=40\%$ & $1.1$ & & $1.000$ & $0.999$ & $0.957$ & & $1.000$ & $0.993$ & $0.964$ & & $1.000$ & $0.978$ & $0.962$ \\
$\mathcal{T}=0.5$ & $1.2$ & & $1.000$ & $0.998$ & $0.937$ & &  $1.000$ & $0.997$ & $0.956$ & & $1.000$ & $0.996$ & $0.981$ \\
\\
(2.2.) & $0.9$ & & $1.000$ & $0.995$ & $0.899$ & & $1.000$ & $0.987$ & $0.909$ & & $1.000$ & $0.974$ & $0.924$ \\
$r=2.25\%$ & $1.0$ & & $1.000$ & $0.995$ & $0.899$ & & $1.000$ & $0.987$ & $0.909$ & & $1.000$ & $0.974$ & $0.924$ \\
$\sigma_{X}=40\%$ & $1.1$ & & $1.000$ & $0.994$ & $0.894$ & & $1.000$ & $0.991$ & $0.911$ & & $1.000$ & $0.986$ & $0.937$ \\
$\mathcal{T}=1.5$ & $1.2$ & & $1.000$ & $0.993$ & $0.877$ & &  $1.000$ & $0.993$ & $0.902$ & & $1.000$ & $0.993$ & $0.942$ \\
\\
(2.3.) & $0.9$ & & $1.000$ & $0.989$ & $0.844$ & & $1.000$ & $0.984$ & $0.863$ & & $1.000$ & $0.978$ & $0.896$ \\
$r=2.25\%$ & $1.0$ & & $1.000$ & $0.989$ & $0.844$ & & $1.000$ & $0.984$ & $0.863$ & & $1.000$ & $0.978$ & $0.896$ \\
$\sigma_{X}=40\%$ & $1.1$ & & $1.000$ & $0.988$ & $0.840$ & & $1.000$ & $0.987$ & $0.863$ & & $1.000$ & $0.984$ & $0.902$ \\
$\mathcal{T}=2.5$ & $1.2$ & & $1.000$ & $0.987$ & $0.825$ & &  $1.000$ & $0.987$ & $0.854$ & & $1.000$ & $0.988$ & $0.903$ \\
\\
(2.4.) & $0.9$ & & $1.000$ & $0.970$ & $0.732$ & & $1.000$ & $0.968$ & $0.761$ & & $1.000$ & $0.968$ & $0.814$ \\
$r=2.25\%$ & $1.0$ & & $1.000$ & $0.970$ & $0.732$ & & $1.000$ & $0.968$ & $0.761$ & & $1.000$ & $0.968$ & $0.814$ \\
$\sigma_{X}=40\%$ & $1.1$ & & $1.000$ & $0.969$ & $0.729$ & & $1.000$ & $0.969$ & $0.760$ & & $1.000$ & $0.970$ & $0.816$ \\
$\mathcal{T}=5$ & $1.2$ & & $1.000$ & $0.969$ & $0.717$ & &  $1.000$ & $0.969$ & $0.752$ & & $1.000$ & $0.972$ & $0.814$ \\
\bottomrule
\end{tabular}
}
\end{center}
$\mbox{}$ \vspace{0.7em} \\
\noindent A brief look at Table~\ref{table 10} and Table~\ref{table 11} reveals that the (relative) tradeability premium under stochastic illiquidity horizon, $\mathfrak{L}^{R, \star}(\cdot)$, has many similarities to its deterministic equivalent $\mathfrak{L}^{\star}(\cdot)$. Indeed, as its deterministic version, $\mathfrak{L}^{R, \star}(\cdot)$ is an increasing function of the (expected) illiquidity horizon $\mathcal{T} = \frac{1}{\vartheta}$ and a decreasing function in the correlation coefficient $\rho$. Additionally, reducing the growth rate in the dynamics of the cash-flow process (i.e.~reducing $b$) leads to an increase in the discount for illiquidity (and hence in the value of tradeability). Finally, varying the asset's volatility $\sigma_{X}$ may also have various effects on the value of tradeability. Indeed, while an increase in $\sigma_{X}$ does not impact the illiquidity factor when $\rho=0$, the same increase induces, for~$\rho>0$, a reduction and, for $\rho <0$,  an increase in the discount for illiquidity. 

\begin{center}
\captionof{table}{Theoretical illiquidity factor, $\mathfrak{I}_{Rel.}^{R}(1,E_{0})$, for $b=-0.04$, $\sigma=0.2$, $\lambda=0.5$ and $\Phi_{X}(1)=0.005$.}
\label{table 11}
\scalebox{0.75}{
\begin{tabular}{lrcccccccccccc}  
\toprule
\multicolumn{14}{c}{\it Illiquidity Factor $\mathfrak{I}_{Rel.}^{R}(1,E_{0})$} \\
\cmidrule{1-14}
  &         &      & \multicolumn{3}{c}{\it No Jump}   &  & \multicolumn{3}{c}{\it Jumps: $\varphi=\log(0.85)$} &  & \multicolumn{3}{c}{\it Jumps: $\varphi=\log(0.7)$}\\
\cmidrule{4-6} \cmidrule{8-10} \cmidrule{12-14}
 \multicolumn{2}{c}{\it Parameters}   &  \multicolumn{12}{c}{\it Correlation $\rho$}  \\
\cmidrule(r){1-2} \cmidrule{4-14} 
  &    $E_{0}$     & &    $\rho=0.5$    &    $\rho=0$  &   $\rho=-0.5$  & &  $\rho=0.5$   & $\rho=0$  & $\rho=-0.5$  & &   $\rho=0.5$      &   $\rho=0$ &  $\rho=-0.5$  \\
\midrule
(1.1.) & $0.9$ & & $0.986$ & $0.963$ & $0.933$ & & $0.967$ & $0.949$ & $0.926$ & & $0.928$ & $0.915$ & $0.899$ \\
$r=2.25\%$ & $1.0$ & & $0.986$ & $0.963$ & $0.933$ & & $0.967$ & $0.949$ & $0.926$ & & $0.928$ & $0.915$ & $0.899$ \\
$\sigma_{X}=20\%$ & $1.1$ & & $0.984$ & $0.957$ & $0.922$ & & $0.983$ & $0.964$ & $0.938$ & & $0.972$ & $0.962$ & $0.948$ \\
$\mathcal{T}=0.5$ & $1.2$ & & $0.977$ & $0.937$ & $0.891$ & &  $0.984$ & $0.956$ & $0.918$ & & $0.991$ & $0.981$ & $0.965$ \\
\\
(1.2.) & $0.9$ & & $0.957$ & $0.899$ & $0.831$ & & $0.957$ & $0.909$ & $0.853$ & & $0.954$ & $0.924$ & $0.886$ \\
$r=2.25\%$ & $1.0$ & & $0.957$ & $0.899$ & $0.831$ & & $0.957$ & $0.909$ & $0.853$ & & $0.954$ & $0.924$ & $0.886$ \\
$\sigma_{X}=20\%$ & $1.1$ & & $0.955$ & $0.894$ & $0.823$ & & $0.960$ & $0.911$ & $0.852$ & & $0.967$ & $0.937$ & $0.900$ \\
$\mathcal{T}=1.5$ & $1.2$ & & $0.949$ & $0.877$ & $0.793$ & &  $0.958$ & $0.902$ & $0.834$ & & $0.974$ & $0.942$ & $0.901$ \\
\\
(1.3.) & $0.9$ & & $0.928$ & $0.844$ & $0.752$ & & $0.934$ & $0.863$ & $0.784$ & & $0.944$ & $0.896$ & $0.840$ \\
$r=2.25\%$ & $1.0$ & & $0.928$ & $0.844$ & $0.752$ & & $0.934$ & $0.863$ & $0.784$ & & $0.944$ & $0.896$ & $0.840$ \\
$\sigma_{X}=20\%$ & $1.1$ & & $0.926$ & $0.840$ & $0.746$ & & $0.935$ & $0.863$ & $0.782$ & & $0.951$ & $0.902$ & $0.846$ \\
$\mathcal{T}=2.5$ & $1.2$ & & $0.921$ & $0.825$ & $0.720$ & &  $0.933$ & $0.854$ & $0.764$ & & $0.954$ & $0.903$ & $0.843$ \\
\\
(1.4.) & $0.9$ & & $0.861$ & $0.732$ & $0.611$ & & $0.874$ & $0.761$ & $0.651$ & & $0.898$ & $0.814$ & $0.728$ \\
$r=2.25\%$ & $1.0$ & & $0.861$ & $0.732$ & $0.611$ & & $0.874$ & $0.761$ & $0.651$ & & $0.898$ & $0.814$ & $0.728$ \\
$\sigma_{X}=20\%$ & $1.1$ & & $0.860$ & $0.729$ & $0.606$ & & $0.874$ & $0.760$ & $0.648$ & & $0.901$ & $0.816$ & $0.730$ \\
$\mathcal{T}=5$ & $1.2$ & & $0.854$ & $0.717$ & $0.586$ & &  $0.871$ & $0.752$ & $0.634$ & & $0.901$ & $0.814$ & $0.724$ \\
\midrule
\midrule
(2.1.) & $0.9$ & & $0.999$ & $0.963$ & $0.898$ & & $0.977$ & $0.949$ & $0.898$ & & $0.937$ & $0.915$ & $0.881$ \\
$r=2.25\%$ & $1.0$ & & $0.999$ & $0.963$ & $0.898$ & & $0.977$ & $0.949$ & $0.898$ & & $0.937$ & $0.915$ & $0.881$ \\
$\sigma_{X}=40\%$ & $1.1$ & & $0.999$ & $0.957$ & $0.880$ & & $0.993$ & $0.964$ & $0.906$ & & $0.978$ & $0.962$ & $0.931$ \\
$\mathcal{T}=0.5$ & $1.2$ & & $0.998$ & $0.937$ & $0.848$ & &  $0.997$ & $0.956$ & $0.876$ & & $0.996$ & $0.981$ & $0.944$ \\
\\
(2.2.) & $0.9$ & & $0.995$ & $0.899$ & $0.761$ & & $0.987$ & $0.909$ & $0.792$ & & $0.974$ & $0.924$ & $0.844$ \\
$r=2.25\%$ & $1.0$ & & $0.995$ & $0.899$ & $0.761$ & & $0.987$ & $0.909$ & $0.792$ & & $0.974$ & $0.924$ & $0.844$ \\
$\sigma_{X}=40\%$ & $1.1$ & & $0.994$ & $0.894$ & $0.748$ & & $0.991$ & $0.911$ & $0.788$ & & $0.986$ & $0.937$ & $0.857$ \\
$\mathcal{T}=1.5$ & $1.2$ & & $0.994$ & $0.877$ & $0.708$ & &  $0.993$ & $0.902$ & $0.752$ & & $0.993$ & $0.942$ & $0.853$ \\
\\
(2.3.) & $0.9$ & & $0.989$ & $0.844$ & $0.664$ & & $0.984$ & $0.863$ & $0.704$ & & $0.978$ & $0.896$ & $0.780$ \\
$r=2.25\%$ & $1.0$ & & $0.989$ & $0.844$ & $0.664$ & & $0.984$ & $0.863$ & $0.704$ & & $0.978$ & $0.896$ & $0.780$ \\
$\sigma_{X}=40\%$ & $1.1$ & & $0.988$ & $0.840$ & $0.654$ & & $0.987$ & $0.863$ & $0.700$ & & $0.984$ & $0.902$ & $0.785$ \\
$\mathcal{T}=2.5$ & $1.2$ & & $0.987$ & $0.825$ & $0.617$ & &  $0.987$ & $0.854$ & $0.674$ & & $0.988$ & $0.903$ & $0.778$ \\
\\
(2.4.) & $0.9$ & & $0.970$ & $0.732$ & $0.506$ & & $0.968$ & $0.761$ & $0.552$ & & $0.968$ & $0.814$ & $0.646$ \\
$r=2.25\%$ & $1.0$ & & $0.970$ & $0.732$ & $0.506$ & & $0.968$ & $0.761$ & $0.552$ & & $0.968$ & $0.814$ & $0.646$ \\
$\sigma_{X}=40\%$ & $1.1$ & & $0.970$ & $0.729$ & $0.499$ & & $0.969$ & $0.760$ & $0.548$ & & $0.970$ & $0.816$ & $0.646$ \\
$\mathcal{T}=5$ & $1.2$ & & $0.969$ & $0.717$ & $0.473$ & &  $0.969$ & $0.752$ & $0.528$ & & $0.972$ & $0.814$ & $0.637$ \\
\bottomrule
\end{tabular}
}
\end{center}
$\mbox{}$ \vspace{0.7em} \\
\noindent Although $\mathfrak{L}^{R,\star}(\cdot)$ resembles in many ways its deterministic version $\mathfrak{L}^{\star}(\cdot)$, the results in Table~\ref{table 10} and Table~\ref{table 11} also indicate clear differences between them. Other than for a deterministic illiquidity horizon, the tradeability premium under stochastic illiquidity horizon is no longer monotone in the initial value of the project, $E_{0}$. Additionally, the discount for illiquidity is not anymore monotone in the jump size $\varphi$.
\begin{figure}
\begin{subfigure}{.5\linewidth}
\centering
\includegraphics[scale=.17]{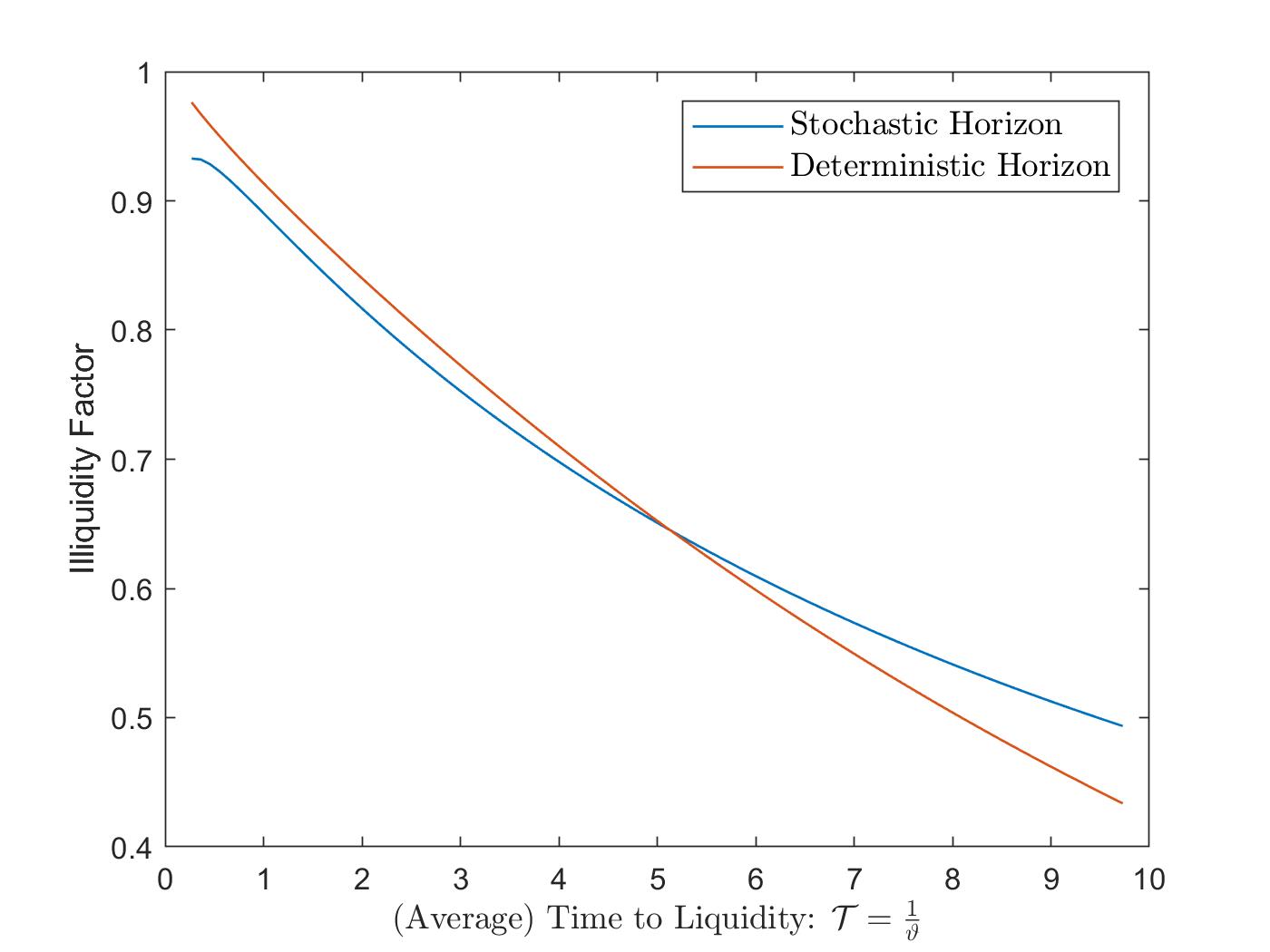}
\caption{$\mathfrak{I}_{Rel.}(\cdot)$ and $\mathfrak{I}^{R}_{Rel.}(\cdot)$ as functions of $\mathcal{T} = \frac{1}{\vartheta}$.}
\label{fig:sub1}
\end{subfigure}%
\begin{subfigure}{.5\linewidth}
\centering
\includegraphics[scale=.17]{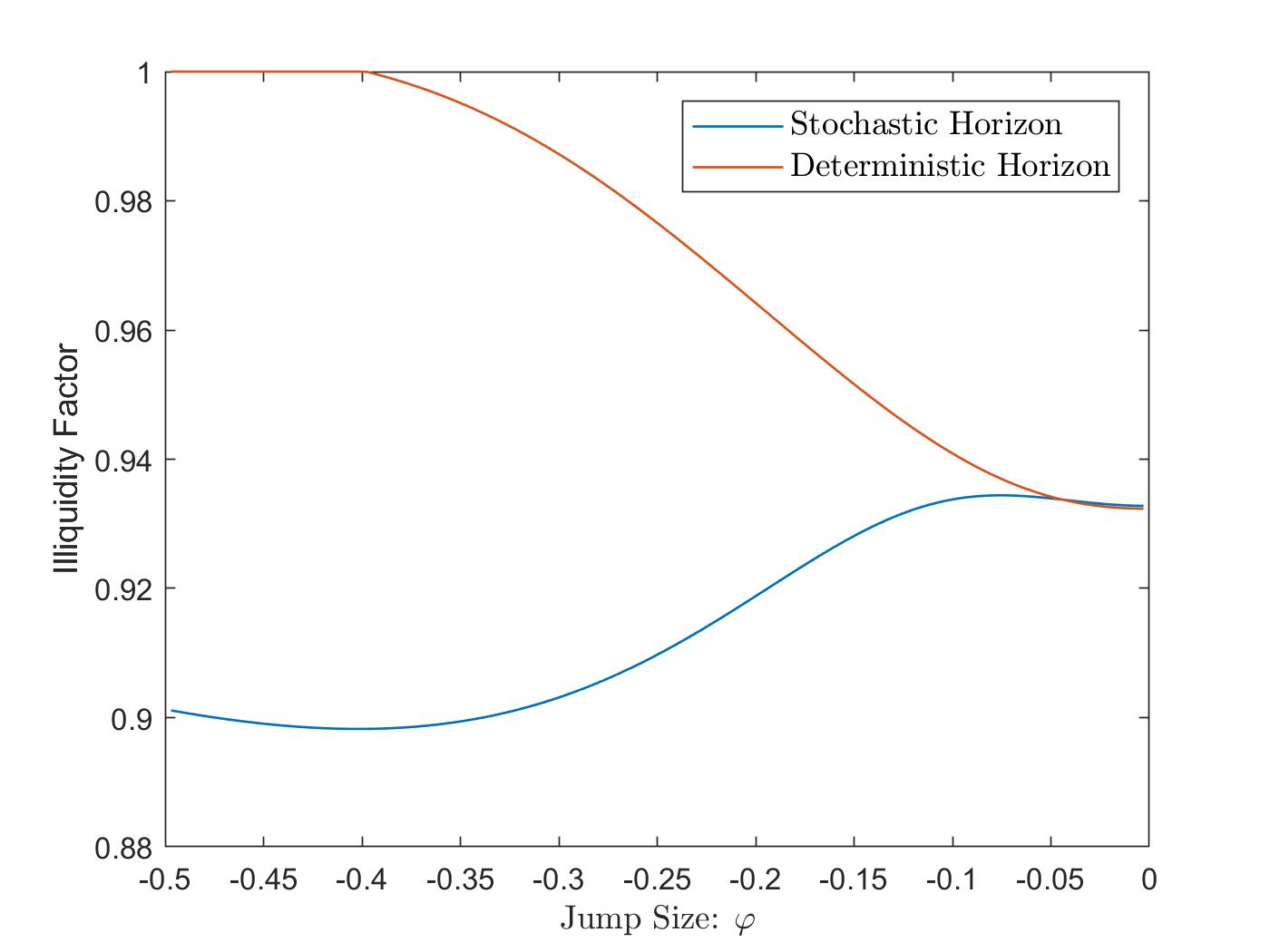}
\caption{$\mathfrak{I}_{Rel.}(\cdot)$ and $\mathfrak{I}^{R}_{Rel.}(\cdot)$ as functions of $\varphi$.}
\label{fig:sub2}
\end{subfigure}\\[1ex]
\begin{subfigure}{\linewidth}
\centering
\includegraphics[scale=.17]{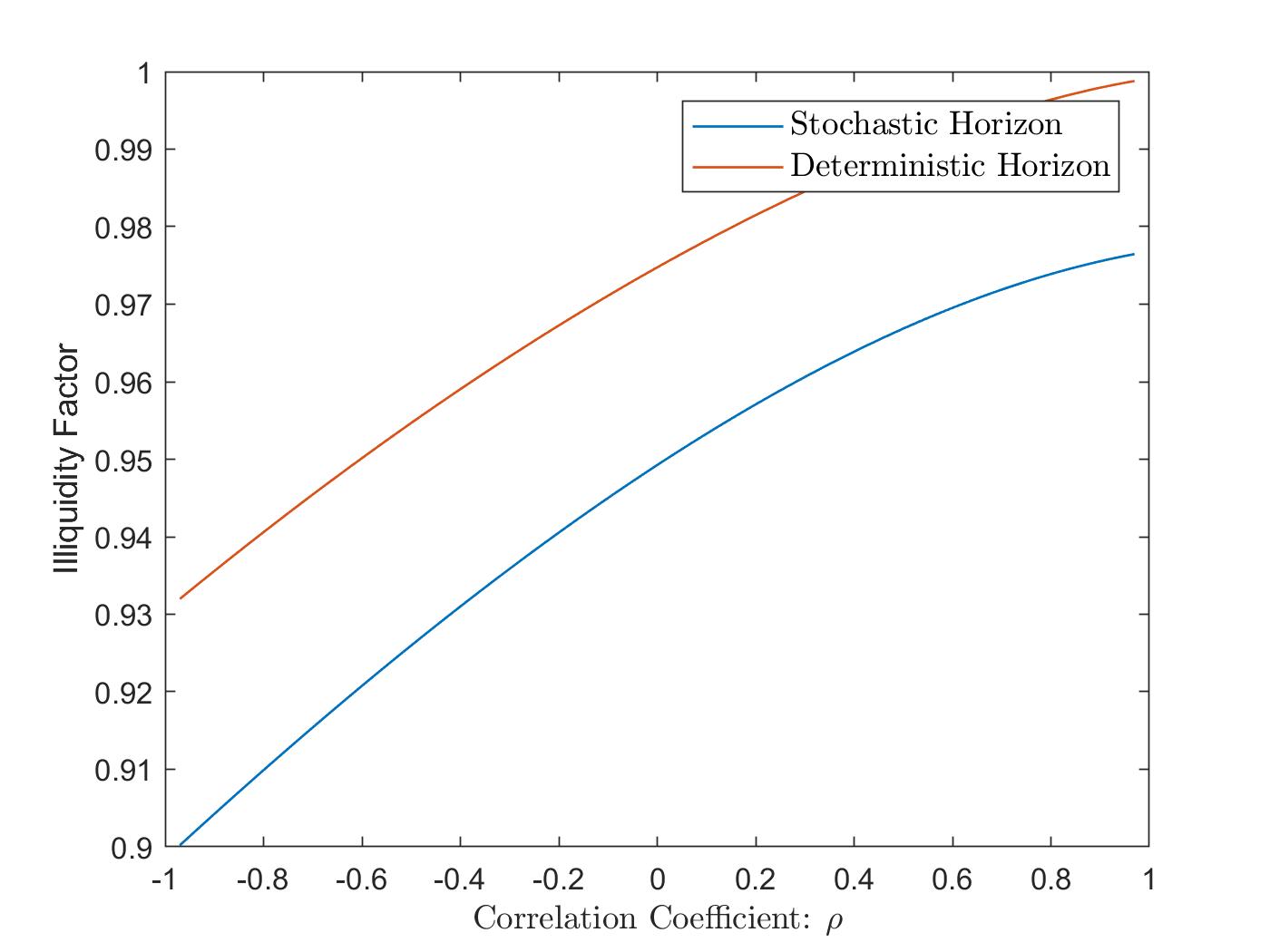}
\caption{$\mathfrak{I}_{Rel.}(\cdot)$ and $\mathfrak{I}^{R}_{Rel.}(\cdot)$ as functions of $\rho$.}
\label{fig:sub3}
\end{subfigure}
\caption{Illiquidity factor under stochastic illiquidity horizon, $\mathfrak{I}^{R}_{Rel.}(\cdot)$, and under deterministic illiquidity horizon, $\mathfrak{I}_{Rel.}(\cdot)$, for $\lambda = 0.5$ and as functions of the (expected) illiquidity horizon $\mathcal{T}=\frac{1}{\vartheta}$, the jump size $\varphi$, or the correlation coefficient $ \rho$. In Figure~\ref{fig:sub2} and Figure~\ref{fig:sub3}, we have chosen $\mathcal{T} = \frac{1}{\vartheta} = 0.5$.}
\label{fig:test1}
\end{figure}

\subsection{Comparison of the Illiquidity Factors}
\begin{figure}
\begin{subfigure}{.5\linewidth}
\centering
\includegraphics[scale=.17]{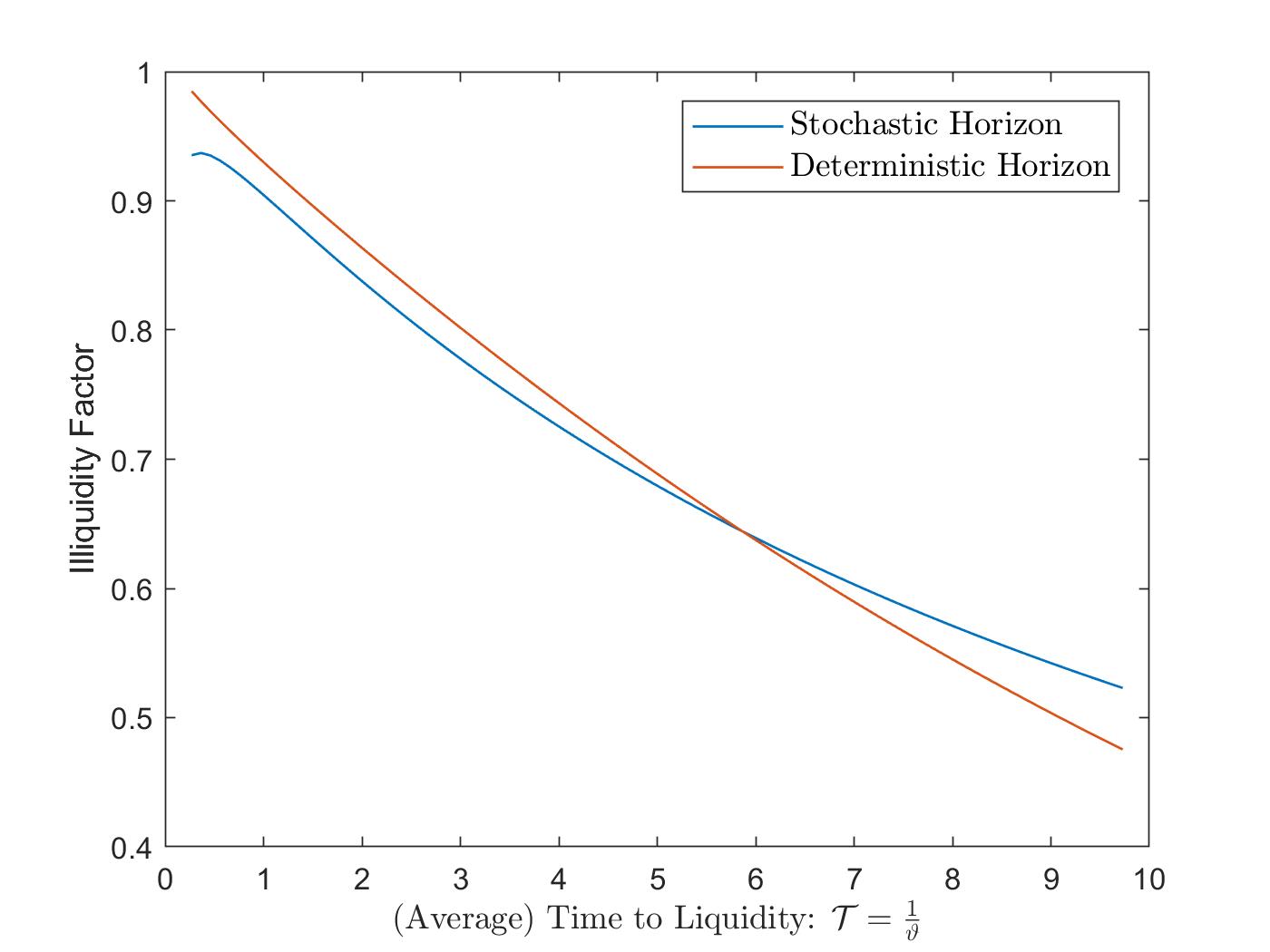}
\caption{$\mathfrak{I}_{Rel.}(\cdot)$ and $\mathfrak{I}^{R}_{Rel.}(\cdot)$ as functions of $\mathcal{T} = \frac{1}{\vartheta}$.}
\label{fig:sub21}
\end{subfigure}%
\begin{subfigure}{.5\linewidth}
\centering
\includegraphics[scale=.17]{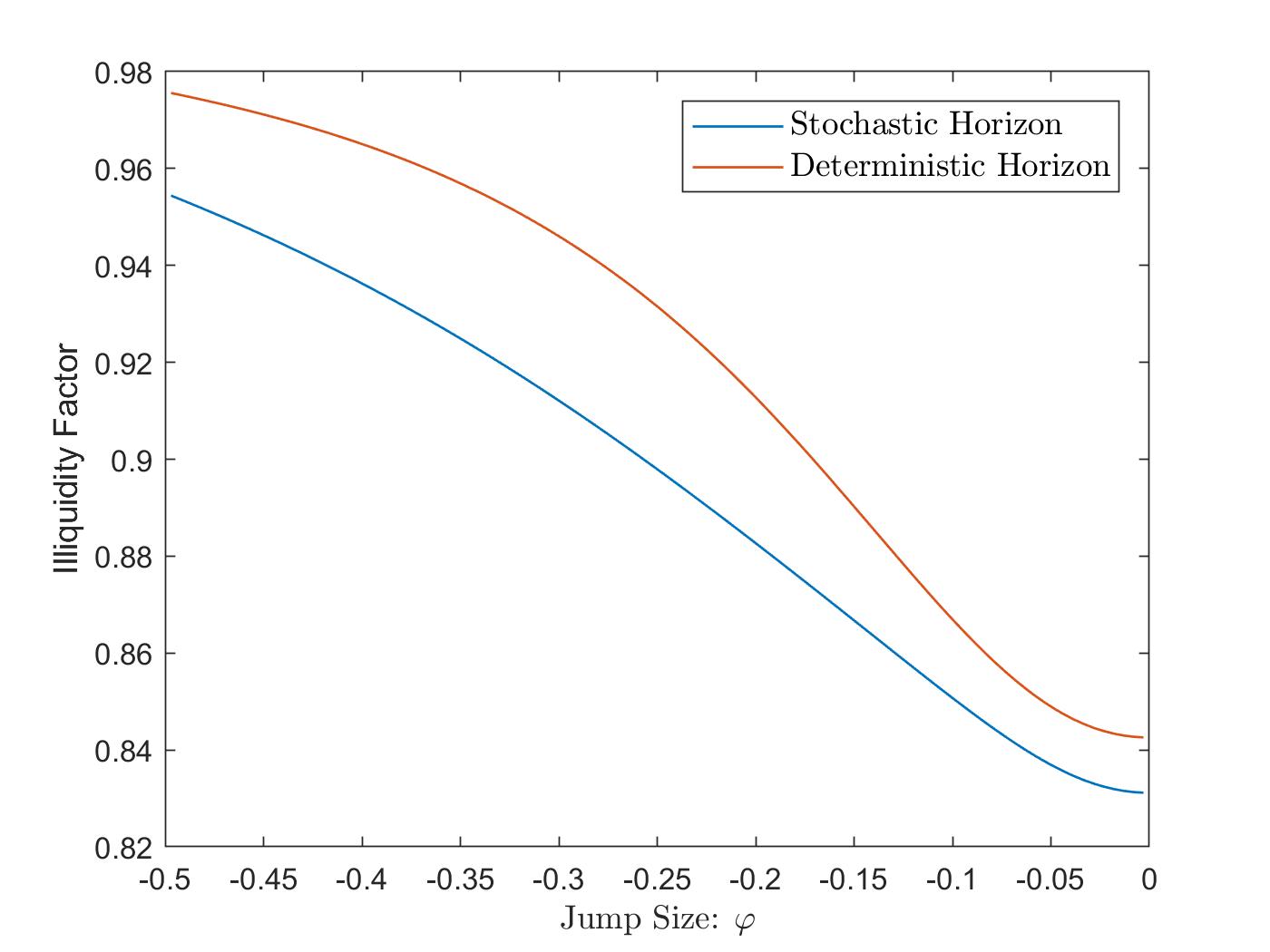}
\caption{$\mathfrak{I}_{Rel.}(\cdot)$ and $\mathfrak{I}^{R}_{Rel.}(\cdot)$ as functions of $\varphi$.}
\label{fig:sub22}
\end{subfigure}\\[1ex]
\begin{subfigure}{\linewidth}
\centering
\includegraphics[scale=.17]{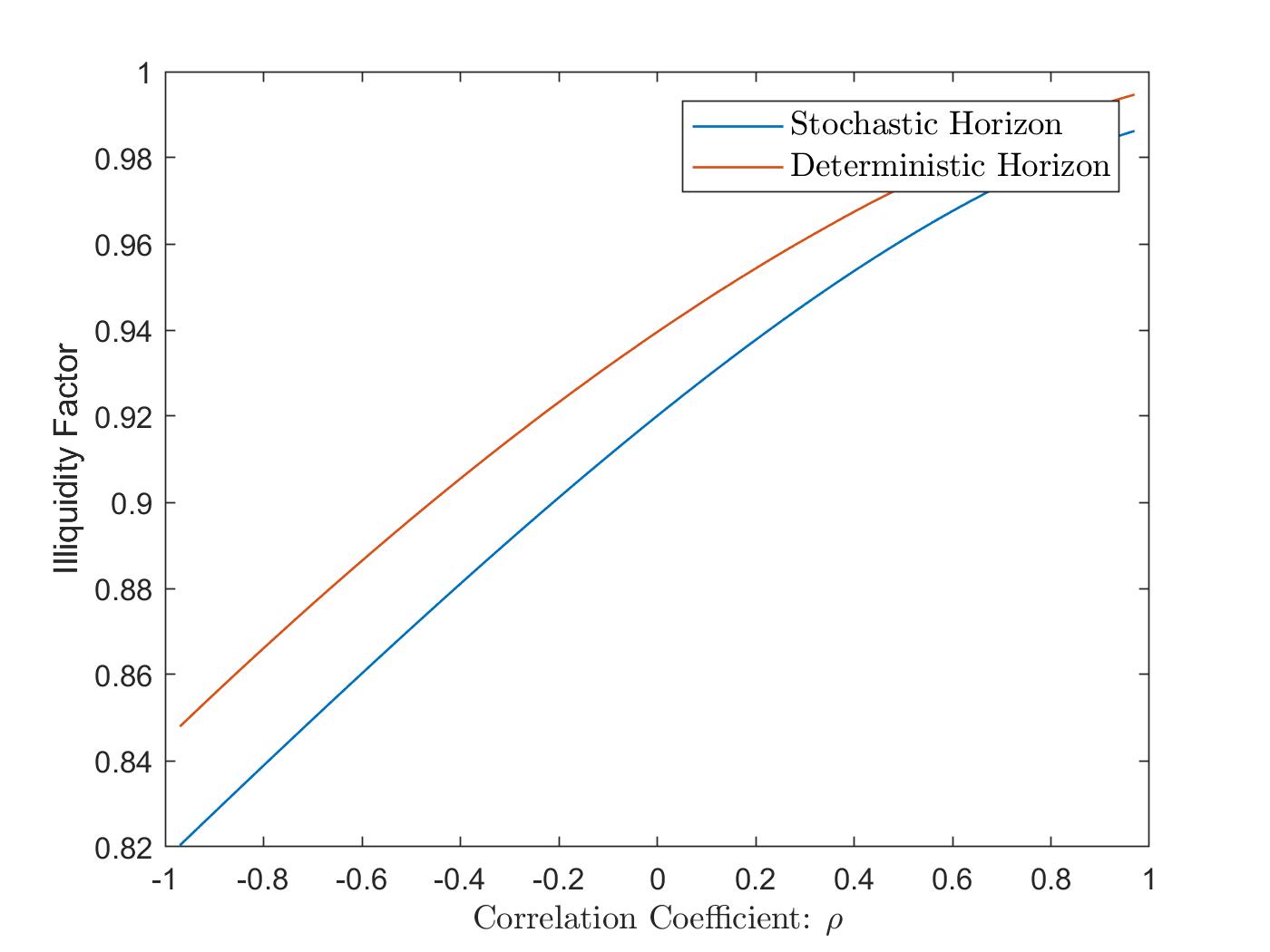}
\caption{$\mathfrak{I}_{Rel.}(\cdot)$ and $\mathfrak{I}^{R}_{Rel.}(\cdot)$ as functions of $\rho$.}
\label{fig:sub23}
\end{subfigure}
\caption{Illiquidity factor under stochastic illiquidity horizon, $\mathfrak{I}^{R}_{Rel.}(\cdot)$, and under deterministic illiquidity horizon, $\mathfrak{I}_{Rel.}(\cdot)$, for $\lambda = 1.0$ and as functions of the (expected) illiquidity horizon $\mathcal{T} = \frac{1}{\vartheta}$, the jump size $\varphi$, or the correlation coefficient $ \rho$. In Figure~\ref{fig:sub22} and Figure~\ref{fig:sub23}, we have chosen $\mathcal{T} = \frac{1}{\vartheta} = 1.5$.}
\label{fig:test2}
\end{figure}

\noindent To finalize the discussion of our numerical results, we provide in Figure~\ref{fig:test1} and Figure~\ref{fig:test2}, comparative plots for the illiquidity factor under deterministic and under stochastic illiquidity horizon. Whenever the parameters are not further specified, the following default values are used: $r= 2.25\%$, $\Phi_{X}(1) = 0.005$, $\sigma_{X} = 0.2$, $\rho = -0.5$, $b=-0.04$, $\sigma= 0.2$, $\varphi = \log(0.85)$, $E_{0} = 1$. \vspace{1em} \\
\noindent The results in Figure~\ref{fig:test1} and Figure~\ref{fig:test2} confirm several of the properties discussed in Section~\ref{NUDE} and Section~\ref{NUAp}. More interestingly, they also show that the tradeability premium under stochastic illiquidity horizon, $\mathfrak{L}^{R,\star}(\cdot)$, may become smaller than its deterministic counterpart. This happens for instance in Figure~\ref{fig:sub1} and Figure~\ref{fig:sub21} when large (expected) illiquidity horizons $\mathcal{T}$ are considered. In such cases, increasing the uncertainty over the duration of the asset's non-tradeability period raises the asset's value. In particular, this means that ``typical market participants'' would prefer, under certain parameter specifications, an asset with stochastic illiquidity horizon over an equivalent one with deterministic illiquidity horizon, i.e.~the market would exhibit a risk-loving behavior. Although this may be at first surprising, it is a well-documented fact that individuals tend to become risk-loving when confronted with negative events and happen to prefer a gamble over a sure (large) loss. Since illiquidity is, in general, an undesirable feature of an asset, it seems reasonable to observe that individuals may try to avoid large non-tradeability periods by gambling over the illiquidity horizon, i.e.~by preferring a stochastic illiquidity horizon over a deterministic illiquidity horizon.
\section{Conclusion}
\label{SEC6}
\noindent We proposed a new framework to evaluate tradeability and discussed it in the context of exponential Lévy markets. We first introduced our tradeability valuation approach under the simplistic assumption of a deterministic illiquidity horizon and subsequently extended our methods to deal with stochastic illiquidity horizons. Our general framework is linked to the asset replacement problem introduced in \cite{ms} and allows for a characterization of (individual) tradeability premiums by means of free-boundary problems. The resulting characterizations are of great practical importance, since they allow for a simple computation of tradeability values via the use of well-established numerical schemes. Using such schemes, we illustrated our approach by deriving numerical results and discussing various properties of the tradeability premiums. In particular, we found that, under certain parameter specifications, ``typical market participants'' may exhibit a risk-loving behavior in the sense that they may prefer an asset with stochastic illiquidity horizon over an equivalent asset with deterministic illiquidity horizon. \vspace{2em} \\
\noindent \acknow{I would like to thank Walter Farkas and Giovanni Barone-Adesi for their advice and constructive comments. I also thank Jérôme Detemple, Sander Willems, Alexander Smirnow, Jakub Rojcek, Matthias Feiler and the participants of the Gerzensee Research Days 2018 for their valuable suggestions.} \vspace{1em} \\
\newpage
\section{Appendices}
\renewcommand{\theequation}{A.\arabic{equation}}
\subsection*{Appendix A: Dynamics of $(Y_t)_{t \geq 0}$ under $\mathbb{Q}^{(1)}$}
%\subsection{Appendix A}
%\noindent \underline{Girsanov's Theorem for Correlated Brownian Motion:} \vspace{1em} \\
%\noindent Let $W$ be an $n$-dimensional Brownian motion with correlation matrix $\Lambda$, i.e. $\langle W^{j},W^{k} \rangle_{t} = \Lambda_{jk}$ for $j, k %\in \{1, \ldots , n \}$. For some $n$-dimensional, adapted process $\gamma$ satisfying $\int \limits_{0}^{t} \| \gamma_{s} \|^{2} ds < \infty$, a.s., %define the Doléan-Dade exponential
%$$ Z_{t}:= \exp \bigg( \int \limits_{0}^{t} \gamma_{s} dW_{s} - \frac{1}{2} \int \limits_{0}^{t} \| \gamma_{s} \|^{2} ds \bigg) .$$

\noindent In this appendix, we derive, for any finite time horizon $T > 0$, the dynamics of the Lévy process $(Y_{t})_{t \in [0,T]}$ under the particular measure transformation (\ref{bubup}). To this end, we denote by $(X_t^{c})_{t \geq 0}$ and $(X_{t}^{d})_{t \geq 0}$ -- and $(Y_{t}^{c})_{t \geq 0}$, $(Y_{t}^{d})_{t \geq 0}$ -- the continuous and discontinuous parts of $(X_{t})_{t \geq 0}$ -- and $(Y_{t})_{t \geq 0}$ respectively --, i.e.~we set
$$ X_{t}^{c}  :=  b_{X}t + \sigma_{X} W_{t}^{X}, \hspace{1.5em} X_{t}^{d}  := X_{t} -X_{t}^{c}, \hspace{1.5em} t \geq 0, $$
\noindent -- and analogously $Y_{t}^{c} := b_{Y}t + \sigma_{Y} W_{t}^{Y}$, $Y_{t}^{d} := Y_{t} - Y_{t}^{c}$, $t \geq 0$. Then, from the independence of the diffusion and jump parts, we note that 
\begin{equation} 
\left. \frac{d \mathbb{Q}^{(1)}}{d \mathbb{Q}} \right|_{\mathcal{F}_{t}}  =  \frac{e^{1 \cdot X_{t}}}{\mathbb{E}^{\mathbb{Q}}\left[ e^{1 \cdot X_{t}}\right]}  = \frac{e^{1 \cdot X_{t}^{c}}e^{1 \cdot X_{t}^{d}}}{\mathbb{E}^{\mathbb{Q}}\left[ e^{1 \cdot X_{t}^{c}}\right] \mathbb{E}^{\mathbb{Q}}\left[ e^{1 \cdot X_{t}^{d}}\right]}  =:  \left. \frac{d \mathbb{Q}^{(1),c}}{d \mathbb{Q}} \right|_{\mathcal{F}_{t}} \left. \frac{d \mathbb{Q}^{(1),d}}{d \mathbb{Q}} \right|_{\mathcal{F}_{t}} .
\end{equation}
\noindent Combining this fact with Girsanov's theorem for multidimensional correlated Brownian motion and the properties of $(X_{t})_{t \geq 0}$ and $(Y_{t})_{t \geq 0}$, we obtain, for $m \in \mathbb{N}$, $(\theta_{0}, \ldots, \theta_{m-1}) \in \mathbb{R}^{m} $ and $ 0 \leq t_{0} < t_{1} < \ldots < t_{m} \leq T$, that
\begin{align}
\mathbb{E}& ^{\mathbb{Q}^{(1)}} \bigg[ \exp\Big\{i \sum \limits_{j=0}^{m-1} \theta_{j} ( Y_{t_{j+1}} - Y_{t_{j}}) \Big\} \bigg] \nonumber \\
&  \hspace{0.5em} = \mathbb{E}^{\mathbb{Q}} \bigg[ \left. \frac{d \mathbb{Q}^{(1)}}{d \mathbb{Q}} \right|_{\mathcal{F}_{T}} \exp \Big \{ i \sum \limits_{j=0}^{m-1} \theta_{j} ( Y_{t_{j+1}} - Y_{t_{j}}) \Big \} \bigg] \nonumber \\
&  \hspace{0.5em} =  \mathbb{E}^{\mathbb{Q}} \bigg[ \left. \frac{d \mathbb{Q}^{(1),c}}{d \mathbb{Q}} \right|_{\mathcal{F}_{T}} \exp \Big \{ i \sum \limits_{j=0}^{m-1} \theta_{j} ( Y_{t_{j+1}}^{c} - Y_{t_{j}}^{c}) \Big \} \bigg] \; \mathbb{E}^\mathbb{Q} \bigg[ \exp \Big \{ i \sum \limits_{j=0}^{m-1} \theta_{j} ( Y_{t_{j+1}}^{d} - Y_{t_{j}}^{d}) \Big \}\bigg] \nonumber \\
& \hspace{0.5em} =  \exp \Big \{ i \sum \limits_{j=0}^{m-1} \theta_{j} \rho \sigma_{X} \sigma_{Y} (t_{j+1}-t_{j}) \Big \} \, \mathbb{E}^{\mathbb{Q}^{(1),c}} \bigg[ \exp \Big \{ i \sum \limits_{j=0}^{m-1} \theta_{j} ( \tilde{Y}_{t_{j+1}}^{c} - \tilde{Y}_{t_{j}}^{c}) \Big \} \bigg] \, \mathbb{E}^\mathbb{Q} \bigg[ \exp \Big \{ i \sum \limits_{j=0}^{m-1} \theta_{j} ( Y_{t_{j+1}}^{d} - Y_{t_{j}}^{d}) \Big \}\bigg] \nonumber \\
& \hspace{0.5em} =  \prod \limits_{j=0}^{m-1} \mathbb{E}^{\mathbb{Q}^{(1)}} \bigg[ e^{i \theta_{j} \big[( \tilde{Y}_{t_{j+1}}^{c} - \tilde{Y}_{t_{j}}^{c}) + \rho \sigma_{X} \sigma_{Y} (t_{j+1} -t_{j})\big]} \bigg] \prod \limits_{j=0}^{m-1} \mathbb{E}^{\mathbb{Q}^{(1)}} \bigg[ e^{i \theta_{j} ( Y_{t_{j+1}}^{d} - Y_{t_{j}}^{d})}\bigg] \nonumber \\
& \hspace{0.5em} =  \prod \limits_{j=0}^{m-1} \mathbb{E}^{\mathbb{Q}^{(1)}} \bigg[ e^{i \theta_{j} ( Y_{t_{j+1}} - Y_{t_{j}})}\bigg],
\label{ap1}
\end{align}
\noindent where
$$ \tilde{Y}_{t}^{c} := Y_{t}^{c} - \rho \sigma_{X} \sigma_{Y}t = b_{Y} t + \sigma_{Y} \tilde{W}_{t}^{Y}, \hspace{1.5em}  \tilde{W}_{t}^{Y} := W_{t}^{Y} - \rho \sigma_{X} t,$$
\noindent and we have used the fact that $(\tilde{W}_{t}^{Y})_{t \in [0,T]}$ is, under $\mathbb{Q}^{(1),c}$, a Brownian motion -- in fact Girsanov's theorem tells us that the processes $(\tilde{W}_{t}^{X})_{t \in [0,T]}$, $\tilde{W}_{t}^{X} := W_{t}^{X} - \sigma_{X}t$, and $(\tilde{W}_{t}^{Y})_{t \in [0,T]}$ are correlated Brownian motions under $\mathbb{Q}^{(1)}$. This shows that $(Y_{t})_{t \in [0,T]}$ has independent increments under $\mathbb{Q}^{(1)}$. \vspace{1em} \\
\noindent Showing that $(Y_{t})_{t \in [0,T]}$ has stationary increments under $\mathbb{Q}^{(1)}$ is easily done and follows from the identity
\begin{align}
\mathbb{E}^{\mathbb{Q}^{(1)}} \bigg[ \exp\Big \{ i \sum \limits_{j=0}^{m-1} \theta_{j} ( Y_{t_{j+1}} - & Y_{t_{j}}) \Big \} \bigg] = \mathbb{E}^{\mathbb{Q}} \bigg[ \left. \frac{d \mathbb{Q}^{(1)}}{d \mathbb{Q}} \right|_{\mathcal{F}_{T}} \exp \Big \{ i \sum \limits_{j=0}^{m-1} \theta_{j} ( Y_{t_{j+1}} - Y_{t_{j}}) \Big \} \bigg] \nonumber \\
& = \mathbb{E}^{\mathbb{Q}} \bigg[ \left. \frac{d \mathbb{Q}^{(1)}}{d \mathbb{Q}} \right|_{\mathcal{F}_{T}} \exp \Big \{ i \sum \limits_{j=0}^{m-1} \theta_{j} Y_{(t_{j+1} - t_{j})} \Big \} \bigg] = \mathbb{E}^{\mathbb{Q}^{(1)}} \bigg[ \exp \Big \{ i \sum \limits_{j=0}^{m-1} \theta_{j} Y_{(t_{j+1} - t_{j})} \Big \} \bigg].
\end{align}
\noindent Finally, it is clear that equivalent measure transformations do not alter both the starting value and the path continuity of processes. Hence, $(Y_{t})_{t \in [0,T]}$ is also under $\mathbb{Q}^{(1)}$ càdlàg and satisfies $Y_{0}=0$. This shows that $(Y_{t})_{t \in [0,T]}$ is under $\mathbb{Q}^{(1)}$ again a Lévy process. \vspace{1em} \\
\noindent Deriving the characteristic exponent of $(Y_{t})_{t \in [0,T]}$ under $\mathbb{Q}^{(1)}$ is now easily done using the equation
\begin{equation}
\mathbb{E}^{\mathbb{Q}^{(1)}} \bigg[ e^{i \theta Y_{t} } \bigg] = e^{i \theta \rho \sigma_{X} \sigma_{Y} t} \; \mathbb{E}^{\mathbb{Q}^{(1),c}} \bigg[ e^{i \theta \tilde{Y}_{t}^{c}} \bigg] \; \mathbb{E}^\mathbb{Q} \bigg[ e^{i \theta Y_{t}^{d}}\bigg] ,
\end{equation}
\noindent which can be derived as in (\ref{ap1}). This gives that the Lévy exponent of $(Y_{t})_{t \in [0,T]}$ under $\mathbb{Q}^{(1)}$, $\Psi_{Y}^{(1)}(\cdot)$, is given by
\begin{equation}
\Psi_{Y}^{(1)}(\theta)= -i(b_{Y}+\rho \sigma_{X} \sigma_{Y}) \theta + \frac{1}{2} \sigma_{Y}^{2} \theta^{2} + \int \limits_{ \mathbb{R}} (1 - e^{i \theta y} + i \theta y \mathds{1}_{\{ | y | \leq 1\}}) \Pi_{Y}( dy) ,
\end{equation}
\noindent i.e.~$(Y_{t})_{t \in [0,T]}$ is under $\mathbb{Q}^{(1)}$ an $\mathbf{F}$-Lévy process with triplet $\big(b_{Y}+\rho \sigma_{X} \sigma_{Y}, \sigma_{Y}^{2}, \Pi_{Y}\big)$.
\subsection*{Appendix B: Proofs - Deterministic Illiquidity Horizon}
\label{AppendixB}
\begin{proof}[\bf Proof of Proposition \ref{Prop1}]
\noindent Due to the discussion preceding Proposition \ref{Prop1}, we only need to show that $\mathfrak{C}_{\bf E}^{\star}(\cdot)$ has enough regularity, i.e.~in particular that
\begin{itemize} \setlength \itemsep{-0.1em}
\item[i)] $x \mapsto \mathfrak{C}_{\bf E}^{\star}(\mathcal{T},x)$ is, for any $\mathcal{T} \in (0,T_{D})$, twice continuously differentiable,
\item[ii)] $t \mapsto e^{-\tilde{r}t} \mathfrak{C}_{\bf E}^{\star}(\mathcal{T},x)$ is, for any $x \in (0,\infty)$, continuously differentiable,
\item[iii)] $(\mathcal{T},x ) \mapsto \mathfrak{C}_{\bf E}^{\star}(\mathcal{T},x)$ is continuous on $[0,T_{D}] \times [0,\infty)$.
\end{itemize}
\noindent We start by briefly outlining the proof of i). Since this part does not involve any martingale arguments, we refer the reader for details to \cite{cv05} and \cite{vo05}. To see i), one first notices that the European-type option $\mathfrak{C}_{\bf E}^{\star}(\cdot)$ can be re-expressed in terms of the function
$$ u(\mathcal{T}, \xi) = \mathbb{E}^{\mathbb{Q}^{(1)}} \left[ e^{-\tilde{r} \mathcal{T}} (e^{\xi + Y_{\mathcal{T}}} -1 )^{+} \right],$$ 
\noindent as 
\begin{align}
\mathfrak{C}_{\bf E}^{\star} (\mathcal{T},x)  = \mathbb{E}^{\mathbb{Q}^{(1)}} \big[ e^{-\tilde{r} \mathcal{T}} (xe^{Y_{\mathcal{T}}}-1)^{+}\big] = \mathbb{E}^{\mathbb{Q}^{(1)}} \left[ e^{-\tilde{r} \mathcal{T}} \left(e^{\log(x)+Y_{\mathcal{T}}} - 1 \right)^{+} \right]  =  u\left(\mathcal{T},\log(x)\right).
\end{align}
Therefore, in order to show the smoothness of $x \mapsto \mathfrak{C}_{\bf E}^{\star}(\mathcal{T},x)$ it is enough to prove the smoothness of $u(\cdot)$ in the log-moneyness coordinate. To this end, two facts can be combined. First, as noted in \cite{cv05} and \cite{vo05}, Condition (\ref{bediwi}) ensures that $Y_{t}$ has, for any $t \in [0,T_{D}]$, a smooth, at least $C^{2}$, ($\mathbb{Q}^{(1)}$-)density with derivatives vanishing at infinity. We denote this density in the following by $q_{t}(\cdot)$. Secondly, setting $\tilde{q}_{t}(y) : = q_{t}(-y)$, we can rewrite $u(\cdot)$ as a convolution of the form
\begin{equation}
u(\mathcal{T}, \xi)  =  e^{-r \mathcal{T}} \int \limits_{\mathbb{R}} \left( e^{\xi + y} -1\right)^{+} q_{\mathcal{T}}(y) \, dy  =  e^{-r \mathcal{T}} \int \limits_{\mathbb{R}} \left( e^{z} -1\right)^{+} \tilde{q}_{\mathcal{T}}(\xi-z) \, dz .
\end{equation}
\noindent Therefore, the decay of $q_{\mathcal{T}}(\cdot)$ and in particular of its derivatives (cf.~\cite{cv05}, \cite{vo05}) allows one to use the dominated convergence theorem to differentiate under the integral sign and to obtain that $x \mapsto \mathfrak{C}_{\bf E}^{\star}(\mathcal{T},x)$ is twice continuously differentiable. \vspace{1em} \\
\noindent We now prove ii) using Fourier methods. This approach was similarly used in \cite{cv05} and relies on a seminal article of Carr and Madan \cite{cm99}. Recall, for an integrable function $f(\cdot)$, the definition of the Fourier transform, $\mathcal{F}$, and Fourier inverse, $\mathcal{F}^{-1}$,
$$ \mathcal{F}f(\xi) := \int \limits_{\mathbb{R}} f(y) e^{iy \xi} \, dy, \; \; \; \; \; \mathcal{F}^{-1} f(y) := \frac{1}{2 \pi} \int \limits_{\mathbb{R}} f(\xi) e^{-i\xi y} \,d\xi ,$$
\noindent and that both operators can be extended to isometries on the space of square-integrable functions. As noted in i), Condition (\ref{bediwi}) ensures that $Y_{t}$ has, for any $t \in [0,T_{D}]$, a smooth, $C^{2}$, ($\mathbb{Q}^{(1)}$-)density which we will denote again by $q_{t}(\cdot)$. Therefore, the characteristic function of $Y_{\mathcal{T}}$ at $\theta$, $\chi_{\mathcal{T}}(\theta)$, can be expressed as
\begin{equation}
e^{-\mathcal{T} \Psi_{Y}^{(1)}(\theta)} = \chi_{\mathcal{T}}(\theta) = \int \limits_{\mathbb{R}} e^{i\theta y} q_{\mathcal{T}}(y) \, dy  .
\end{equation}
\noindent We now consider, for $k \in \mathbb{R}$, the modified call price defined by 
\begin{equation}
\mathfrak{c}_{\mathcal{T}}(k) := e^{k} \int \limits_{k}^{\infty} e^{-\tilde{r}\mathcal{T}} (e^{y}-e^{k}) q_{\mathcal{T}}(y) \, dy,
\label{kaka}
\end{equation}
\noindent and easily see that with $k := \log\big(\frac{K}{x}\big)$, $x \in (0,\infty)$ and $K \in (0,\infty)$, it satisfies that
\begin{equation}
x \cdot \mathfrak{c}_{\mathcal{T}}(k) = x \cdot e^{k} \, \mathbb{E}^{\mathbb{Q}^{(1)}} \left[ e^{-\tilde{r} \mathcal{T}} \big( e^{Y_{\mathcal{T}}}-e^{k} \big)^{+} \right] = e^{k} \, \mathbb{E}_{x}^{\mathbb{Q}^{(1)}} \left[ e^{-\tilde{r} \mathcal{T}} \big( E_{\mathcal{T}}-K \big)^{+} \right] .
\label{IMrela}
\end{equation}
\noindent Additionally, we set $ \mathfrak{c}^{\ast}_{\mathcal{T}}(k) := e^{-\tilde{r}t} \mathfrak{c}_{\mathcal{T}}(k)$. Arguing as in \cite{cm99} one sees that Condition (\ref{condidi}) implies both the integrability and square-integrability of the discounted modified call price $k \mapsto \mathfrak{c}^{\ast}_{\mathcal{T}}(k)$. Furthermore one readily derives, using (\ref{kaka}), that
\begin{align}
\mathcal{F} \mathfrak{c}_{\mathcal{T}}^{\ast}(v)  = \int \limits_{\mathbb{R}} \mathfrak{c}_{\mathcal{T}}^{\ast}(k) e^{ikv} \, dk = \frac{e^{-\tilde{r} T_{D}} \, \chi_{\mathcal{T}}(v-2i)}{(iv+1)(iv+2)}.
\label{hahiha}
\end{align}
\noindent Notice that this expression is clearly differentiable with respect to $t$ and that one obtains 
\begin{equation}
\partial_{t} \mathcal{F} \mathfrak{c}_{\mathcal{T}}^{\ast}(v) =  \frac{e^{-\tilde{r} T_{D}} \, \chi_{\mathcal{T}}(v-2i) \, \Psi_{Y}^{(1)}(v-2i)}{(iv+1)(iv+2)}.
\label{zuzu}
\end{equation}
%% From the integrability and the square integrability of the discounted modified price and the boundedness of $\mathcal{T}$, one obtains that (\ref{hahiha}) is square-integrable and bounded. Furthermore, from...
\noindent From the Lévy-Khintchine formula/representation, one additionally sees that $\Psi^{(1)}_{Y}(v-2i) = \mathcal{O}(|v|^{2})$ (as $|v| \rightarrow \infty$) -- hence at $\infty$ the denominator compensates $\Psi_{Y}^{(1)}(v-2i)$.  Combining these arguments with the fact that, under (\ref{bediwi}),
\begin{equation}
|\chi_{\mathcal{T}}(z) | \leq C(\mathcal{T}) \exp(-c(\mathcal{T})|z|^{\gamma}) \hspace{1em} \mbox{for some} \;  \;\; \gamma>0\footnote{$\gamma = 2$ if $\sigma \neq 0$ and $\gamma = \alpha$ if $\sigma = 0$ and the second condition is satisfied. This was already noted in \cite{vo05} (cf.~\cite{sa}).} \; \;\;  \mbox{and ``constants'' } \; \; C(\mathcal{T}), c(\mathcal{T}) >0
\label{CoProof}
\end{equation}
\noindent and, in particular, that $\mathcal{T} \mapsto C(\mathcal{T})$, $\mathcal{T} \mapsto c(\mathcal{T})$ can be chosen to be continuous (by the continuity of $\mathcal{T} \mapsto \chi_{\mathcal{T}}(z)$), tells us that (\ref{zuzu}) is in any case dominated (locally in $\mathcal{T}$) by an integrable function that does not have any $\mathcal{T}$-dependency.\footnote{It suffices to take, for a given (compact) $\mathcal{T}$-neighborhood $U$, $C^{\star} := \max \limits_{t \in U} C(t)$ and $c^{\star} := \min \limits_{t \in U} c(t)$ in (\ref{CoProof}).} Finally, this allows us to use the dominated convergence theorem in order to conclude that
\begin{equation} \partial_{t} \mathfrak{c}_{\mathcal{T}}^{\ast}(k) = \partial_{t} \mathcal{F}^{-1} \mathcal{F} \mathfrak{c}_{\mathcal{T}}^{\ast}(k) = \mathcal{F}^{-1} \partial_{t} \mathcal{F} \mathfrak{c}_{\mathcal{T}}^{\ast}(k),
\label{zuiz}
\end{equation}
\noindent which shows, in particular by means of Relation (\ref{IMrela}) with $K = 1$, that $t \mapsto e^{-\tilde{r}t} \mathfrak{C}_{\bf E}^{\star}(\mathcal{T},x)$ is for any $x \in (0,\infty)$ differentiable. The continuity of the derivative is easily seen from (\ref{zuiz}) and (\ref{zuzu}) and the dominated convergence theorem, by noting that $t \mapsto \chi_{\mathcal{T}}(v-2i)$ is continuous (recall that $\mathcal{T}=T_{D}-t$). \vspace{1em} \\
\noindent Finally, iii) is a direct consequence of Relation (\ref{IMrela}) and the continuity of $(\mathcal{T},k) \mapsto \mathfrak{c}_{\mathcal{T}}(k)$, which follows again from (\ref{hahiha}) by means of Fourier inversion and the dominated convergence theorem. This finalizes the proof.
\end{proof}
\begin{proof}[\bf Proof of Lemma \ref{lemma1}]
\noindent The first part of $a)$, i.e.~the non-decreasing property follows directly from the path properties of exponential Lévy models. As this is easily proved, we focus on showing the convexity of the American-type option. To start, let $\mathcal{T} \in [0,T_{D}]$ be arbitrary but fixed. We define, for any initial value $x \in [0,\infty)$ and any stopping time $\tau \in \mathfrak{T}_{[0,\mathcal{T}]}$, the two value functions $V(\cdot)$ and $V^{\ast}(\cdot)$ by
\begin{equation}
V(\tau,x) := \mathbb{E}_{x}^{\mathbb{Q}^{(1)}} \left[ e^{-\tilde{r} \tau} \big( E_{\tau}-1 \big)^{+} \right]
\end{equation}
\noindent and 
\begin{equation}
V^{\ast}(x) := \sup \limits_{\tau \in \mathfrak{T}_{[0,\mathcal{T}]}} V(\tau,x) ,
\end{equation}
\noindent and note that $V^{\ast}(x) = \mathfrak{C}_{{\bf A}}^{\star}(\mathcal{T}, x)$.
\noindent Given two initial values $x_{1}$ and $x_{2}$ and an arbitrary $\lambda \in [0,1]$, we set $ \tilde{x} := \lambda x_{1} + (1-\lambda) x_{2} $ and fix some $\epsilon >0$. By definition of $V^{\ast}(\cdot)$, we can find a stopping time $\tau_{\epsilon}$ satisfying $ V^{\ast}(\tilde{x}) \leq V(\tau_{\epsilon},\tilde{x}) + \epsilon $. Furthermore, from the (strong) Markov property of $(E_{t})_{t \in [0,T_{D}]}$ and the properties of the pay-off function, we have that
\begin{equation}
V(\tau_{\epsilon}, \tilde{x}) \leq \lambda V(\tau_{\epsilon}, x_{1}) + (1- \lambda) V(\tau_{\epsilon},x_{2}) ,
\end{equation}
\noindent which implies that
\begin{equation}
V^{\ast}(\tilde{x}) \leq V(\tau_{\epsilon},\tilde{x}) + \epsilon \leq \lambda V(\tau_{\epsilon}, x_{1}) + (1- \lambda) V(\tau_{\epsilon},x_{2}) + \epsilon \leq \lambda V^{\ast}(x_{1}) + (1- \lambda) V^{\ast}(x_{2}) + \epsilon .
\end{equation}
\noindent Since $\epsilon$ was arbitrary, this gives the convexity of the American-type option. \vspace{1em} \\
\noindent Property $b)$ follows directly by noting that, for $0\leq \mathcal{T}_{1} \leq \mathcal{T}_{2} \leq T_{D}$, any stopping time $\tau \in \mathfrak{T}_{[0,\mathcal{T}_{1}]}$ also satisfies $\tau \in \mathfrak{T}_{[0,\mathcal{T}_{2}]}$. Therefore, we are left with Part $c)$. To prove this last part, we use the (strong) Markov property of $(E_{t})_{t \in [0,T_{D}]}$ as well as the property that, for $x,y \in [0,\infty)$, $ \big|(x-1)^{+} - (y-1)^{+} \big| \leq |x-y|$ holds. We then obtain, for a fixed $\mathcal{T} \in [0,T_{D}]$, that
\begin{eqnarray}
\Big| \sup \limits_{\tau \in \mathfrak{T}_{[0,\mathcal{T}]}}  \mathbb{E}_{x}^{\mathbb{Q}^{(1)}} \left[ e^{-\tilde{r} \tau} \big( E_{\tau}-1 \big)^{+} \right] - \sup \limits_{\tau \in \mathfrak{T}_{[0,\mathcal{T}]}}  \mathbb{E}_{y}^{\mathbb{Q}^{(1)}} \left[ e^{-\tilde{r} \tau} \big( E_{\tau}-1 \big)^{+} \right] \Big| \hspace{8em} \nonumber \\
\leq \;  \sup \limits_{\tau \in \mathfrak{T}_{[0,\mathcal{T}]}} \Big| \mathbb{E}_{x}^{\mathbb{Q}^{(1)}} \left[ e^{-\tilde{r} \tau} \big( E_{\tau}-1 \big)^{+} \right]- \mathbb{E}_{y}^{\mathbb{Q}^{(1)}} \left[ e^{-\tilde{r} \tau} \big( E_{\tau}-1 \big)^{+} \right] \Big| \nonumber \\
\leq \;\big| x-y \big | \cdot \sup \limits_{\tau \in \mathfrak{T}_{[0,\mathcal{T}]}} \mathbb{E}^{\mathbb{Q}^{(1)}} \left[ e^{-(\tilde{r}-\Phi_{Y}^{(1)}(1)) \tau} e^{Y_{\tau}-\tau \Phi_{Y}^{(1)}(1)} \right]. \hspace{3.6em}
\label{tripleA}
\end{eqnarray}
\noindent Since the process $\left(e^{Y_{t}-t \Phi_{Y}^{(1)}(1)}\right)_{t \in [0,T_{D}]}$ is known to be a ($\mathbb{Q}^{(1)}$-)martingale, we can take
$$ C := \left\{ \begin{array}{ll}
1, & \mbox{if} \;\;  \tilde{r} \geq \Phi_{Y}^{(1)}(1), \\
e^{-(\tilde{r}-\Phi_{Y}^{(1)}(1)) \mathcal{T}}, & \mbox{otherwise},
\end{array} \right.$$
\noindent and obtain from (\ref{tripleA}) that
$$ | \mathfrak{C}_{{\bf A}}^{\star}(\mathcal{T},x) - \mathfrak{C}_{{\bf A}}^{\star}(\mathcal{T},y) |  \leq C | x -y | .$$
\end{proof}
\begin{proof}[\bf Proof of the smooth-fit property in Proposition \ref{prop3}]
\noindent This part provides a proof of Equation (\ref{equazzzion}), i.e.~we show that, for all $\mathcal{T} \in (0,T_{D}]$, we have
\begin{equation}
\partial_{x} \mathfrak{L}^{\star}( \mathcal{T}, \mathfrak{b}_{s}(\mathcal{T}))  =  1 - \partial_{x} \mathfrak{C}_{{\bf E}}^{\star} (\mathcal{T}, \mathfrak{b}_{s}(\mathcal{T})) .
\end{equation}
\noindent For this equation to hold, it is sufficient to have that, for any $\mathcal{T} \in (0,T_{D}]$, the function $x \mapsto \mathfrak{C}_{\bf A}^{\star}( \mathcal{T}, x ) $ is in $\mathfrak{b}_{s}(\mathcal{T})$ differentiable with $\partial_{x} \mathfrak{C}_{\bf A}^{\star} ( \mathcal{T}, \mathfrak{b}_{s}(\mathcal{T})) = 1$. We show that this is true. \vspace{1em}\\
\noindent First, we recall that for a Lévy process $(Z_{t})_{t \geq 0}$ on a probability space $ ( \Omega, \mathcal{F}, \mathbb{P})$ a fixed level $z \in \mathbb{R}$ is said to be regular for $(z,\infty)$, if we have that
$$ \mathbb{P}_{z}( \tau_{z}^{+} =0 ) =1 ,$$ 
\noindent where $\tau_{z}^{+}$ is given by
$$ \tau_{z}^{+} := \inf \{ t \geq 0: \; Z_{t} \in (z,\infty) \} ,$$
\noindent and we set as usual $\inf \emptyset = \infty$. As noted for instance in \cite{kyp}, Theorem 6.5.,~any Lévy process of infinite variation has the particularity that the point $0$ is regular for the interval $(0,\infty) $. Since we have assumed that $\sigma_{Y} \neq 0$, the ($\mathbb{Q}^{(1)}$-)Lévy process $\left( Y_{t} \right)_{t \geq 0}$ has clearly infinite variation (c.f.~\cite{sa}, \cite{Ap}). Therefore, it suffices to show that the regularity of $0$ for $(0,\infty)$ and $\left( Y_{t} \right)_{t \geq 0}$ implies the smooth-fit property of $\mathfrak{C}_{\bf A}^{\star}(\cdot)$. We show it by adapting the proof of Theorem 4.1.~in \cite{lm11}: \vspace{0.1em} \\
Let us fix $\mathcal{T} \in (0,T_D]$. We start by noting that 
\begin{equation}
\lim \limits_{h \downarrow 0 } \frac{\mathfrak{C}_{\bf A}^{\star} (\mathcal{T}, \mathfrak{b}_{s}(\mathcal{T}) + h)-\mathfrak{C}_{\bf A}^{\star} (\mathcal{T}, \mathfrak{b}_{s}(\mathcal{T}))}{h} =  1 .
\end{equation}
\noindent This directly follows since any $x \geq \mathfrak{b}_{s}(\mathcal{T})$ satisfies that $ \mathfrak{C}_{\bf A}^{\star}(\mathcal{T},x) =  x-1 $. Therefore, we only need to show that
\begin{equation}
\lim \limits_{h \uparrow 0 } \frac{\mathfrak{C}_{\bf A}^{\star} (\mathcal{T}, \mathfrak{b}_{s}(\mathcal{T}) + h)-\mathfrak{C}_{\bf A}^{\star} (\mathcal{T}, \mathfrak{b}_{s}(\mathcal{T}))}{h} =  1 .
\end{equation}
\noindent First, we obtain from $\mathfrak{C}_{\bf A}^{\star} (\mathcal{T}, \mathfrak{b}_{s}(\mathcal{T})) = \left( \mathfrak{b}_{s}(\mathcal{T}) -1\right)^{+}$ and $\mathfrak{C}_{\bf A}^{\star} (\mathcal{T}, x) \geq \left(x-1 \right)^{+}$ that, for any $h <0$,
$$ \frac{\mathfrak{C}_{\bf A}^{\star} (\mathcal{T}, \mathfrak{b}_{s}(\mathcal{T}) + h)-\mathfrak{C}_{\bf A}^{\star} (\mathcal{T}, \mathfrak{b}_{s}(\mathcal{T}))}{h} \geq \frac{\left( \mathfrak{b}_{s}(\mathcal{T}) + h - 1\right)^{+}-\left( \mathfrak{b}_{s}(\mathcal{T}) - 1\right)^{+}}{h} .$$
\noindent This gives that
\begin{equation}
\liminf \limits_{h \uparrow 0} \frac{\mathfrak{C}_{\bf A}^{\star} (\mathcal{T}, \mathfrak{b}_{s}(\mathcal{T}) + h)-\mathfrak{C}_{\bf A}^{\star} (\mathcal{T}, \mathfrak{b}_{s}(\mathcal{T}))}{h} \geq 1 .
\end{equation}
\noindent To show that 
\begin{equation}
\limsup \limits_{h \uparrow 0} \frac{\mathfrak{C}_{\bf A}^{\star} (\mathcal{T}, \mathfrak{b}_{s}(\mathcal{T}) + h)-\mathfrak{C}_{\bf A}^{\star} (\mathcal{T}, \mathfrak{b}_{s}(\mathcal{T}))}{h} \leq 1 ,
\end{equation}
\noindent we consider, for $h<0$, the optimal stopping problem related to $\mathfrak{C}_{\bf A}^{\star} ( \mathcal{T}, \mathfrak{b}_{s}(\mathcal{T})+h)$: First, we define the stopping time
\begin{align}
\tau_{h} & :=  \inf  \{ t \in [0,\mathcal{T}): \; \left( \mathfrak{b}_{s}(\mathcal{T})+h \right)e^{Y_{t}} \geq  \mathfrak{b}_{s}( \mathcal{T}) \} \nonumber\\
& \hspace{0.3em} =  \inf \left \{ t \in [0,\mathcal{T}): \;Y_{t} \geq  \log \left( \frac{\mathfrak{b}_{s}( \mathcal{T})}{\mathfrak{b}_{s}(\mathcal{T})+h} \right) \right \}
\end{align}
\noindent and note from the regularity of $0$ for the set $(0,\infty)$ that $\tau_{h} \rightarrow 0$ a.s.~when $h \uparrow 0$. This can be seen by the following argument: On the almost sure set $\{ \tau_{0}^{+} =0 \}$, we can find for any $t_{0} \in (0,\mathcal{T})$ a point $u \in [0,t_{0}]$ such that $Y_{u} >0$. Then, taking $h <0$ small enough (i.e.~near enough to zero) gives that $Y_{u} > \log \left( \frac{\mathfrak{b}_{s}( \mathcal{T})}{\mathfrak{b}_{s}(\mathcal{T})+h} \right)$. Consequently, $\lim \limits_{h \uparrow 0} \tau_{h} \leq t_{0}$ a.s.~and from the arbitrariness of $t_{0} \in (0,\mathcal{T})$ this already gives that $\lim \limits_{h \uparrow 0} \tau_{h} =0$. \vspace{0.2em} \\
\noindent Next, noting that
$$\mathfrak{C}_{\bf A}^{\star}(\mathcal{T}, \mathfrak{b}_{s}(\mathcal{T})) \geq \mathbb{E}_{\mathfrak{b}_{s}(\mathcal{T})}^{\mathbb{Q}^{(1)}} \left[ e^{-\tilde{r} \tau_{h}} \left( E_{\tau_{h}} - 1 \right)^{+}\right] $$
\noindent and combining this inequality with the optimality of the stopping time $\tau_{h}$ for the starting value $\mathfrak{b}_{s}(\mathcal{T})+h$ gives, for $h <0$, that
\begin{align}
\frac{\mathfrak{C}_{\bf A}^{\star}(\mathcal{T}, \mathfrak{b}_{s}(\mathcal{T})+h) - \mathfrak{C}_{\bf A}^{\star}(\mathcal{T}, \mathfrak{b}_{s}(\mathcal{T}))}{h}  =  \frac{\mathbb{E}_{\mathfrak{b}_{s}(\mathcal{T})+h}^{\mathbb{Q}^{(1)}} \left[ e^{-\tilde{r} \tau_{h}} \left( E_{\tau_{h}} - 1 \right)^{+}\right]-\mathfrak{C}_{\bf A}^{\star}(\mathcal{T}, \mathfrak{b}_{s}(\mathcal{T}))}{h} \hspace{3em}  \nonumber \\
  \leq  \mathbb{E}^{\mathbb{Q}^{(1)}} \left[ e^{-\tilde{r} \tau_{h}} \frac{\left((\mathfrak{b}_{s}(\mathcal{T})+h)e^{Y_{\tau_{h}}} -1 \right)^{+} - \left(\mathfrak{b}_{s}(\mathcal{T})e^{Y_{\tau_{h}}} -1 \right)^{+}}{h}\right].
\end{align}
\noindent Since $x \mapsto \left( x-1 \right)^{+}$ is continuously differentiable in a neighbourhood of $\mathfrak{b}_{s}(\mathcal{T})$, we have that
\begin{equation}
\lim \limits_{h \uparrow 0} \frac{\left((\mathfrak{b}_{s}(\mathcal{T})+h)e^{Y_{\tau_{h}}} -1 \right)^{+} - \left(\mathfrak{b}_{s}(\mathcal{T})e^{Y_{\tau_{h}}} -1 \right)^{+}}{h} = 1 .
\end{equation}
\noindent Finally, using Lemma \ref{lemma1}.$c)$ allows us to apply the dominated convergence theorem, to obtain that
$$ \limsup \limits_{h \uparrow 0} \frac{\mathfrak{C}_{\bf A}^{\star} (\mathcal{T}, \mathfrak{b}_{s}(\mathcal{T}) + h)-\mathfrak{C}_{\bf A}^{\star} (\mathcal{T}, \mathfrak{b}_{s}(\mathcal{T}))}{h} \leq 1 , $$
\noindent which gives the result.
\end{proof}

\subsection*{Appendix C: Proofs - Stochastic Illiquidity Horizon}
\label{AppendixC}
\begin{proof}[\bf Proof of Proposition \ref{Prop4}]
\noindent First, we note that the continuity of $x \mapsto \mathfrak{C}_{\bf E}^{R,\star}(x)$ on $[0,\infty)$ follows from the dominated convergence theorem, by combining Condition (\ref{CONDinf}) with Representations (\ref{morca1}) and (\ref{Zutzut}). Additionally, the continuity of $x \mapsto \partial_{x}\mathfrak{C}_{\bf E}^{R,\star}(x)$ on $(0,\infty)$ follows analogously using (\ref{morca1}), the continuity of $x \mapsto \mathfrak{C}_{\bf E}^{\star}(t_{R},x)$ for all $t_{R} >0$, and the inequality
$$ \big| \mathfrak{C}_{\bf E}^{\star}(t_{R},x) - \mathfrak{C}_{\bf E}^{\star}(t_{R},y) \big| \leq e^{-(\tilde{r}  - \Phi_{Y}^{(1)}(1))t_{R}} | x-y | , \hspace{2em} \forall x,y \in (0,\infty) .$$
\noindent Therefore, we are left with the proof of Equations (\ref{OIDE1}), (\ref{OIDE2}). Here, we start by re-considering the $\tilde{r}$-killed version of $(E_{t})_{t \geq 0}$, $(\bar{E}_{t})_{t \geq 0}$, i.e.~the process whose transition probabilities are given by
\begin{equation}
\mathbb{Q}_{x}^{(1)}\big(\bar{E}_{t} \in A \big ) = \mathbb{E}^{\mathbb{Q}^{(1)}}_{x} \left[ e^{-\tilde{r}t} \, \mathds{1}_{A}(E_{t}) \right],
\end{equation}
\noindent and identify, without loss of generality, its cemetery state with $\partial \equiv 0$. We then re-express $\mathfrak{C}_{\bf E}^{R,\star}(\cdot)$ as solution to an optimal stopping problem: We view the stochastic illiquidity horizon $T_{R}$ as jump time of a corresponding Poisson process\footnote{Our assumptions on $T_{R}$ clearly imply that the Poisson process is independent of $(\bar{E}_{t})_{t \geq 0}$.}~$(N_{t})_{t \geq 0}$ with intensity $\vartheta > 0$ and consider, for any $z = (n,x)\in \mathbb{N}_{0} \times [0,\infty)$, the (strong) Markov process $(Z_{t})_{t \geq 0}$ defined by means of $Z_{t} := (n+N_{t}, \bar{E}_{t})$, $\bar{E}_{0} = x$, on the state domain $\mathcal{D} := \mathbb{N}_{0} \times [0,\infty)$. Then, $\mathfrak{C}_{\bf E}^{R,\star}(\cdot)$ can be equivalently written as
\begin{equation}
\mathfrak{C}_{\bf E}^{R,\star}(x) = \widetilde{V}_{E}\big((0,x)\big),
\label{RUApre}
\end{equation}
\noindent where, for $z = (n,x) \in \mathcal{D}$, the value function $\widetilde{V}_{E}(\cdot)$ is defined, under the measure $\mathbb{Q}^{(1),Z}_{z}$ having initial distribution $Z_{0} = z$, by
\begin{equation}
\widetilde{V}_{E}(z) := \mathbb{E}_{z}^{\mathbb{Q}^{(1),Z}} \left[ G\big(Z_{\tau_{\mathcal{S}}}\big)\right], \hspace{1.5em} G(z) := (x-1)^{+},
\end{equation}
\noindent and $\tau_{\mathcal{S}} := \inf \{t \geq 0: \, Z_{t} \in \mathcal{S} \}$, $\mathcal{S} := \big(\mathbb{N} \times (0,\infty) \big) \cup \big( \mathbb{N}_{0} \times \{ 0 \} \big)$, is a stopping time that is $\mathbb{Q}^{(1),Z}_{z}$-almost surely finite for any $z = (n,x)$.\footnote{The finiteness of this stopping time directly follows from the properties (e.g.~finiteness of the first moment) of the exponential distribution of any intensity $\vartheta >0$.} Furthermore, the stopping domain $\mathcal{S}$ forms (under an appropriate product-metric) a closed set in $\mathcal{D}$.\footnote{We note that several choices of a product-metric on $\mathcal{D}$ give the closedness of the set $\mathcal{S}$. In particular, one may choose on $\mathbb{N}_{0}$ the following metric
$$ d_{\mathbb{N}_{0}}(m,n) := \left \{ \begin{array}{lc}
1 + |2^{-m} - 2^{-n} |, & m\neq n, \\
0, & m =n,
\end{array} \right. $$
\noindent and consider the product-metric on $\mathcal{D}$ obtained by combining $d_{\mathbb{N}_{0}}(\cdot,\cdot)$ on $\mathbb{N}_{0}$ with the Euclidean metric on $[0,\infty)$. \label{GENLIQfootnoteMETRIC}}~Therefore, standard arguments based on the strong Markov property of $(Z_{t})_{t \geq 0}$ (cf.~\cite{pe06}) imply that $\widetilde{V}_{E}(\cdot)$ solves the following problem
\begin{align}
\mathcal{A}_{Z} \widetilde{V}_{E}(z) & = 0, \hspace{2em} \mbox{on} \;  \mathcal{D} \setminus \mathcal{S}, \\
\widetilde{V}_{E}(z) & = G(z), \hspace{1.5em} \mbox{on} \;  \mathcal{S},
\end{align}
\noindent where $\mathcal{A}_{Z}$ denotes the infinitesimal generator of the process $(Z_{t})_{t \geq 0}$. To complete the proof, it therefore suffices to note that (for any suitable function $V:\mathcal{D} \rightarrow \mathbb{R}$) the infinitesimal generator $\mathcal{A}_{Z}$ can be re-expressed as
\begin{align}
\mathcal{A}_{Z} V\big((n,x)\big) & = \mathcal{A}_{N}^{n} V\big((n,x) \big) + \mathcal{A}_{\bar{E}}^{x} V \big((n,x) \big) \nonumber \\
& = \vartheta \left( V\big( (n+1,x) \big) - V\big( (n,x) \big) \right) + \mathcal{A}_{E}^{x} V\big( (n,x) \big) - \tilde{r} V\big( (n,x) \big) , \label{IgEn}
\end{align}
\noindent where $\mathcal{A}_{N}$ denotes the infinitesimal generator of the Poisson process $(N_{t})_{ t \geq 0}$ and the notation $\mathcal{A}_{N}^{n}$, $\mathcal{A}_{\bar{E}}^{x}$, and $\mathcal{A}_{E}^{x}$ is used to indicate that the generators are applied to $n$ and $x$ respectively. Indeed, recovering $\mathfrak{C}_{\bf E}^{R,\star}(\cdot)$ via (\ref{RUApre}) while noting Relation (\ref{IgEn}) and the fact that for any $x \in [0,\infty)$ we have
\begin{equation}
\widetilde{V}_{E}\big((1,x)\big) = G\big((1,x)\big) = (x-1)^{+} 
\end{equation}
\noindent finally gives the claim.
\end{proof}

\begin{proof}[\bf Proof of Proposition \ref{Prop5}]
\noindent To start, we note that, under $\tilde{r} \leq \Phi_{Y}^{(1)}(1) $, the American-type switching option $\mathfrak{C}_{\bf A}^{R,\star}(\cdot)$ reduces to its European counterpart $\mathfrak{C}_{\bf E}^{R,\star}(\cdot)$. As earlier, this is a direct consequence of the fact that the process $\left( e^{-\tilde{r}t} E_{t}\right)_{t \geq 0} $ then becomes a ($\mathbb{Q}^{(1)}$-)submartingale. In this case, the result directly follows via Proposition~\ref{Prop4}, with $\mathfrak{b}_{s}^{R} = \infty$. \vspace{1em} \\
\noindent For $\tilde{r} > \Phi_{Y}^{(1)}(1)$, we first note that Theorem 1 in \cite{mo02} implies the existence of a finite optimal stopping boundary $\mathfrak{b}_{s}^{R} >0$. Indeed, this follows by combining Lemma \ref{lemma2} with the fact that, with
$$ \mathfrak{C}_{\bf A}^{\infty,\star}(x) := \sup \limits_{\tau \in \mathfrak{T}_{[0,\infty)}} \mathfrak{C}^{\star}(\tau,x),$$ 
$$ \big \{x \in [0,\infty): \, \mathfrak{C}_{\bf A}^{\infty,\star}(x) = (x-1)^{+} \big \} \; \subseteq \; \big \{x \in [0,\infty): \, \mathfrak{C}_{\bf A}^{R,\star}(x) = (x-1)^{+} \big \}$$
\noindent and by arguing as in Section \ref{sec22}.\ref{BectionIII}~Therefore, by viewing the stochastic illiquidity horizon $T_{R}$ as jump time of a corresponding Poisson process $(N_{t})_{t \geq 0}$ with intensity $\vartheta>0$, we can re-express our optimal stopping problem in the following form: We consider, for any $z = (n,x)\in \mathbb{N}_{0} \times [0,\infty)$, the (strong) Markov process $(Z_{t})_{t \geq 0}$ defined by means of $Z_{t} := (n+N_{t}, \bar{E}_{t})$, $\bar{E}_{0} = x$, on the state domain $\mathcal{D} := \mathbb{N}_{0} \times [0,\infty)$ and identify again its cemetery state with $\partial \equiv 0$. Then, we note that
\begin{equation}
\mathfrak{C}_{\bf A}^{R,\star}(x) = \widetilde{V}_{A}\big((0,x)\big),
\label{RUApreAME}
\end{equation}
\noindent where, for $ z =(x,n) \in \mathcal{D}$, the value function $\widetilde{V}_{A}(\cdot)$ is defined, under the measure $\mathbb{Q}^{(1),Z}_{z}$ having initial distribution $Z_{0} = z$, by
\begin{equation}
\widetilde{V}_{A}(z) := \mathbb{E}_{z}^{\mathbb{Q}^{(1),Z}} \left[ G\big(Z_{\tau_{\mathcal{S}}}\big)\right], \hspace{1.5em} G(z) := (x-1)^{+},
\end{equation}
\noindent and $\tau_{\mathcal{S}} := \inf \{t \geq 0: \, Z_{t} \in \mathcal{S} \}$, $\mathcal{S} := \big(\mathbb{N} \times (0,\infty)\big) \cup \big( \mathbb{N}_{0} \times \{ 0 \} \big) \cup \big( \{0 \} \times [\mathfrak{b}_{s}^{R},\infty) \big)$ is a stopping time that is $\mathbb{Q}^{(1),Z}_{z}$-almost surely finite for any $z = (n,x)$. Furthermore, the stopping domain $\mathcal{S}$ forms (under an appropriate product-metric) a closed set in~$\mathcal{D}$.\footnote{As earlier, this property can be obtained under the product-metric considered in Footnote~\ref{GENLIQfootnoteMETRIC}.}~Therefore, standard arguments based on the strong Markov property of $(Z_{t})_{t \geq 0}$ (cf.~\cite{pe06}) imply that $\widetilde{V}_{A}(\cdot)$ solves the following problem
\begin{align}
\mathcal{A}_{Z} \widetilde{V}_{A}(z) & = 0, \hspace{2em} \mbox{on} \; \mathcal{D} \setminus \mathcal{S}, \\
\widetilde{V}_{A}(z) & = G(z), \hspace{1.5em} \mbox{on} \; \mathcal{S},
\end{align}
\noindent where $\mathcal{A}_{Z}$ denotes the infinitesimal generator of the process $(Z_{t})_{t \geq 0}$. To complete the proof, we therefore argue as in the proof of Proposition \ref{Prop4}, i.e.~we recover $\mathfrak{C}_{\bf A}^{R,\star}(\cdot)$ via (\ref{RUApreAME}) and combine Relation (\ref{IgEn}) with the fact that for any $x \in [0,\infty)$ we have
\begin{equation}
\widetilde{V}_{A}\big( (1,x) \big) = G \big( (1,x) \big) = (x-1)^{+}.
\end{equation}
\noindent Since Equation (\ref{OIDEinI}) is naturally satisfied, this leads to the required problem. The continuity of the function $x \mapsto \mathfrak{C}_{\bf A}^{R,\star}(\cdot)$ directly follows from its convexity (cf.~Lemma \ref{lemma2}). Therefore, the proof is complete.
\end{proof}

\subsection*{Appendix D}
\noindent In this appendix, we briefly derive a semi-analytical solution to the free-boundary problem of Proposition~\ref{prop6}, when the dynamics of $(S_{t})_{t \geq 0}$ and $(E_{t})_{t \geq 0}$ are given by (\ref{salope}) and (\ref{eq2}), (\ref{Dzzyn1}) and assuming non-positive jumps, i.e.~$\varphi \leq 0$. This is used to obtain numerical results in Section \ref{numres}.\ref{NUAp} \vspace{1em} \\
\noindent To start, we first note that, under the given dynamics and with $\tilde{b} := b + \rho \sigma_{X} \sigma$, the free-boundary problem reads:
\begin{itemize} \setlength \itemsep{-0.5em}
\item[1.] If $\tilde{r} \leq \tilde{b} $, the (absolute) tradeability premium $\mathfrak{L}^{R,\star}(\cdot)$ satisfies
$$ \mathfrak{L}^{R,\star}(x) = 0, \hspace{2em} \forall x \in [0,\infty) .$$
\item[2.] If $\tilde{r} > \tilde{b} $, the pair $\big(\mathfrak{L}^{R,\star}(\cdot),\mathfrak{b}_{s}^{R}\big)$ solves the following free-boundary problem:
\begin{equation}
\label{ZZZ1}
\frac{1}{2}\sigma^{2}x^{2} \partial_{x}^{2} \mathfrak{L}^{R,\star} (x) + \big( \tilde{b} -\lambda (e^{\varphi}-1) \big)x \partial_{x}\mathfrak{L}^{R,\star} (x) + \lambda \left( \mathfrak{L}^{R,\star} (xe^{\varphi}) -\mathfrak{L}^{R,\star} (x) \right)  - (\tilde{r} + \vartheta) \mathfrak{L}^{R,\star}(x) = 0 ,
\end{equation}
\noindent on $x \in (0,\mathfrak{b}_{s}^{R})$ and subject to the boundary conditions
\begin{align}
\mathfrak{L}^{R,\star}(\mathfrak{b}_{s}^{R}) & = \mathfrak{b}_{s}^{R} - 1 - \mathfrak{C}_{{\bf E}}^{R,\star}(\mathfrak{b}_{s}^{R}), \label{ZZZ2}\\
\partial_{x} \mathfrak{L}^{R,\star}(\mathfrak{b}_{s}^R) & = 1 - \partial_{x} \mathfrak{C}_{{\bf E}}^{R,\star} (\mathfrak{b}_{s}^R),  \label{ZZZ3} \\
\mathfrak{L}^{R,\star}(0) & = 0. \label{ZZZ4}
\end{align} 
\end{itemize}
\noindent Therefore, it is sufficient to focus on the non-trivial case, i.e.~we assume from now on that $\tilde{r} > \tilde{b} $.
\noindent Here, we decompose the full domain $[0,\infty)$ into two intervals, $I_{1} := [0,\mathfrak{b}_{s}^{R})$ and $I_{2} := [\mathfrak{b}_{s}^{R}, \infty)$, derive solutions $V_{1}(\cdot)$ and $V_{2}(\cdot)$ on these respective domains and combine them to recover $\mathfrak{L}^{R,\star}(\cdot)$ via 
\begin{equation}
\label{RECov}
\mathfrak{L}^{R,\star} (x) = \left \{ \begin{array}{cc}
V_{1}(x), & x \in I_{1}, \\
V_{2}(x), & x \in I_{2}. \\
\end{array} \right. 
\end{equation}
\noindent We now turn to the derivation of these solutions. First, it is clear that, on $I_{2}$, $V_{2}(x) = x - 1 - \mathfrak{C}_{{\bf E}}^{R,\star}(x)$ must hold. Hence, we only need to derive an expression for $V_{1}(\cdot)$. Here, we start by noting that $\Phi_{Y}^{(1)}(\theta)$, the Laplace exponent of $(Y_{t})_{t \geq 0}$ under $\mathbb{Q}^{(1)}$, is well-defined for all $\theta \in \mathbb{R}$. Furthermore, it can be easily seen that $\theta \mapsto \Phi_{Y}^{(1)}(\theta)$ is convex and satisfies $\Phi_{Y}^{(1)}(0)=0$ and $\lim \limits_{ |\theta| \rightarrow \infty} \Phi_{Y}^{(1)}(\theta) \; = \; \infty $. Consequently, the equation $ \Phi_{Y}^{(1)}(\theta) = y$ has, for any $y >0$, two solutions, a positive and a negative root. In the sequel, we denote by $\Big(\Phi_{Y}^{(1)} \Big)^{-1,+} \big(y\big)$ its positive root and by $\Big( \Phi_{Y}^{(1)} \Big) ^{-1, -} \big(y\big)$ its negative root. Using this notation, one easily shows that, under $\varphi \leq 0$, the general solution of the homogeneous equation (\ref{ZZZ1}) on $I_{1}$ takes the form
\begin{equation}
V_{1}(x) = c_{1}^{+} x^{\gamma_{+}} + c_{1}^{-} x^{\gamma_{-}} ,
\end{equation}
\noindent where $\gamma_{+} = \Big(\Phi_{Y}^{(1)}\Big)^{-1,+} \big(\tilde{r} + \vartheta\big)$, $\gamma_{-} = \Big(\Phi_{Y}^{(1)}\Big)^{-1,-} \big(\tilde{r} + \vartheta\big)$ and $c_{1}^{+}$, $c_{1}^{-}$ are constants to be determined. Therefore, to conclude, we only need to derive $c_{1}^{+}$, $c_{1}^{-}$ and $\mathfrak{b}_{s}^{R}$ and make use of Conditions (\ref{ZZZ2})-(\ref{ZZZ4}). First, we note that (\ref{ZZZ4}) implies that $c_{1}^{-} \equiv 0$. Additionally, Conditions (\ref{ZZZ2}) and (\ref{ZZZ3}) give the following equations:
\begin{align}
c_{1}^{+} \big( \mathfrak{b}_{s}^{R} \big)^{\gamma_{+}}  & = \mathfrak{b}_{s}^{R} -1 - \mathfrak{C}_{\bf E}^{R,\star}(\mathfrak{b}_{s}^{R}), \label{SysEq1}\\
\gamma_{+} c_{1}^{+} \big( \mathfrak{b}_{s}^{R} \big)^{ \gamma_{+} -1}  & =  1- \partial_{x} \mathfrak{C}_{\bf E}^{R,\star}(\mathfrak{b}_{s}^{R}) . \label{SysEq2}
\end{align} 
\noindent The latter system can now be solved to obtain $c_{1}^{+}$ and $\mathfrak{b}_{s}^{R}$. First, rewritting (\ref{SysEq2}) gives that
\begin{equation}
c_{1}^{+} = \frac{\big(\mathfrak{b}_{s}^{R} \big)^{1-\gamma_{+}}}{\gamma_{+}} \big(1- \partial_{x} \mathfrak{C}_{\bf E}^{R,\star}(\mathfrak{b}_{s}^{R})  \big).
\end{equation}
\noindent Then, inserting this result in (\ref{SysEq1}) leads to the following non-linear equation in $\mathfrak{b}_{s}^{R}$:
\begin{equation}
\mathfrak{b}_{s}^{R} = 1 + \mathfrak{C}_{\bf E}^{R,\star}(\mathfrak{b}_{s}^{R}) + \frac{\mathfrak{b}_{s}^{R} }{\gamma_{+}} \big(1- \partial_{x} \mathfrak{C}_{\bf E}^{R,\star}(\mathfrak{b}_{s}^{R}) \big).
\end{equation}
\noindent Therefore, solving the latter equation for $\mathfrak{b}_{s}^{R}$ allows us to subsequently derive $c_{1}^{+}$. This finally allows us to recover the tradeability premium $\mathfrak{L}^{R,\star}(\cdot)$ via (\ref{RECov}). \vspace{2em} \\
% \acknow{The author thanks Walter Farkas and Giovanni Barone-Adesi for their advice, as well as Jérôme Detemple and Sander Willems for useful comments.} \vspace{-0.5em} \\

\end{document}